\documentclass[a4paper,12pt,twoside]{article}
\usepackage{psfig,epsfig,graphicx}
\usepackage{amssymb,amsmath,bbm,rotating,subfigure,here,citesort}
\newcommand{\befig}{\begin{figure}}
\newcommand{\efig}{\end{figure}}
\newcommand{\betab}{\begin{table}}
\newcommand{\etab}{\end{table}}
\newcommand{\barray}{\begin{array}}
\newcommand{\earray}{\end{array}}
\newcommand{\be}{\begin{equation}}
\newcommand{\ee}{\end{equation}}
\newcommand{\bea}{\begin{eqnarray}}
\newcommand{\eea}{\end{eqnarray}}
\newcommand{\benn}{\begin{displaymath}}
\newcommand{\eenn}{\end{displaymath}}
\newcommand{\beann}{\begin{eqnarray*}}
\newcommand{\eeann}{\end{eqnarray*}}
\newcommand{\inv}{\frac{1}}
\newcommand{\gtsim}{\gtrsim}
\newcommand{\ltsim}{\lesssim}
\newcommand{\Order}{{\cal O}}   
\newcommand{\fm}{\mbox{fm}}

\newcommand{\GeV}{\mbox{GeV}}

\newcommand{\alphaS}{\alpha_s}

\newcommand{\SVM}{SVM}
\newcommand{\pert}{P}
\newcommand{\nprt}{N\!P}
\newcommand{\glueball}{G\!B}
\newcommand{\WW}{Wegner-Wilson}
\newcommand{\pot}{\mbox{\scriptsize pot}}
\newcommand{\self}{\mbox{\scriptsize self}}
\newcommand{\actiondensity}{\mathit{s}}

\newcommand{\qbar}{{\bar{q}}}

\newcommand{\G}{{\cal G}}       
\newcommand{\GG}{\hat{\cal{G}}} 
\newcommand{\Identity}{{1\!\rm l}}

\newcommand{\Pc}{{\cal P}}      
\newcommand{\Ps}{{\cal P}_S}    
\newcommand{\Tr}{\mbox{Tr}}             
\newcommand{\rTr}{\Tr_{r}}        
\newcommand{\ronexrtwoTr}{\Tr_{r_1 \otimes r_2}}  
\newcommand{\nrTr}{\tilde{\Tr}_{r}}             
\newcommand{\nroneTr}{\tilde{\Tr}_{r_1}}              
\newcommand{\nrtwoTr}{\tilde{\Tr}_{r_2}}              
\newcommand{\nronexrtwoTr}{\tilde{\Tr}_{r_1 \otimes r_2}}
\newcommand{\roneIdentity}{\Identity_{r_1}}
\newcommand{\rtwoIdentity}{\Identity_{r_2}}
\newcommand{\ronexrtwoIdentity}{\Identity_{r_1 \otimes r_2}}
\newcommand{\Projector}{\mbox{P}} 
\newcommand{\tensor}{t_{r_1 \otimes r_2}} 
\newcommand{\fundamental}{\mbox{\scriptsize N}_c}
\newcommand{\adjoint}{\mbox{\scriptsize N}_c^2\!-\!1}
\newcommand{\Fundamental}{\mbox{N}_c}
\newcommand{\Adjoint}{\mbox{N}_c^2\!-\!1}
\begin{document}
\pagenumbering{arabic}
%
%
%
\pagestyle{empty}
%
%
\title{ \vspace*{-2.cm} {\normalsize\rightline{HD-THEP-02-22}}
  {\normalsize\rightline{hep-ph/0211287}}  \vspace*{0.cm} 
{\Large
\bf  
\boldmath 
Confining QCD Strings, Casimir Scaling,
and a Euclidean Approach to High-Energy Scattering
}}
%
%
\author{}
\date{} \maketitle
\vspace*{-2.5cm}
\begin{center}
\renewcommand{\thefootnote}{\alph{footnote}}
{\large 
A.~I.~Shoshi$^{1,}$\footnote{shoshi@tphys.uni-heidelberg.de},
F.~D.~Steffen$^{1,}$\footnote{Frank.D.Steffen@thphys.uni-heidelberg.de},
H.~G.~Dosch$^{1,}$\footnote{H.G.Dosch@thphys.uni-heidelberg.de}, and
H.~J.~Pirner$^{1,2,}$\footnote{pir@tphys.uni-heidelberg.de}}

{\it $^1$Institut f\"ur Theoretische Physik, Universit\"at Heidelberg,\\
Philosophenweg 16 {\sl \&}\,19, D-69120 Heidelberg, Germany}

{\it $^2$Max-Planck-Institut f\"ur Kernphysik, Postfach 103980, \\
D-69029 Heidelberg, Germany}
\end{center}
%
%
\begin{abstract}
  
  We compute the chromo-field distributions of static color dipoles in
  the fundamental and adjoint representation of $SU(N_c)$ in the
  loop-loop correlation model and find Casimir scaling in agreement
  with recent lattice results. Our model combines perturbative gluon
  exchange with the non-perturbative stochastic vacuum model which
  leads to confinement of the color charges in the dipole via a string
  of color fields. We compute the energy stored in the confining
  string and use low-energy theorems to show consistency with the
  static quark-antiquark potential. We generalize Meggiolaro's
  analytic continuation from parton-parton to gauge-invariant
  dipole-dipole scattering and obtain a Euclidean approach to
  high-energy scattering that allows us in principle to calculate
  $S$-matrix elements directly in lattice simulations of QCD. We apply
  this approach and compute the $S$-matrix element for high-energy
  dipole-dipole scattering with the presented Euclidean loop-loop
  correlation model. The result confirms the analytic continuation of
  the gluon field strength correlator used in all earlier applications
  of the stochastic vacuum model to high-energy scattering.

\vspace{1.cm}


\medskip

\noindent
{\it PACS numbers}: 
%
%
12.38.-t,  
%
%
12.38.Lg,  
%
%
11.80.Fv  
%
%
\end{abstract}
%
%
\newpage
\pagestyle{empty}
\tableofcontents
\addtocontents{toc}{\protect\enlargethispage{1.cm}}
%
%
\newpage
\pagenumbering{arabic}
\pagestyle{plain}
\setcounter{footnote}{0}
%
\makeatletter
\@addtoreset{equation}{section}
\makeatother
\renewcommand{\theequation}{\thesection.\arabic{equation}}
%
%
\section{Introduction}
\label{Sec_Introduction}


The structure of the {\em QCD vacuum} is responsible for color
confinement, spontaneous chiral symmetry breaking, and dynamical mass
generation~\cite{Smilga:ck}. Hadronic reactions are expected to show
further manifestations of a non-trivial QCD vacuum. It is indeed a key
issue to unravel the effects of confinement and topologically
non-trivial gauge field configurations (such as instantons) on such
reactions~\cite{Dosch:JHW2002,Ringwald:2002iy}. Moreover, it would be
a significant breakthrough to understand the size, behavior and growth
of hadronic cross sections with increasing c.m.~energy from the QCD
Lagrangian.


{\em Lattice QCD} is the principal theoretical tool to study the QCD
vacuum from first principles. Numerical simulations of QCD on
Euclidean lattices give strong evidence for color confinement and
spontaneous chiral symmetry breaking and describe dynamical mass
generation from the QCD
Lagrangian~\cite{Rothe:1997kp,Kronfeld:2002pi,Lattice2002}.
However, since lattice QCD is limited to the Euclidean formulation of
QCD, it cannot be applied in Minkowski space-time to simulate
high-energy reactions in which particles are moving near the light
cone. Furthermore, although lattice investigations have significantly
enhanced our understanding of non-perturbative phenomena and
particularly confinement, one can quote the concluding sentence of
Greensite's recent review~\cite{Greensite:2003bk}: ``The confinement
problem is still open, and remains a major intellectual challenge in
our field.'' Here (phenomenological) models that allow analytic
calculations are important as they provide valuable complementary
insights.


In this work we introduce the Euclidean version of the {\em loop-loop
  correlation model} (LLCM) which has been developed in Minkowski
space-time to describe high-energy reactions of hadrons and
photons~\cite{Shoshi:2002in} on the basis of a functional integral
approach~\cite{Nachtmann:1991ua+X,Nachtmann:ed.kt,Kramer:1990tr,Dosch:1994ym,Dosch:RioLecture}.
The central element in our approach is the gauge-invariant
Wegner-Wilson loop~\cite{Wegner:1971qt,Wilson:1974sk}: The physical
quantities considered are obtained from the vacuum expectation value
(VEV) of one {\WW} loop, $\langle W_{r}[C] \rangle$, and the
correlation of two {\WW} loops, $\langle W_{r_1}[C_1] W_{r_2}[C_2]
\rangle$. Here $r_{(i)}$ indicates the $SU(N_c)$ representation of the
{\WW} loops which we keep as general as possible. We express $\langle
W_{r}[C] \rangle$ and $\langle W_{r_1}[C_1] W_{r_2}[C_2] \rangle$ in
terms of the gauge-invariant bilocal gluon field strength correlator
integrated over {\em minimal surfaces} by using the non-Abelian Stokes
theorem~\cite{Arefeva:dp+X} and a matrix cumulant
expansion~\cite{VAN_KAMPEN_1974_1976+X} in the Gaussian approximation.
The latter approximation relies on the assumption of a Gaussian
dominance in the correlations of gauge-invariant non-local gluon field
strengths, i.e.\ the dominance of the bilocal correlator over higher
ones, and is supported by lattice investigations~\cite{Bali:1998aj}.
In our model this Gaussian approximation leads directly to the Casimir
scaling of the static quark-antiquark potential which for $SU(3)$ has
clearly been confirmed on the
lattice~\cite{Deldar:1999vi,Bali:2000un}. We decompose the
gauge-invariant bilocal gluon field strength correlator into a
perturbative and a non-perturbative component: The {\em stochastic
  vacuum model} (\SVM)~\cite{Dosch:1987sk+X} is used for the
non-perturbative low-frequency background field and {\em perturbative
  gluon exchange} for the additional high-frequency contributions.
This combination allows us to describe long and short distance
correlations in agreement with lattice calculations of the gluon field
strength
correlator~\cite{DiGiacomo:1992df+X,D'Elia:1997ne,Bali:1998aj,Meggiolaro:1999yn}.
Moreover, it leads to a static quark-antiquark potential with color
Coulomb behavior for small and confining linear rise for large source
separations. We calculate the static quark-antiquark potential with
the LLCM parameters determined in fits to high-energy scattering
data~\cite{Shoshi:2002in} and find good agreement with lattice data.
We thus have one model that describes both static hadronic properties
and high-energy reactions of hadrons and photons in good agreement
with experimental and lattice QCD data.


We apply the LLCM to compute the {\em chromo-electric fields}
generated by a static color dipole in the fundamental and adjoint
representation of $SU(N_c)$. The non-perturbative SVM component
describes the formation of a color flux tube that confines the two
color sources in the dipole~\cite{Rueter:1994cn} while the
perturbative component leads to color Coulomb fields. We find {\em
  Casimir scaling} for both the perturbative and non-perturbative
contributions to the chromo-electric fields again as a direct
consequence of the Gaussian appoximation in the gluon field strengths.
The mean squared radius of the confining QCD string is calculated as a
function of the dipole size. Transverse and longitudinal energy
density profiles are provided to study the interplay between
perturbative and non-perturbative physics for different dipole sizes.
The transition from perturbative to string behavior is found at source
separations of about $0.5\,\fm$ in agreement with the recent results
of L\"uscher and Weisz~\cite{Luscher:2002qv}.


The {\em low-energy theorems}, known in lattice QCD as Michael sum
rules~\cite{Michael:1986yi}, relate the energy and action stored in
the chromo-fields of a static color dipole to the corresponding ground
state energy. The Michael sum rules, however, are incomplete in their
original form~\cite{Michael:1986yi}. We present the complete energy
and action sum rules~\cite{Rothe:1995hu+X,Michael:1995pv,Green:1996be}
in continuum theory taking into account the contributions to the
action sum rule found in~\cite{Dosch:1995fz} and the trace anomaly
contribution to the energy sum rule~\cite{Rothe:1995hu+X}. Using these
low-energy theorems, we compare the energy and action stored in the
confining string with the confining part of the static quark-antiquark
potential. This allows us to confirm consistency of the model results
and to determine the values of the Callan-Symanzik $\beta$ function
and the strong coupling $\alphaS$ at the renormalization scale at
which the non-perturbative SVM component is working. The values
obtained for $\beta$ and $\alphaS$ are compared to model independent
QCD results for the Callan-Symanzik function. Earlier investigations
along these lines have been incomplete since only the contribution
from the traceless part of the energy-momentum tensor has been
considered in the energy sum rule.


To study the effect of the confining QCD string examined in Euclidean
space-time on high-energy reactions in Minkowski space-time, an {\em
  analytic continuation} from Euclidean to Minkowski space-time is
needed. For investigations of high-energy reactions in our Euclidean
model, the gauge-invariant bilocal gluon field strength correlator can
be analytically continued from Euclidean to Minkowski space-time.
This analytic continuation has been introduced for applications of the
SVM to high-energy
reactions~\cite{Kramer:1990tr,Dosch:1994ym,Dosch:RioLecture} and is
used in our Minkowskian applications of the
LLCM~\cite{Shoshi:2002in,Shoshi:2002ri,Shoshi:2002fq,Shoshi:2002mt}.
Recently, an alternative analytic continuation for parton-parton
scattering has been established in the perturbative context by
Meggiolaro~\cite{Meggiolaro:1996hf+X}. This analytic continuation has
already been used to access high-energy scattering from the
supergravity side of the AdS/CFT correspondence~\cite{Janik:2000zk+X},
which requires a positive definite metric in the definition of the
minimal surface~\cite{Rho:1999jm}, and to examine the effect of
instantons on high-energy scattering~\cite{Shuryak:2000df+X}.


In this work we generalize Meggiolaro's analytic
continuation~\cite{Meggiolaro:1996hf+X} from parton-parton to
gauge-invariant {\em dipole-dipole scattering} such that $S$-matrix
elements for high-energy reactions can be computed from configurations
of {\WW} loops in Euclidean space-time and with {\em Euclidean}
functional integrals. This shows how one can access high-energy
reactions directly in lattice QCD. First attempts in this direction
have already been carried out but only very few signals could be
extracted, while most of the data was dominated by
noise~\cite{DiGiacomo:2002PC}. We apply this approach to compute the
scattering of dipoles at high-energy in the Euclidean LLCM. We recover
exactly the result derived with the analytic continuation of the gluon
field strength correlator~\cite{Shoshi:2002in}. This confirms the
analytic continuation used in all earlier applications of the
stochastic vacuum model to high-energy
scattering~\cite{Kramer:1990tr,Dosch:1994ym,Dosch:RioLecture,Rueter:1996yb,Dosch:1998nw,Rueter:1998qy+X,Rueter:1998up,Dosch:1997ss,Berger:1999gu,Donnachie:2000kp+X,Dosch:2001jg,Kulzinger:2002iu}
including the Minkowskian applications of the
LLCM~\cite{Shoshi:2002in,Shoshi:2002ri,Shoshi:2002fq,Shoshi:2002mt}.
In fact, the $S$-matrix element obtained has already been used as the
basis for a unified description of hadronic high-energy
reactions~\cite{Shoshi:2002in}, to study saturation effects in
hadronic cross
sections~\cite{Shoshi:2002in,Shoshi:2002ri,Shoshi:2002mt}, and to
investigate manifestations of the confining QCD string in high-energy
reactions of photons and hadrons~\cite{Shoshi:2002fq}.


The outline of the paper is as follows. In Sec.~\ref{Sec_The_Model}
the LLCM is introduced in its Euclidean version and the general
computations of $\langle W_{r}[C] \rangle$ and $\langle W_{r_1}[C_1]
W_{r_2}[C_2] \rangle$ are presented. Based on these evaluations, we
compute the potential of a static color dipole in
Sec.~\ref{Sec_Static_Potential} and the associated chromo-field
distributions in Sec.~\ref{Sec_Flux_Tube} with emphasis on Casimir
scaling and the interplay between perturbative color Coulomb behavior
and non-perturbative formation of the confining QCD string.  In
Sec.~\ref{Sec_Low_Energy_Theorems} low-energy theorems are discussed
and used to show consistency of the model results and to determine the
values of $\beta$ and $\alphaS$ at the renormalization scale at which
the non-perturbative SVM component is working. In
Sec.~\ref{Sec_DD_Scattering} the Euclidean approach to high-energy
scattering is presented and applied to compute high-energy
dipole-dipole scattering in our Euclidean model. In the Appendixes we
review the derivation of the non-Abelian Stokes theorem, give
parametrizations of the loops and the minimal surfaces, and provide
the detailed computations for the results in the main text.

%
%
%
%
\section{The Loop-Loop Correlation Model}
\label{Sec_The_Model}

In this section the vacuum expectation value of one {\WW} loop and the
correlation of two {\WW} loops are computed for arbitrary loop
geometries within a Gaussian approximation in the gluon field
strengths. The results are applied in the following sections. We
describe our model for the QCD vacuum in which the stochastic vacuum
model~\cite{Dosch:1987sk+X} is used for the non-perturbative
low-frequency background field (long-distance correlations) and
perturbative gluon exchange for the additional high-frequency
contributions (short-distance correlations).

\subsection{Vacuum Expectation Value of one {\WW} Loop}
\label{Sec_<W[C]>}

A crucial quantity in gauge theories is the Wegner-Wilson loop
operator~\cite{Wegner:1971qt,Wilson:1974sk}
\be
        W_r[C] = 
        \nrTr\,\Pc
        \exp\!\left[
        -i g \oint_{\scriptsize C} dZ_{\mu}\,\G_{\mu}^a(Z)\,t_r^a 
        \right]      
        \ .
\label{Eq_WW_loop}        
\ee
Concentrating on $SU(N_c)$ {\WW} loops, where $N_c$ is the number of
colors, the subscript $r$ indicates a representation of $SU(N_c)$,
$\nrTr = \rTr(\cdots)/\Tr \Identity_r$ is the normalized trace in the
corresponding color space with unit element $\Identity_r$, $g$ is the
strong coupling, and $\G_{\mu}(Z) = \G_{\mu}^a(Z) t_r^a$ represents
the gluon field with the $SU(N_c)$ group generators in the
corresponding representation, $t_r^a$, that demand the path ordering
indicated by $\Pc$ on the closed path $C$ in space-time. A
distinguishing theoretical feature of the {\WW} loop is its invariance
under local gauge transformations in color space. Therefore, it is the
basic object in lattice gauge
theories~\cite{Wegner:1971qt,Wilson:1974sk,Rothe:1997kp} and has been
considered as the fundamental building block for a gauge theory in
terms of gauge invariant variables~\cite{Migdal:1983gj}.
Phenomenologically, the {\WW} loop represents the phase factor
associated with the propagation of a very massive color source in the
representation $r$ of the gauge group $SU(N_c)$.

To compute the expectation value of the {\WW} loop~(\ref{Eq_WW_loop})
in the QCD vacuum
\be
        \Big\langle W_r[C] \Big\rangle_G
        = \Big\langle
        \nrTr\,\Pc
        \exp\!\left[-i\,g \oint_{\scriptsize C} dZ_{\mu}\,\G_{\mu}^a(Z)\,t_r^a \right]      
        \Big\rangle_G
        \ ,
\label{Eq_<W[C]>}
\ee
we transform the line integral over the loop $C$ into an integral over
the surface $S$ with $\partial S = C$ by applying the {\em non-Abelian
  Stokes theorem}~\cite{Arefeva:dp+X}
\be
        \Big\langle W_r[C] \Big\rangle_G
        = \Big\langle
        \nrTr\,\Ps
          \exp \left[-i\,\frac{g}{2} 
                \int_{S} \! d\sigma_{\mu\nu}(Z) 
                \G^a_{\mu\nu}(O,Z;C_{ZO})\,t_r^a 
          \right] 
        \Big\rangle_G
        \ ,
\label{Eq_Non-Abelian_Stokes_<W[C]>}
\ee
where $\Ps$ indicates surface ordering and $O$ is an arbitrary
reference point on the surface $S$. In
Eq.~(\ref{Eq_Non-Abelian_Stokes_<W[C]>}) the gluon field strength
tensor, $\G_{\mu\nu}(Z) = \G_{\mu\nu}^{a}(Z)\,t_r^{a}$, is parallel
transported to the reference point $O$ along the path $C_{ZO}$,
\be
        \G_{\mu\nu}(O,Z;C_{ZO}) 
        = \Phi_r(O,Z;C_{ZO}) \G_{\mu\nu}(Z) \Phi_r(O,Z;C_{ZO})^{-1}
        \ ,
\label{Eq_gluon_field_strength_tensor}
\ee
with the QCD Schwinger string
\be
        \Phi_r(O,Z;C_{ZO}) 
        = \Pc \exp 
        \left[-i\,g \int_{C_{ZO}} \!\!dZ_{\mu}\G^a_{\mu}(Z)\,t_r^a \right] 
        \ .
\label{Eq_parallel_transport}
\ee
A more detailed explanation of the non-Abelian Stokes theorem and the
associated surface ordering is given in
Appendix~\ref{Sec_Surface_Ordering}.

The QCD vacuum expectation value $\langle \ldots \rangle_G$ represents
functional integrals in which the functional integration over the
fermion fields has already been carried out as indicated by the
subscript $G$~\cite{Nachtmann:ed.kt}.  The model we use for the
evaluation of $\langle \ldots \rangle_G$ is based on the {\em quenched
  approximation} which does not allow string breaking through
dynamical quark-antiquark production. So far, it is not clear how to
introduce dynamical quarks into this model. One suggestion is
presented in Appendix A of Ref.~\cite{Nachtmann:ed.kt}.

Due to the linearity of the functional
integral, $\langle \nrTr \ldots \rangle = \nrTr \langle \ldots
\rangle$, we can write
\be
        \Big\langle W_r[C] \Big\rangle_G
        = \nrTr 
        \Big\langle
        \Ps \exp \left[-i\,\frac{g}{2} 
                \int_{S} \! d\sigma_{\mu\nu}(Z) 
                \G^a_{\mu\nu}(O,Z;C_{ZO})\,t_r^a 
          \right] 
        \Big\rangle_G
        \ .
\label{Eq_Tr_<W[C]>}
\ee
For the evaluation of~(\ref{Eq_Tr_<W[C]>}), a {\em matrix cumulant
  expansion} is used as explained in~\cite{Nachtmann:ed.kt} (cf.\ 
also~\cite{VAN_KAMPEN_1974_1976+X})
\bea
        && \Big\langle 
                \Ps \, \exp 
                \left[-i\,\frac{g}{2} 
                \int_{S} \! d\sigma(Z) \G(O,Z;C_{ZO})
                \right] 
           \Big\rangle_G \nonumber \\
        && 
        = \exp[\,\,\sum_{n=1}^{\infty}\frac{1}{n !}(-i\,\frac{g}{2})^n
        \int d\sigma(X_1)\cdots d\sigma(X_n)\,K_n(X_1,\cdots,X_n)]
        \ ,
\label{Eq_matrix_cumulant_expansion}
\eea
where space-time indices are suppressed to simplify notation. The
cumulants $K_n$ consist of expectation values of {\em ordered}
products of the non-commuting matrices $\G(O,Z;C_{ZO})$. The leading
matrix cumulants are
\bea
        K_1(X)       
        & = & \langle \G(O,X;C_X) \rangle_G, 
\label{Eq_K_1_matrix_cumulant}\\
        K_2(X_1,X_2) 
        & = & \langle \Ps
        [\G(O,X_1;C_{X_1})\G(O,X_2;C_{X_2})]\rangle_G 
        \nonumber\\
        &   & - \frac{1}{2}
        \left[\langle \G(O,X_1;C_{X_1})\rangle_G 
        \langle \G(O,X_2;C_{X_2})\rangle_G 
        + (1 \leftrightarrow 2)\,\right] \ .
\label{Eq_K_2_matrix_cumulant}
\eea
Since the vacuum does not prefer a specific color direction, $K_1$
vanishes and $K_2$ becomes
\be
        K_2(X_1,X_2) 
        = \langle\Ps [\G(O,X_1;C_{X_1})\G(O,X_2;C_{X_2})]\rangle_G
        \ .
\label{Eq_K_2_matrix_cumulant<-no_color_direction_preferred}
\ee
Now, we approximate the functional integral associated with the
expectation values $\langle \ldots \rangle_G$ as a {\em Gaussian
  integral} in the parallel transported gluon field
strength~(\ref{Eq_gluon_field_strength_tensor}). This Gaussian
approximation is supported by lattice
investigations~\cite{Bali:1998aj} that show a dominance of the bilocal
gauge-invariant gluon field strength correlator over higher-point
non-local correlators. As a consequence of the Gaussian approximation,
the cumulants factorize into two-point field correlators such that all
higher cumulants, $K_n$ with $n>2$, vanish.\footnote{We are going to
  use the cumulant expansion in the Gaussian approximation also for
  perturbative gluon exchange.  Here certainly the higher cumulants
  are non-zero.} Thus, $\langle W_r[C] \rangle_G$ can be expressed in
terms of $K_2$
\bea
&& \!\!\!\!\!\!\!\!\!\!\!
        \Big\langle W_r[C] \Big\rangle_G = 
        \nrTr 
        \exp\!\left[-\frac{g^2}{8} \!
          \int_{S} \! d\sigma_{\mu\nu}(X_1) \!
          \int_{S} \! d\sigma_{\rho\sigma}(X_2) 
        \right.
        \nonumber \\
&& \!\!\!\!\!\!\!\!\!\!\!
        \hphantom{\Big\langle W_r[C] \Big\rangle_G = \nrTr \exp}
        \left.
          \Big\langle \Ps 
          [\G^a_{\mu\nu}(O,X_1;C_{X_1 O})\,t_r^a\,\,
          \G^b_{\rho\sigma}(O,X_2;C_{X_2 O})\,t_r^b] 
          \Big\rangle_G 
        \right]
\label{Eq_matrix_cumulant_expansion_<W[C]>}
\eea
Due to the color neutrality of the vacuum, the gauge-invariant bilocal
gluon field strength correlator contains a $\delta$ function in
color space,
\be
        \Big\langle
        \frac{g^2}{4\pi^2}
        \left[\G^a_{\mu\nu}(O,X_1;C_{X_1 O})
        \G^b_{\rho\sigma}(O,X_2;C_{X_2 O})\right]
        \Big\rangle_G
        =: \inv{4}\delta^{ab} 
        F_{\mu\nu\rho\sigma}(X_1,X_2,O;C_{X_1 O},C_{X_2 O}) 
\label{Eq_Ansatz}
\ee
which makes the surface ordering $\Ps$
in~(\ref{Eq_matrix_cumulant_expansion_<W[C]>}) irrelevant. The tensor
$F_{\mu\nu\rho\sigma}$ will be specified in
Sec.~\ref{Sec_QCD_Components}.  With~(\ref{Eq_Ansatz}) and the
quadratic Casimir operator $C_2(r)$,
\be
        t_r^a\,t_r^a = t_r^2 = C_2(r)\,\Identity_r
        \ ,
\label{Eq_quadratic_Casimir_operator}
\ee
Eq.~(\ref{Eq_matrix_cumulant_expansion_<W[C]>}) reads
\be
        \Big\langle W_r[C] \Big\rangle_G
        = \nrTr 
        \exp\left[
        - \frac{C_2(r)}{2}\,\chi_{SS}\,\Identity_r
        \right] 
        = \exp \left[-\frac{C_2(r)}{2}\,\chi_{SS}\right] 
        \ ,
\label{Eq_final_result_<W[C]>}
\ee
where
\be
        \chi_{SS}
        :=  \frac{\pi^2}{4} 
        \int_{S} \! d\sigma_{\mu\nu}(X_1) 
        \int_{S} \! d\sigma_{\rho\sigma}(X_2)
        F_{\mu\nu\rho\sigma}(X_1,X_2,O;C_{X_1 O},C_{X_2 O}) 
        \ .
\label{Eq_chi_SS}        
\ee
In this rather general result~(\ref{Eq_final_result_<W[C]>}) obtained
directly from the color neutrality of the QCD vacuum and the Gaussian
approximation in the gluon field strengths, the more detailed aspects
of the QCD vacuum and the geometry of the considered {\WW} loop are
encoded in the function $\chi_{SS}$ which is computed in
Appendix~\ref{Sec_Chi_Computation} for a rectangular loop.

In explicit computations we use for $S$ the {\em minimal surface},
which is the planar surface spanned by the loop ($C = \partial S$)
that leads most naturally to Wilson's area law~\cite{Dosch:1987sk+X}.
Of course, the results should not depend on the choice of the surface.
In our model the perturbative and non-perturbative non-confining
components satisfy this requirement. The non-perturbative confining
component in $F_{\mu\nu\rho\sigma}$ depends on the choice of the
surface due to the Gaussian approximation and the associated
truncation of the cumulant expansion.  Since the minimal surface leads
to a static quark-antiquark potential that is in good agreement with
lattice data (see Sec.~\ref{Sec_Static_Potential}), we think that the
minimal surface reduces the contribution from higher cumulants.
Within bosonic string theory, our minimal surface represents the
world-sheet of a {\em rigid} string: Our model does not describe
fluctuations or excitations of the string and thus cannot reproduce
the L\"uscher term which has recently been confirmed by L\"uscher and
Weisz~\cite{Luscher:2002qv}.

\subsection{The Loop-Loop Correlation Function}
\label{Sec_<W[C_1]W[C_2]>}

The computation of the {\em loop-loop correlation function} $\langle
W_{r_1}[C_{1}] W_{r_2}[C_{2}] \rangle_G$ starts again with the
application of the {\em non-Abelian Stokes
  theorem}~\cite{Arefeva:dp+X} which allows us to transform the line
integrals over the loops $C_{1,2}$ into integrals over surfaces
$S_{1,2}$ with $\partial S_{1,2} = C_{1,2}$
\bea
        &&
        \Big\langle W_{r_1}[C_{1}] W_{r_2}[C_{2}] \Big\rangle_G 
        = \Big\langle 
        \nroneTr\,\Ps
          \exp \left[-i\,\frac{g}{2} 
                \int_{S_1} \! d\sigma_{\mu\nu}(X_1) 
                \G^a_{\mu\nu}(O_1,X_1;C_{X_1 O_1})\,t_{r_1}^a 
          \right] 
        \nonumber \\
        &&
        \quad\quad\quad\quad\quad
        \times\,\nrtwoTr\,\Ps
          \exp \left[-i\,\frac{g}{2} 
                \int_{S_2} \! d\sigma_{\rho\sigma}(X_2) 
                \G^b_{\rho\sigma}(O_2,X_2;C_{X_2 O_2})\,t_{r_2}^b 
          \right] 
        \Big\rangle_G
\label{Eq_Non-Abelian_Stokes_<W[C1]W[C2]>}
\eea
where $O_{1}$ and $O_{2}$ are the reference points on the surfaces
$S_{1}$ and $S_{2}$, respectively, that enter through the non-Abelian
Stokes theorem. In order to ensure gauge invariance in our model, the
gluon field strengths associated with the loops must be compared at
{\em one} reference point $O$. Due to this physical constraint, the
surfaces $S_{1}$ and $S_{2}$ are required to touch at a common
reference point $O_{1} = O_{2} = O$.

To treat the product of the two traces
in~(\ref{Eq_Non-Abelian_Stokes_<W[C1]W[C2]>}), we transfer the
approach of Berger and Nachtmann~\cite{Berger:1999gu} (cf.\ 
also~\cite{Shoshi:2002in}) to Euclidean space-time. Accordingly, the
product of the two traces, $\nroneTr(\cdots)\,\nrtwoTr(\cdots)$, over
$SU(N_c)$ matrices in the $r_1$ and $r_2$ representations,
respectively, is interpreted as one trace
$\nronexrtwoTr(\cdots):=\ronexrtwoTr(\cdots)/\ronexrtwoTr(\ronexrtwoIdentity)$
that acts in the tensor product space built from the $r_1$ and $r_2$
representations,
\bea
        \Big\langle W_{\!r_1}[C_{1}] W_{\!r_2}[C_{2}] \Big\rangle_G 
        \!\!\!\! & \!\!\! = \!\!\! &\!
        \Big\langle 
        \nronexrtwoTr
          \!\left\{\!\!
            \Big[\Ps \exp\!\big[\!-\!i\frac{g}{2} \!
                \int_{S_{1}} \!\!\!\! d\sigma_{\mu\nu}(X_{1}) 
                \G^a_{\mu\nu}(O,X_{1};C_{X_{1} O})\,t_{r_1}^a \big] 
                \,\otimes\,\Identity_{r_2}\Big]
          \right.
        \nonumber \\
        &&\!\!\!\!\!\!\!\!\!\!\!\!\!
        \times
          \left. 
          \Big[\Identity_{r_1}\,\otimes\,
          \Ps \exp\! \big[\!-\!i\frac{g}{2} \!
                \int_{S_{2}} \!\!\!\! d\sigma_{\rho\sigma}(X_{2}) 
                \G^b_{\rho\sigma}(O,X_{2};C_{X_{2} O})\,t_{r_2}^b \big]
          \Big]
          \!\!\right\}\!
        \Big\rangle_G 
\label{Eq_trace_trick_<W[C1]W[C2]>}
\eea
With the identities
\bea
        \exp\left(\,t_{r_1}^a\,\right) \,\otimes\, \Identity_{r_2} 
        & = & \exp\left(\,t_{r_1}^a \,\otimes\, \Identity_{r_2}\,\right) \ , \\
        \Identity_{r_1} \,\otimes\, \exp\left(\,t_{r_2}^a\,\right) 
        & = & \exp\left(\,\Identity_{r_1} \,\otimes\, t_{r_2}^a\,\right) \ ,
\label{Eq_exp(t^a)_times_1_identities}
\eea
the tensor products can be shifted into the exponents. Using the
matrix multiplication relations in the tensor product space
\bea
        \big( t_{r_1}^a \,\otimes\, \Identity_{r_2} \big)
        \big( t_{r_1}^b \,\otimes\, \Identity_{r_2} \big) 
        & = & t_{r_1}^a t_{r_1}^b \,\otimes\, \Identity_{r_2} \ ,
        \nonumber
        \\
        \big( t_{r_1}^a \,\otimes\, \Identity_{r_2} \big)
        \big( \Identity_{r_1} \,\otimes\, t_{r_2}^b \big) 
        & = & t_{r_1}^a \,\otimes\, t_{r_2}^b \ ,
\label{Eq_matrix_multiplication_in_tensor_product_space}
\eea
and the vanishing of the commutator
\be
        \left[t_{r_1}^a \otimes \Identity_{r_2}, 
        \Identity_{r_1} \otimes t_{r_2}^b\right] 
        = 0
        \ ,
\label{Eq_[t_x_1,1_x_t]}
\ee
the two exponentials in (\ref{Eq_trace_trick_<W[C1]W[C2]>}) commute
and can be written as one exponential
\be
        \Big\langle W[C_{1}] W[C_{2}] \Big\rangle_G =
        \Big\langle 
        \nronexrtwoTr\,\Ps \exp\!
            \left[-i\,\frac{g}{2} 
                \int_{S} \! d\sigma_{\mu\nu}(X) 
                \GG_{\mu\nu}(O,X;C_{XO}) 
            \right]
        \Big\rangle_G    
\label{Eq_<W[C1]W[C2]>_analogous_to_<W[C]>}
\ee
with the following gluon field strength tensor acting in the tensor
product space:
\be
        \GG_{\mu\nu}(O,X;C_{XO})
        := \left\{ \begin{array}{cc}
            \G_{\mu\nu}^a(O,X;C_{XO})
                \big( t_{r_1}^a \,\otimes\, \Identity_{r_2} \big)
            & \mbox{for $\,\,\, X\,\,\, \in \,\,\, S_1$}  \ , \\
            \G_{\mu\nu}^a(O,X;C_{XO})
                \big( \Identity_{r_1} \,\otimes\, t_{r_2}^a \big)
            & \mbox{for $\,\,\, X\,\,\, \in \,\,\, S_2$} \ .
        \end{array}\right.
\label{Eq_GG} 
\ee
In Eq.~(\ref{Eq_<W[C1]W[C2]>_analogous_to_<W[C]>}) the surface
integrals over $S_1$ and $S_2$ are written as one integral over the
combined surface $S = S_1 + S_2$ so that the right-hand side (RHS)
of~(\ref{Eq_<W[C1]W[C2]>_analogous_to_<W[C]>}) becomes very similar to
the RHS of~(\ref{Eq_Non-Abelian_Stokes_<W[C]>}). This allows us to
proceed analogously to the computation of $\langle W_r[C] \rangle_G$
in the previous section. After exploiting the linearity of the
functional integral, the matrix cumulant expansion is applied, which
holds for $\GG_{\mu\nu}(O,X;C_{XO})$ as well. Then, with the color
neutrality of the vacuum and by imposing the Gaussian approximation
now in the color components of the gluon field strength
tensor,\footnote{Note that the Gaussian approximation on the level of
  the color components of the gluon field strength tensor (component
  factorization) differs from the one on the level of the gluon field
  strength tensor (matrix factorization) used to compute $\langle
  W_{r}[C] \rangle$ in the original version of the
  SVM~\cite{Dosch:1987sk+X}. Nevertheless, with the additional
  ordering rule~\cite{Rueter:1994cn} explained in detail in Sec.~2.4
  of~\cite{Dosch:2000va}, a modified component factorization is
  obtained that leads to the same area law as the matrix
  factorization.} only the $n=2$ term of the matrix cumulant expansion
survives, which leads to
\bea
&& \!\!\!\!\!\!\!\!\!\!\!
        \Big\langle W_{r_1}[C_{1}] W_{r_2}[C_{2}] \Big\rangle_G 
\label{Eq_matrix_cumulant_expansion_<W[C1]W[C2]>}\\
&& \!\!\!\!\!\!\!\!\!\!\!
        = \nronexrtwoTr
        \exp\!\left[\!-\frac{g^2}{8} \!\!
          \int_{S} \!\! d\sigma_{\mu\nu}(X_1) \!
          \int_{S} \!\! d\sigma_{\rho\sigma}(X_2) 
          \Big\langle\!\Ps 
          [\GG_{\mu\nu}(O,X_1;C_{X_1 O}) 
          \GG_{\rho\sigma}(O,X_2;C_{X_2 O})] 
          \!\Big\rangle_{\!\!G}
        \right] \ .
\nonumber  
\eea

Using definition~(\ref{Eq_GG}) and
relations~(\ref{Eq_matrix_multiplication_in_tensor_product_space}), we
now redivide the exponent
in~(\ref{Eq_matrix_cumulant_expansion_<W[C1]W[C2]>}) into integrals of
the ordinary parallel transported gluon field strengths over the
separate surfaces $S_{1}$ and $S_{2}$
\bea
        && \Big\langle W_{r_1}[C_{1}] W_{r_2}[C_{2}] \Big\rangle_G = 
        \nronexrtwoTr
        \exp \Bigg[-\frac{g^2}{8}
\label{Eq_exponent_decomposition_<W[C1]W[C2]>}
\\ &&\hspace{-0.8cm}
          \times \Bigg\{
          \int_{S_1} \!\!\! d\sigma_{\mu\nu}(X_1) \!\! 
          \int_{S_2} \!\!\! d\sigma_{\rho\sigma}(X_2)\, 
          \Ps \!\left [ \Big\langle\! 
          \G^a_{\mu\nu}(O,X_1;C_{X_1 O}) 
          \G^b_{\rho\sigma}(O,X_2;C_{X_2 O})\!\Big\rangle_{\!\!G}
                \big(t_{r_1}^a  \,\otimes\, t_{r_2}^b\big) \right ] 
          \nonumber \\
        &&\hspace{-0.8cm}
          + \,
          \int_{S_2} \!\!\! d\sigma_{\mu\nu}(X_1) \!\! 
          \int_{S_1} \!\!\! d\sigma_{\rho\sigma}(X_2)\, 
          \Ps \!\left [ \Big\langle\! 
            \G^a_{\mu\nu}(O,X_1;C_{X_1 O}) 
            \G^b_{\rho\sigma}(O,X_2;C_{X_2 O})\!\Big\rangle_{\!\!G}
                \big(t_{r_1}^a  \,\otimes\, t_{r_2}^b\big) \right ] 
            \nonumber \\
        &&\hspace{-0.8cm}
          + \,
          \int_{S_1} \!\!\! d\sigma_{\mu\nu}(X_1) \!\! 
          \int_{S_1} \!\!\! d\sigma_{\rho\sigma}(X_2)\, 
          \Ps \!\left [ \Big\langle \!
            \G^a_{\mu\nu}(O,X_1;C_{X_1 O}) 
            \G^b_{\rho\sigma}(O,X_2;C_{X_2 O})\!\Big\rangle_{\!\!G}
                \big(t_{r_1}^a t_{r_1}^b \,\otimes\, \Identity_{r_2}\big)\right ] 
        \nonumber \\
        &&\hspace{-0.8cm}
        \left.
          + \,
          \int_{S_2} \!\!\! d\sigma_{\mu\nu}(X_1) \!\! 
          \int_{S_2} \!\!\! d\sigma_{\rho\sigma}(X_2)\, 
           \Ps \!\left [ \Big\langle\! 
             \G^a_{\mu\nu}(O,X_1;C_{X_1 O}) 
             \G^b_{\rho\sigma}(O,X_2;C_{X_2 O})\!\Big\rangle_{\!\!G}
                \big(\Identity_{r_1} \,\otimes\, t_{r_2}^a t_{r_2}^b\big) \right ]
              \!\Bigg\}\right]
        \nonumber
\eea
Here the surface ordering $\Ps$ is again irrelevant due to the
color neutrality of the vacuum~(\ref{Eq_Ansatz}), and
(\ref{Eq_exponent_decomposition_<W[C1]W[C2]>}) becomes
\bea
        && \Big\langle W_{r_1}[C_{1}] W_{r_2}[C_{2}] \Big\rangle_G 
        = \nronexrtwoTr
        \exp\!\Bigg[
                - \frac{\chi_{S_1 S_2}+\chi_{S_2 S_1}}{2}\,
                \big(t_{r_1}^a \,\otimes\, t_{r_2}^a\big) 
        \nonumber \\
        && \quad\quad\quad\quad  
        - \,\frac{\chi_{S_1 S_1}}{2}
                \big(t_{r_1}^a t_{r_1}^a\,\otimes\,\rtwoIdentity\big) 
            - \frac{\chi_{S_2 S_2}}{2}
                \big(\roneIdentity\,\otimes\,t_{r_2}^a t_{r_2}^a\big) 
              \Bigg]
\label{Eq_eikonal_functions_<W[C1]W[C2]>}
\eea
with
\be
        \chi_{S_i S_j}
        := \frac{\pi^2}{4} 
        \int_{S_i} \! d\sigma_{\mu\nu}(X_1) 
        \int_{S_j} \! d\sigma_{\rho\sigma}(X_2)
        F_{\mu\nu\rho\sigma}(X_1,X_2,O;C_{X_1 O},C_{X_2 O}) 
        \ .
\label{Eq_chi_Si_Sj}        
\ee
The symmetries in the tensor structure of $F_{\mu\nu\rho\sigma}$ [see
Eqs.~(\ref{Eq_F_decomposition}), (\ref{Eq_PGE_Ansatz_F}), and
(\ref{Eq_MSV_Ansatz_F})] lead to $\chi_{S_1 S_2} = \chi_{S_2 S_1}$.
With the quadratic Casimir
operator~(\ref{Eq_quadratic_Casimir_operator}), our final Euclidean
result for general $SU(N_c)$ representations $r_1$ and $r_2$
becomes\footnote{Note that the Euclidean $\chi_{S_i S_i} \neq 0$ in
  contrast to $\chi_{S_i S_i} = 0$ for Minkowskian light-like loops
  $C_i$ considered in the original version of the Berger-Nachtmann
  approach~\cite{Berger:1999gu,Shoshi:2002in}.}
\bea
        &&\Big\langle W_{r_1}[C_{1}] W_{r_2}[C_{2}] \Big\rangle_{\!G} 
        \\
        && = \nronexrtwoTr
        \exp\!\Bigg[
                - \chi_{S_1 S_2}\,
                \big(t_{r_1}^a \,\otimes\, t_{r_2}^a\big) 
                - \Big(\frac{C_2(r_1)}{2}\,\chi_{S_1 S_1} + \frac{C_2(r_2)}{2}\,\chi_{S_2 S_2}\Big)\,
                \ronexrtwoIdentity
        \Bigg]
        \nonumber
\label{Eq_final_general_Euclidean_result_<W[C1]W[C2]>}
\eea
where $\ronexrtwoIdentity := \roneIdentity\otimes\rtwoIdentity$. After
specifying the representations $r_1$ and $r_2$, the tensor product
$\tensor:=t_{r_1}^a \,\otimes\, t_{r_2}^a$ can be expressed as a sum
of projection operators $\Projector_i$ with the property $\Projector_i
\,\tensor = \lambda_i \,\Projector_i$
\be
        \tensor = \sum \lambda_i\,\Projector_i
        \quad\quad \mbox{with} \quad\quad 
        \lambda_i = 
        \frac{\nronexrtwoTr\big(\Projector_i \,\tensor\big)}
        {\nronexrtwoTr \big(\Projector_i\big)}
        \ ,
\label{Eq_tensor_decomposition}
\ee
which corresponds to the decomposition of the tensor product space
into irreducible representations.

For two {\WW} loops in the {\em fundamental representation} of
$SU(N_c)$, $r_1 = r_2 = \Fundamental$, which could describe the
trajectories of two quark-antiquark pairs, the
decomposition~(\ref{Eq_tensor_decomposition}) becomes trivial:
\be
        t_{\fundamental}^a \,\otimes\, t_{\fundamental}^a
        = \frac{N_c-1}{2N_c} \Projector_s - \frac{N_c+1}{2N_c} \Projector_a
        \ ,
\label{Eq_projector_tFa_x_tFa_relation}
\ee
with the projection operators
\bea
        &&
        (\Projector_s)_{(\alpha_1 \alpha_2)( \beta_1 \beta_2)} =
        \frac{1}{2}
        (\delta_{\alpha_1 \beta_1} \delta_{\alpha_2 \beta_2} 
        +\delta_{\alpha_1 \beta_2} \delta_{\alpha_2 \beta_1}) \ ,
        \\
        &&
        (\Projector_a)_{(\alpha_1 \alpha_2)( \beta_1 \beta_2)} =
        \frac{1}{2}
        (\delta_{\alpha_1 \beta_1} \delta_{\alpha_2 \beta_2} 
        -\delta_{\alpha_1 \beta_2} \delta_{\alpha_2 \beta_1}) \ ,
\label{Eq_projectors}
\eea
which decompose the direct product space of two fundamental $SU(N_c)$
representations into the irreducible representations
\be
        \mbox{N}_c \,\otimes\, \mbox{N}_c 
        = (\mbox{N}_c + 1)\mbox{N}_c/2 \,\oplus\, \overline{\mbox{N}_c(\mbox{N}_c - 1)/2}
        \ .
\label{Eq_tensor_product_fundamental_decomposition}
\ee
With
$\Tr_{\fundamental\otimes\fundamental}\,\Identity_{\fundamental\otimes\fundamental}
= N_c^2$ and the projector properties
\be
        \Projector^2_{s,a} = \Projector_{s,a}
        \, , \!\!\!\quad 
        \Tr_{\fundamental\otimes\fundamental} \,\Projector_s = (N_c + 1)N_c/2
        \, , \!\!\!\quad \mbox{and} \!\!\!\quad
        \Tr_{\fundamental\otimes\fundamental} \,\Projector_a = (N_c - 1)N_c/2
        \, ,
\label{Eq_projector_properties_fundamental}
\ee 
we find for the loop-loop correlation function with both loops in the
fundamental $SU(N_c)$ representation
\bea
        && \Big\langle W_{\fundamental}[C_{1}] W_{\fundamental}[C_{2}] \Big\rangle_G 
        = \exp\!\left[- \frac{C_2(\!\Fundamental\!)}{2}\Big(\chi_{S_1 S_1} + \chi_{S_2 S_2}\Big)\right]
\label{Eq_final_Euclidean_result_<W[C1]W[C2]>_fundamental}\\
        && \quad\quad\quad\quad
        \times
        \Bigg(\frac{N_c+1}{2N_c}\exp\!\left[-\frac{N_c-1}{2N_c}\chi_{S_1 S_2}\right]
        + \frac{N_c-1}{2N_c}\exp\!\left[ \frac{N_c+1}{2N_c}\chi_{S_1 S_2}\right]\Bigg)
        \nonumber
\eea
where
\be
        C_2(\!\Fundamental\!) = \frac{N_c^2-1}{2N_c}
        \ .
\label{Eq_Casimir_fundamental}
\ee

For one {\WW} loop in the {\em fundamental} and one in the {\em adjoint
  representation} of $SU(N_c)$, $r_1 = \Fundamental$ and $r_2 =
\Adjoint$, which is needed in Sec.~\ref{Sec_Flux_Tube} to investigate
the chromo-field distributions around color sources in the adjoint
representation, the decomposition~(\ref{Eq_tensor_decomposition})
reads
\be
        t_{\fundamental}^a \,\otimes\, t_{\adjoint}^a
        \,\,=\,\, 
        -\,\frac{N_c}{2}\,\Projector_1
        \,+\, \inv{2}\,\Projector_2 
        \,-\, \inv{2}\,\Projector_3 
\label{Eq_projector_tFa_x_tAa_relation}
\ee
with the projection operators\footnote{The explicit form of the
  projection operators $\Projector_1$, $\Projector_2$, and
  $\Projector_3$ can be found in~\cite{Cvitanovic_1984} but note that
  we use the Gell-Mann (conventional) normalization of the gluons. The
  eigenvalues, $\lambda_i$, of the projection operators
  in~(\ref{Eq_projector_tFa_x_tAa_relation}) can be evaluated
  conveniently with the computer program
  COLOUR~\cite{Hakkinen:1996bb}.}  $\Projector_1$, $\Projector_2$, and
$\Projector_3$ that decompose the direct product space of one
fundamental and one adjoint representation of $SU(N_c)$ into the
irreducible representations
\be
        \Fundamental\,\otimes\,\Adjoint
        \,\,=\,\,
        \mbox{N}_c
        \,\,\oplus\,\,\inv{2}\,\mbox{N}_c(\mbox{N}_c-1)(\mbox{N}_c+2)
        \,\,\oplus\,\,\inv{2}\,\mbox{N}_c(\mbox{N}_c+1)(\mbox{N}_c-2) \, ,
\label{Eq_tensor_product_f_x_a_decomposition}
\ee
which reduces for $N_c = 3$ to the well-known $SU(3)$ decomposition
\be
        3\,\otimes\,8 
        = 3\,\oplus\,15\,\oplus\,\bar{6}
        \ .
\label{Eq_tensor_product_f_x_a_SU(3)_decomposition}
\ee
With
$\Tr_{\fundamental\otimes\adjoint}\,\Identity_{\fundamental\otimes\adjoint}
= N_c(N_c^2 - 1)$ and projector properties analogous
to~(\ref{Eq_projector_properties_fundamental}), we obtain the
loop-loop correlation function for one loop in the fundamental and one
in the adjoint representation of $SU(N_c)$
\bea
        && \!\!\!\!\!\!\!\!
        \Big\langle W_{\fundamental}[C_{1}] W_{\adjoint}[C_{2}] \Big\rangle_G 
        = \exp\!\left[- \Big(\frac{C_2(\!\Fundamental\!)}{2}\,\chi_{S_1 S_1} 
        + \frac{C_2(\!\Adjoint\!)}{2}\,\chi_{S_2 S_2}\Big)\right]
\label{Eq_final_Euclidean_result_<Wf[C1]Wa[C2]>}\\
        && \!\!\!\!\!\!\!\!
        \times\,
        \Bigg(\!\inv{N_c^2\!-\!1}\exp\!\Big[\frac{N_c}{2}\chi_{S_1 S_2}\Big]
        +\frac{N_c\!+\!2}{2(N_c\!+\!1)}\exp\!\Big[\!-\inv{2}\chi_{S_1 S_2}\Big]
        +\frac{N_c\!-\!2}{2(N_c\!-\!1)}\exp\!\Big[\inv{2}\chi_{S_1 S_2}\Big]
        \!\Bigg)
\nonumber
\eea
where
\be
        C_2(\!\Adjoint\!) = N_c
        \ .
\label{Eq_Casimir_adjoint}
\ee

Note that our expressions for the loop-loop correlation
function~(\ref{Eq_final_general_Euclidean_result_<W[C1]W[C2]>}) and,
more specifically,
(\ref{Eq_final_Euclidean_result_<W[C1]W[C2]>_fundamental})
and~(\ref{Eq_final_Euclidean_result_<Wf[C1]Wa[C2]>}), are rather
general results---as is our result for the VEV of one \WW\ 
loop~(\ref{Eq_final_result_<W[C]>})---obtained directly from the color
neutrality of the QCD vacuum and the Gaussian approximation in the
gluon field strengths. The loop geometries, which characterize the
problem under investigation, are again encoded in the functions
$\chi_{S_i S_j}$, where also more detailed aspects of the QCD vacuum
enter in terms of $F_{\mu\nu\rho\sigma}$, i.e., the gauge-invariant
bilocal gluon field strength correlator~(\ref{Eq_Ansatz}).

For the explicit computations of $\chi_{S_1 S_2}$ presented in
Appendix~\ref{Sec_Chi_Computation}, one has to specify surfaces
$S_{1,2}$ with the restriction $\partial S_{1,2} = C_{1,2}$ according
to the non-Abelian Stokes theorem. We choose for $S_{1,2}$ {\em
  minimal surfaces} that are built from the plane areas spanned by the
corresponding loops $C_{1,2}$ and the infinitesimally thin tube which
connects the two surfaces $S_1$ and $S_2$. This is in line with our
surface choice in applications of the LLCM to high-energy
reactions~\cite{Shoshi:2002in,Shoshi:2002ri,Shoshi:2002fq,Shoshi:2002mt}.
The thin tube allows us to compare the field strengths in surface
$S_1$ with the field strengths in surface $S_2$.

Due to the Gaussian approximation and the associated truncation of the
cumulant expansion, the non-perturbative confining contribution to the
loop-loop correlation function depends on the surface choice. For
example, our results for the chromo-field distributions of color
dipoles obtained with the minimal surfaces differ {\em quantitatively}
from the ones obtained with the pyramid mantle choice for the
surfaces~\cite{Rueter:1994cn} even if the same parameters are used.
The {\em qualitative} main features of the non-perturbative SVM
component (such as confinement via flux tube formation), however,
emerge very similarly in both scenarios. From a comparison of the
static quark-antiquark potential to the energy stored in the
chromo-electric fields presented in
Sec.~\ref{Sec_Low_Energy_Theorems}, we infer that the minimal surfaces
are more compatible with the Gaussian approximation.  Indeed, the
application of low-energy theorems in
Sec.~\ref{Sec_Low_Energy_Theorems} will show that the minimal surfaces
are important for the consistency between the results for the VEV of
one loop, $\langle W_{r}[C] \rangle$, and the loop-loop correlation
function, $\langle W_{r_1}[C_1] W_{r_2}[C_2] \rangle$. In addition,
the simplicity of the minimal surfaces gives definitive advantages in
analytical computations. For example, it has allowed us to represent
the confining string as an integral over stringless dipoles with a
given dipole number density~\cite{Shoshi:2002fq}.

In applications of the model to high-energy
scattering~\cite{Shoshi:2002in,Shoshi:2002ri,Shoshi:2002fq,Shoshi:2002mt}
the surfaces are interpreted as the world-sheets of the confining QCD
strings in line with the picture obtained for the static dipole
potential from the VEV of one loop. The minimal surfaces are the most
natural choice to examine the scattering of two rigid strings without
any fluctuations or excitations. Our model does not choose the surface
dynamically and, thus, cannot describe string flips between two
non-perturbative color dipoles.  Recently, new developments toward a
dynamical surface choice and a theory for the dynamics of the
confining strings have been reported~\cite{Shevchenko:2002xi}.

\subsection{Perturbative and Non-Perturbative QCD Components}
\label{Sec_QCD_Components}

We decompose the gauge-invariant bilocal gluon field strength
correlator~(\ref{Eq_Ansatz})---as in the Minkowskian version of our
model~\cite{Shoshi:2002in}---into a perturbative ($\pert$) and
non-perturbative ($\nprt$) component
\be
        F_{\mu\nu\rho\sigma} 
        = F_{\mu\nu\rho\sigma}^{\pert} + F_{\mu\nu\rho\sigma}^{\nprt} 
        \ ,
\label{Eq_F_decomposition}
\ee
where $F_{\mu\nu\rho\sigma}^{\nprt}$ gives the low-frequency
background field contribution modeled by the non-perturbative {\em
  stochastic vacuum model}~\cite{Dosch:1987sk+X} and
$F_{\mu\nu\rho\sigma}^{\pert}$ the additional high-frequency
contribution described by {\em perturbative gluon exchange}. This
combination allows us to describe long and short distance correlations
in agreement with lattice calculations of the gluon field strength
correlator~\cite{DiGiacomo:1992df+X,D'Elia:1997ne,Bali:1998aj,Meggiolaro:1999yn}.
Moreover, this two component ansatz leads to the static
quark-antiquark potential with color Coulomb behavior for small and
confining linear rise for large source separations in good agreement
with lattice data as shown in Sec.~\ref{Sec_Static_Potential}. Note
that in addition to our two component ansatz an ongoing effort to
reconcile the non-perturbative SVM with perturbative gluon exchange
has led to complementary
methods~\cite{Simonov:kt,Shevchenko:1998ej,Shevchenko:2002xi}.

We compute the perturbative correlator $F_{\mu\nu\rho\sigma}^{\pert}$
from the Euclidean gluon propagator in the Feynman--'t~Hooft gauge:
\be
        \Big\langle  \G^a_{\mu}(X_1)\G^b_{\nu}(X_2) \Big\rangle
        = \int
        \frac{d^4K}{(2\pi)^4}
        \,\frac{\delta^{ab}\delta_{\mu\nu}}{K^2+m_G^2}
        \, e^{-iK(X_1-X_2)}
        \ ,
\label{Eq_massive_gluon_propagator}
\ee
where we introduce an {\em effective gluon mass} of $m_G = m_{\rho} =
0.77\,\GeV$ to limit the range of the perturbative interaction in the
infrared (IR) region. This IR cutoff for the perturbative component is
important in applications of our model to high-energy
scattering~\cite{Shoshi:2002in}. Its value has been chosen such that
the unintegrated gluon distribution for transverse momenta below
$|\vec{k}_{\perp}| \approx 1\,\GeV$ is dominated by non-perturbative
physics~\cite{Shoshi:2002fq}. Of course, the parameter $m_G$ is also
important for the interplay between the perturbative and
non-perturbative components in the presented Euclidean applications.
Furthermore, our value for $m_G$ gives the ``perturbative glueball''
($GB$) generated by our perturbative component a finite mass of
$M_{\glueball}^{\pert} = 2m_G = 1.54\,\GeV$, which is larger than that
of its non-perturbative counterpart discussed below. This ensures that
long-range correlations are dominated by non-perturbative physics.

In leading order in the strong coupling $g$, the resulting bilocal
gluon field strength correlator is gauge invariant already without the
parallel transport to a common reference point so that
$F_{\mu\nu\rho\sigma}^{\pert}$ depends only on the difference $Z= X_1
- X_2$:
\bea
        F_{\mu\nu\rho\sigma}^{\pert}(Z)
        \!\!&=&\!\!\frac{g^2}{\pi^2}\, \inv{2}\Bigl[
                       \frac{\partial}{\partial Z_\nu}
                         \left(Z_\sigma \delta_{\mu\rho}
                         -Z_\rho \delta_{\mu\sigma}\right)
                       +\frac{\partial}{\partial Z_\mu}
                         \left(Z_\rho \delta_{\nu\sigma}
                         -Z_\sigma \delta_{\nu\rho}\right)\Bigr]\,
              D_{\pert}(Z^2)
\label{Eq_PGE_Ansatz_F}\\ 
        \!\!& &\!\! 
        \hspace{-2cm}
        = \,\,
        -\,\frac{g^2}{\pi^2}\!
                \int \!\!\frac{d^4K}{(2\pi)^4} \,e^{-iKZ}\,\Bigl[
                K_\nu K_\sigma \delta_{\mu\rho}  - K_\nu K_\rho   \delta_{\mu\sigma}
              + K_\mu K_\rho  \delta_{\nu\sigma} - K_\mu K_\sigma \delta_{\nu\rho} \Bigr]\,
           \tilde{D}_{\pert}^{\prime}(K^2)
        \nonumber
\eea
with the perturbative correlation function
\bea
        D_{\pert}(Z^2)
        & = & \frac{m_G^2}{2\,\pi^2 Z^2}\,K_2(m_G\,|Z|)
\label{Eq_Dp(z,mg)}\\
        \tilde{D}_{\pert}^{\prime}(K^2)
        &:= & 
        \frac{d}{dK^2}\int d^4Z\,e^{iKZ}\,D_{\pert}(Z^2)
        \,\, =\,\, 
        -\,\inv{K^2 + m_G^2}
         \ .
\label{Eq_D'p(K,mg)}
\eea

The perturbative gluon field strength correlator has also been
considered at next-to-leading order, where the dependence of the
correlator on both the renormalization scale and the renormalization
scheme becomes explicit and an additional tensor structure arises
together with a path dependence of the
correlator~\cite{Eidemuller:1997bb}. However, cancellations of
contributions from this additional tensor structure have been
shown~\cite{Shevchenko:1998ej}. We refer to Sec.~3.3 of
Ref.~\cite{Dosch:2000va} for a more detailed discussion of this issue.

We describe the perturbative correlations in our phenomenological
applications only with the leading tensor
structure~(\ref{Eq_PGE_Ansatz_F}) and take into account radiative
corrections by replacing the constant coupling $g^2$ with the running
coupling
\be
        g^2(Z^2)
        = 4 \pi \alphaS(Z^2)
        = \frac{48 \pi^2}
        {(33-2 N_f) 
        \ln\left[
                (Z^{-2} + M^2)/\Lambda_{QCD}^2
        \right]}
\label{Eq_g2(z_perp)}
\ee
in the final step of the computation of the $\chi$ function, where the
Euclidean distance $|Z|$ over which the correlation occurs provides
the renormalization scale. In Eq.~(\ref{Eq_g2(z_perp)}) $N_f$ denotes
the number of dynamical quark flavors, which is set to $N_f = 0$ in
agreement with the quenched approximation, $\Lambda_{QCD} =
0.25\;\GeV$, and $M$ allows us to freeze $g^2$ for
$|Z|\rightarrow\infty$. Relying on low-energy theorems, we freeze the
running coupling at the value $g^2 = 10.2$ ($\equiv \alphaS = 0.81$),
i.e.\ $M = 0.488\,\GeV$, at which our non-perturbative results for the
confining potential and the total flux tube energy of a static
quark-antiquark pair coincide (see
Sec.~\ref{Sec_Low_Energy_Theorems}).

The tensor structure~(\ref{Eq_PGE_Ansatz_F}) together with the
perturbative correlation function~(\ref{Eq_Dp(z,mg)})
or~(\ref{Eq_D'p(K,mg)}) leads to the color Yukawa potential (which
reduces for $m_G = 0$ to the color Coulomb potential) as shown in
Sec.~\ref{Sec_Static_Potential}. The perturbative contribution
thus dominates the full potential at small quark-antiquark
separations.

If the path connecting the points $X_1$ and $X_2$ is a straight line,
the non-perturbative correlator $F_{\mu\nu\rho\sigma}^{\nprt}$ also
depends only on the difference $Z=X_1-X_2$. Then, the most general
form of the correlator that respects translational, Lorentz, and
parity invariance reads~\cite{Dosch:1987sk+X}
\bea
        F_{\mu\nu\rho\sigma}^{\nprt}(Z) 
        & = & F_{\mu\nu\rho\sigma}^{\nprt\,c}(Z) +  
               F_{\mu\nu\rho\sigma}^{\nprt\,nc}(Z)
\label{Eq_MSV_Ansatz_F}\\
        &  = & \inv{3(N_c^2-1)}\,G_2\, \Bigl\{
        \kappa\, \left(\delta_{\mu\rho}\delta_{\nu\sigma}
          -\delta_{\mu\sigma}\delta_{\nu\rho}\right) \,
        D(Z^2)                                
        \nonumber\\
        &   &  
        +\,(1-\kappa)\,\inv{2}\Bigl[
        \frac{\partial}{\partial Z_\nu}
        \left(Z_\sigma \delta_{\mu\rho}
          -Z_\rho \delta_{\mu\sigma}\right)
        +\frac{\partial}{\partial Z_\mu}
        \left(Z_\rho \delta_{\nu\sigma}
          -Z_\sigma \delta_{\nu\rho}\right)\Bigr]\,
        D_1(Z^2) \Bigr\}
        \nonumber\\
        &  = & \inv{3(N_c^2-1)}\,G_2 
        \int \frac{d^4K}{(2\pi)^4} \,e^{-iKZ}\,\Bigl\{
        \kappa \,\left(\delta_{\mu\rho}\delta_{\nu\sigma}
          -\delta_{\mu\sigma}\delta_{\nu\rho}\right)\, 
        \tilde{D}(K^2)                                
        \nonumber\\
        &   &  -\,(1-\kappa)\,\Bigl[
        K_\nu K_\sigma \delta_{\mu\rho}   - K_\nu K_\rho   \delta_{\mu\sigma}
        + K_\mu K_\rho  \delta_{\nu\sigma} - K_\mu K_\sigma \delta_{\nu\rho} \Bigr]\,
               \tilde{D}_{1}^{\prime}(K^2) \Bigr\} 
        \nonumber \ ,
\eea
where
\be
        \tilde{D}_{1}^{\prime}(K^2) 
        := \frac{d}{dK^2} \int d^4Z\, D_{1}(Z^2)\, e^{iKZ} 
        \ .
\label{Eq_D1_prime}
\ee
In all previous applications of the SVM, this form, depending only on
$Z=X_1-X_2$, has been used. New lattice results on the path dependence
of the correlator~\cite{DiGiacomo:2002mq} show a dominance of the
shortest path. This result is effectively incorporated in the model
since the straight paths dominate in the averaging over all paths.

Let us emphasize that the non-perturbative
correlator~(\ref{Eq_MSV_Ansatz_F}) is a sum of the two different
tensor structures, $F_{\mu\nu\rho\sigma}^{\nprt\,nc}$ and
$F_{\mu\nu\rho\sigma}^{\nprt\,c}$, with characteristic behavior: The
tensor structure $F_{\mu\nu\rho\sigma}^{\nprt\,nc}$ is characteristic
for Abelian gauge theories, exhibits the same tensor structure as the
perturbative correlator~(\ref{Eq_PGE_Ansatz_F}) and does not lead to
confinement~\cite{Dosch:1987sk+X}. In contrast, the tensor structure
$F_{\mu\nu\rho\sigma}^{\nprt\,c}$ can occur only in non-Abelian gauge
theories and Abelian gauge theories with monopoles and leads to
confinement~\cite{Dosch:1987sk+X}. Therefore, we call the tensor
structure multiplied by $(1-\kappa)$ non-confining ($nc$) and the one
multiplied by $\kappa$ confining ($c$).

The non-perturbative correlator~(\ref{Eq_MSV_Ansatz_F}) involves the
gluon condensate $G_2 := \langle \frac{g^2}{4\pi^2} \G^a_{\mu\nu}(0)
\G^a_{\mu\nu}(0) \rangle$~\cite{Shifman:1979bx+X}, the weight
parameter $\kappa$, and the correlation length $a$ which enters
through the non-perturbative correlation functions $D$ and $D_1$.
While the perturbative correlation function $D_{\pert}$ given
in~(\ref{Eq_Dp(z,mg)}) is computed from the gluon propagator (with a
finite effective gluon mass), the non-perturbative correlation
functions $D$ and $D_1$ can be studied rigorously in lattice QCD
investigations~\cite{DiGiacomo:1992df+X,D'Elia:1997ne,Bali:1998aj,Meggiolaro:1999yn}.
In addition, the non-perturbative correlation functions are
constrained by the following physical considerations. (i) The
correlations at large distances should decrease exponentially so that
the interaction range is determined by the glueball mass. (ii) Toward
small distances, the non-perturbative correlation functions must
satisfy $D(0)=D_1(0)=1$ to ensure the correct relation between the VEV
of infinitesimal plaquettes and the gluon condensate $G_2$. (iii) The
correlation functions must stay positive at all distances to be
compatible with a spectral representation~\cite{Dosch:1998th}.

We adopt for our calculations a
simple {\em exponential correlation function}
\be
        D(Z^2) = D_1(Z^2) = \exp(-|Z|/a)
        \ ,
\label{Eq_SVM_correlation_functions}
\ee
which is consistent with the physical constraints discussed and has
been successfully tested in fits to lattice data of the gluon field
strength correlator~\cite{Bali:1998aj,Meggiolaro:1999yn}.  The
exponential correlation function stays positive for all Euclidean
distances $Z$ and is compatible with a spectral representation of the
correlation function~\cite{Dosch:1998th}. This is a conceptual
improvement since the correlation function used in several earlier
applications of the SVM becomes negative at large
distances~\cite{Dosch:1994ym,Rueter:1994cn,Dosch:1995fz,Rueter:1996yb,Dosch:1998nw,Rueter:1998qy+X,Rueter:1998up,Dosch:1997ss,Berger:1999gu,Dosch:2001jg}.

With the exponential correlation
function~(\ref{Eq_SVM_correlation_functions}), fits to the lattice
data of the gluon field strength correlator down to distances of
$0.4\,\fm$ give the following values for the parameters of the
non-perturbative correlator~\cite{Meggiolaro:1999yn}: $G_2 =
0.173\,\GeV^4$, $\kappa = 0.746$, and $a = 0.219\,\fm$. We have
optimized these parameters in a fit to high-energy scattering
data\footnote{Since we describe both lattice QCD data obtained in the
  quenched approximation and high-energy scattering data taken in the
  presence of light quarks, our value for the gluon condensate, $G_2=
  0.074\,\GeV^4$, interpolates between $G_2= 0.173\,\GeV^4$ found in
  quenched lattice QCD
  investigations~\cite{DiGiacomo:1992df+X,Meggiolaro:1999yn} and $G_2=
  0.024 \pm 0.011\, \GeV^4$ found in
  phenomenology~\cite{Shifman:1979bx+X,Narison:1995tw} and full
  lattice QCD investigations~\cite{D'Elia:1997ne,Meggiolaro:1999yn}.
  It is known that the effect of light quarks reduces the value of
  $G_2$ substantially~\cite{Novikov:xi+X}.}~\cite{Shoshi:2002in}:
\be
        a =  0.302\,\fm, \quad 
        \kappa = 0.74, \quad 
        G_2 = 0.074\,\GeV^4
        \ .
\label{Eq_MSV_scattering_fit_parameter_results}
\ee
We use these optimized
parameters~(\ref{Eq_MSV_scattering_fit_parameter_results}) throughout
this work. They lead to a static quark-antiquark potential that is in
good agreement with lattice data (see Sec.~\ref{Sec_Static_Potential})
and, in particular, give a QCD string
tension~(\ref{Eq_sting_tension_from_exp_correlation}) of $\sigma_3 =
0.22\,\GeV^2 \equiv 1.12 \,\GeV/\fm$ which is consistent with hadron
spectroscopy~\cite{Kwong:1987mj}, Regge
theory~\cite{Goddard:1973qh+X}, and lattice QCD
investigations~\cite{Bali:2001gf}. Moreover, the non-perturbative
component with $a=0.302\,\fm$ generates a ``non-perturbative
glueball'' with a mass of $M_{\glueball}^{\nprt} = 2/a = 1.31\,\GeV$
which is smaller than $M_{\glueball}^{\pert}=1.54\,\GeV$ and thus
governs the long-range correlations as expected. We thus have one
model that describes both static hadronic properties and high-energy
reactions of hadrons and photons in good agreement with experimental
and lattice QCD data.

At this point, we would like to comment on the model parameters and
the accuracy of the results. Although there are strong hints for the
choice of the integration surface and physical constraints on the
non-perturbative correlation functions, we have no criteria from first
principles that fix these model ingredients unambigiously.  Therefore,
we have checked different integration surfaces and different
non-perturbative correlation functions: While the analytic result for
the string tension changes, the general picture (e.g.\ the confining
linear rise of the static dipole potential and flux tube formation) is
reproduced by readjusting the parameters $a$, $\kappa$, and $G_2$.
Therefore, the model parameters are meaningful only within about 20\%
accuracy, which estimates possible errors incurred by chosing a
certain combination of integration surfaces and correlation functions.
In Table~\ref{Tab_SVM_parameters} we show different sets of parameters
used together with different surfaces and different correlation
functions in applications of the SVM to high-energy scattering. The
table documents the stability of the SVM parameters within 20\%.
However, after the Gaussian approximation (or truncation of the
cumulant expansion) and the specification of the integration surface
and the correlation functions, the quantitative results depend
sensitively on some of the model parameters. To achieve a good fit to
high-energy cross sections~\cite{Shoshi:2002in}, a fine tuning of $a$
and $G_2$ is necessary.
\begin{table}
\caption{\small Different sets of parameters used together with different surfaces and different correlation
functions in applications of the SVM to high-energy scattering.}
\label{Tab_SVM_parameters}
\begin{center}
\begin{tabular}{|c||c|c|c|c|}\hline
Reference & \cite{Dosch:1994ym} & \cite{Dosch:1998nw} & \cite{Rueter:1998up} & \cite{Shoshi:2002in}\\ \hline
$a\,\,(\fm)$ & 0.350 & 0.346 & 0.346 & 0.302 \\ \hline 
$G_2\,\,(\GeV^4)$ & 0.0605 & 0.0631 & 0.0631 & 0.074 \\ \hline
$\kappa$ & 0.74 & 0.74 & 0.74 & 0.74 \\ \hline
$m_G\,\,(\GeV)$ & --- & --- & 0.571 & 0.77 \\ \hline
Integration surface & pyramid & pyramid & pyramid & minimal \\ \hline
Correlation function & Bessel & Bessel & Bessel & exponential \\ \hline
Perturbative component & no & no & yes & yes \\ \hline
\end{tabular}
\end{center} \vspace*{-0.6cm}
\end{table}

Finally, we should discuss the pragmatic treatment of renormalization
of the perturbative component~(\ref{Eq_PGE_Ansatz_F}) that dominates
the small distance correlations. Only the lowest order result is
adopted in which the strong coupling is promoted to a 1-loop effective
running coupling.  The mass renormalization of the considered heavy
quarks and antiquarks is also taken into account by subtracting the
self-energy of the sources in the computation of the static color
dipole potential. Although phenomenologically successful, one needs to
refine the treatment of renormalization, for example, by explicitly
taking into account counterterms for the cusps and by introducing a
factorization scale in order to put the model on a more solid basis.
We defer this task to future work and turn now to the phenomenological
performance of our pragmatic approach.

%
%
%
%
%
\section{The Static Color Dipole Potential}
\label{Sec_Static_Potential}

In this section the QCD potential of static color dipoles in the
fundamental and adjoint representation of $SU(N_c)$ is computed in our
model.  Color Coulomb behavior is found for small dipole sizes and the
confining linear rise for large dipole sizes. Casimir scaling is
obtained in agreement with lattice QCD investigations.

\begin{figure}[b!]
  \centerline{\epsfig{figure=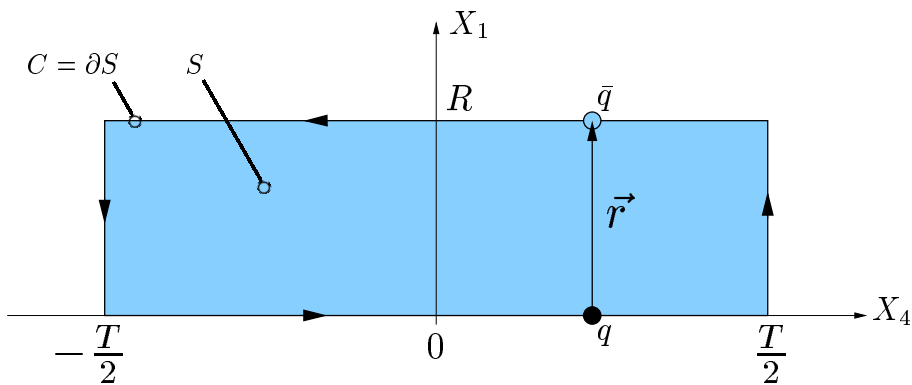,width=12.cm}}
\caption{\small 
  A static color dipole of size $R$ in the fundamental representation.
  The rectangular path $C$ of spatial extension $R$ and temporal
  extension $T$ indicates the world-line of the dipole described the
  {\WW} loop $W_{\fundamental}[C]$. The shaded area bounded by the
  loop $C=\partial S$ represents the minimal surface $S$ used to
  compute the static dipole potential.}
\label{Fig_ONE_WWL}
\end{figure}

The static color dipole---two static color sources separated by a
distance $R$ in a net color singlet state---is described by a {\WW}
loop $W_r[C]$ with a rectangular path $C$ of spatial extension $R$ and
temporal extension $T\to\infty$ where $r$ indicates the $SU(N_c)$
representation of the sources considered.  Figure~\ref{Fig_ONE_WWL}
illustrates a static color dipole in the fundamental representation
$r=\Fundamental$.  The potential of the static color dipole is
obtained from the VEV of the corresponding Wegner-Wilson
loop~\cite{Wilson:1974sk,Brown:1979ya}
\be
        V_r(R) 
        = - \lim_{T \to \infty} \inv{T} 
        \ln \langle W_r[C] \rangle_{\pot}
        \ ,
\label{Eq_static_potential}
\ee
where ``pot'' indicates the subtraction of the self-energy of the
color sources. This subtraction corresponds to the mass
renormalization of the heavy color sources as discussed above. The
static quark-antiquark potential $V_{\fundamental}$ is obtained from a
loop in the fundamental rep\-re\-sen\-ta\-tion ($r\!=\!\Fundamental$)
and the potential of a static gluino pair $V_{\adjoint}$ from a loop
in the adjoint representation ($r\!=\!\Adjoint$).

With our result for $\langle W_r[C] \rangle$,
(\ref{Eq_final_result_<W[C]>}), obtained with the Gaussian
approximation in the gluon field strength, the static potential reads
\be
        V_r(R) = \frac{C_2(r)}{2}\,\lim_{T \to \infty} \inv{T}\,\chi_{SS\,\pot}
\label{Eq_Vr(R)_Gaussian_approximation}
\ee
with the self-energy subtracted, i.e.\ $\chi_{SS\,\pot} := \chi_{SS} -
\chi_{SS\,\self}$ (see Appendix~\ref{Sec_Chi_Computation}). According
to the structure of the gluon field strength correlator,
(\ref{Eq_Ansatz}) and~(\ref{Eq_F_decomposition}), there are
perturbative ($\pert$) and non-perturbative ($\nprt$) contributions to
the static potential:
\be
        V_r(R) = \frac{C_2(r)}{2}\,\lim_{T \to \infty} \inv{T}\,
                 \left\{\chi_{SS\,\pot}^{\pert}
                   +\left(\chi_{SS\,\pot}^{\nprt\,\,nc} +
                     \chi_{SS\,\pot}^{\nprt\,\,c} \right) \right\}
        \ ,
\label{Eq_Vr(R)_P+NP}
\ee
where the explicit form of the $\chi$ functions is given
in~(\ref{Eq_chi_SS_NP_c_T->infty_V_E}),
(\ref{Eq_chi_SS_NP_nc_T->infty_pot_E}),
and~(\ref{Eq_chi_SS_P_T->infty_pot_E}).

The perturbative contribution to the static potential describes the
{\em color Yukawa potential} (which reduces to the {\em color Coulomb
  potential}~\cite{Kogut:1979wt} for $m_G=0$)
\be
        V_r^{\pert}(R) 
        = - C_2(r)\,\frac{g^2(R)}{4 \pi R} \exp[-m_G R] 
        \ .
\label{Eq_Vr(R)_color-Yukawa}
\ee
Here we have used the result for $\chi_{SS\,\pot}^{\pert}$ given
in~(\ref{Eq_chi_SS_P_T->infty_pot_E}) and the perturbative correlation
function
\be
        D^{\prime\,(3)}_{\pert}(\vec{Z}^2)
        := \int \frac{d^4K}{(2\pi)^3}\,e^{iKZ}\,
        \tilde{D}^{\prime\,(3)}_{\pert}(K^2)\,\delta(K_4)
         = -\,\frac{\exp[-\,m_G\,|\vec{Z}|]}{4\pi|\vec{Z}|}
\label{Eq_D'(3)p(z,mg)}
\ee
which is obtained from the massive gluon
propagator~(\ref{Eq_massive_gluon_propagator}). As shown below, the
perturbative contribution dominates the static potential for small
dipole sizes $R$.

The non-perturbative contributions to the static potential, the {\em
  non-confining} component ($nc$) and the {\em confining} component
($c$), read
\bea
        V_r^{\nprt\,\,nc}(R)
        & = & 
        C_2(r)\,\,
        \frac{\pi^2 G_2 (1-\kappa)}{3(N_c^2-1)}\,\,
        D_1^{\prime\,(3)}(R^2)
        \ ,
\label{Eq_Vr(R)_NP_nc}\\
        V_r^{\nprt\,\,c}(R) 
        & = & 
        C_2(r)\,\,
        \frac{\pi^2 G_2 \kappa}{3(N_c^2-1)}\,\,
        \int_0^R \!\! d\rho\,
        (R-\rho)\,
        D^{(3)}(\rho^2)
        \ ,
\label{Eq_Vr(R)_NP_c}
\eea
where we have used the results for $\chi_{SS\,\pot}^{\nprt\,\,nc}$ and
$\chi_{SS\,\pot}^{\nprt\,\,c}=\chi_{SS}^{\nprt\,\,c}$ given
respectively in~(\ref{Eq_chi_SS_NP_nc_T->infty_pot_E})
and~(\ref{Eq_chi_SS_NP_c_T->infty_V_E}) obtained with the minimal
surface, i.e.\ the planar surface bounded by the loop as indicated by
the shaded area in Fig.~\ref{Fig_ONE_WWL}. With the exponential
correlation function~(\ref{Eq_SVM_correlation_functions}), the
correlation functions in~(\ref{Eq_Vr(R)_NP_nc})
and~(\ref{Eq_Vr(R)_NP_c}) read
\bea
        D^{\prime\,(3)}_1(\vec{Z}^2)
        &\!\!:=\!\!&
        \int \frac{d^4K}{(2\pi)^3}\,e^{iKZ}\,
        \tilde{D}^{\prime\,(3)}_1(K^2)\,\delta(K_4)
        \,\,=\,\, -\,a\,|\vec{Z}|^2\,K_2[|\vec{Z}|/a]
        \ ,
\label{Eq_D'(3)np_nc(z,a)}\\
        D^{(3)}(\vec{Z}^2)      
        &\!\!:=\!\! & \int \frac{d^4K}{(2\pi)^3}\,e^{iKZ}\,\tilde{D}(K^2)\,\delta(K_4)
        \,\,=\,\, 2\,|\vec{Z}|\,K_1[|\vec{Z}|/a]
        \ .
\label{Eq_D(3)np_c(z,a)}
\eea
For large dipole sizes, $R \gtsim 0.5\ \fm$, the non-confining
contribution~(\ref{Eq_Vr(R)_NP_nc}) vanishes exponentially while the
confining contribution~(\ref{Eq_Vr(R)_NP_c})---as anticipated---leads
to {\em confinement}~\cite{Dosch:1987sk+X}, i.e.\ the confining linear
increase,
\be
        V_r^{\nprt\,\,c}(R)\Big|_{R\,\gtsim\,0.5\,\mbox{\scriptsize fm}} 
       = \sigma_r R  + \mbox{const.}
\label{Eq_Vr(R)_NP_c_linear}
\ee
Thus, the QCD {\em string tension} is given by the confining SVM
component~\cite{Dosch:1987sk+X}: For a color dipole in the $SU(N_c)$
representation $r$, it reads
\be
        \sigma_r 
        = C_2(r)\,\,\frac{\pi^3 G_2 \kappa}{48} 
          \int_0^\infty dZ^2 D(Z^2) 
        = C_2(r)\,\,\frac{\pi^3 \kappa G_2 a^2}{24}
        \ ,
\label{Eq_string_tension}
\ee
where the exponential correlation
function~(\ref{Eq_SVM_correlation_functions}) is used in the final
step.  Since the string tension can be computed from first principles
within lattice QCD~\cite{Bali:2001gf},
relation~(\ref{Eq_string_tension}) puts an important constraint on the
three parameters of the non-perturbative QCD vacuum $a$, $G_2$, and
$\kappa$. With the values for $a$, $G_2$, and $\kappa$ given
in~(\ref{Eq_MSV_scattering_fit_parameter_results}), which are used
throughout this work, one obtains for the string tension of the
$SU(3)$ quark-antiquark potential ($r=3$) a reasonable value of
\be
        \sigma_3 
        = 0.22\,\GeV^2 \equiv 1.12 \,\GeV/\fm
        \ .
\label{Eq_sting_tension_from_exp_correlation}
\ee 

\begin{figure}[t!]
  \centerline{\epsfig{figure=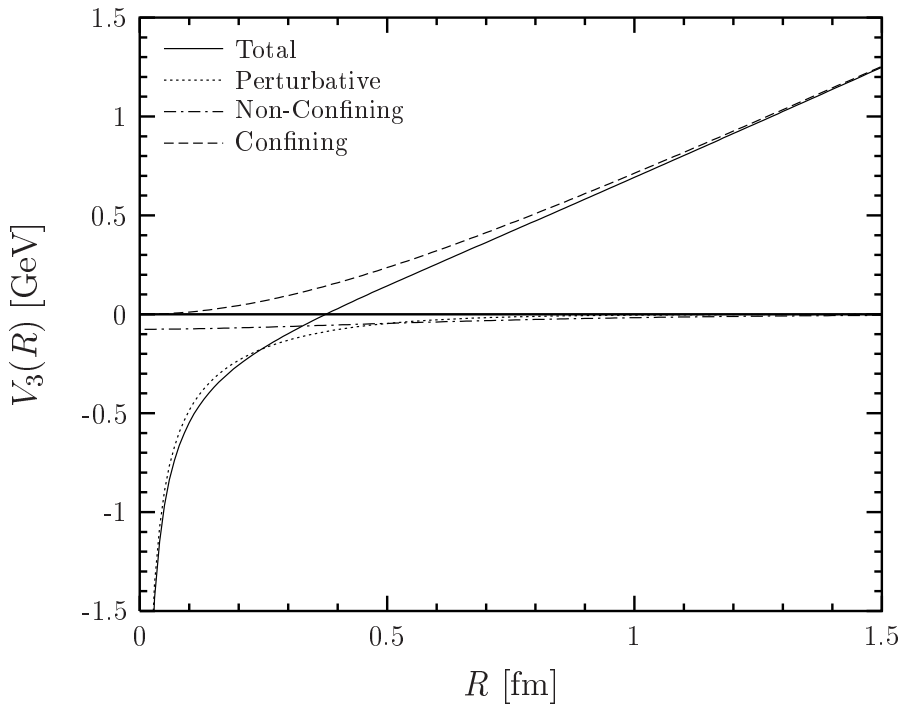,width=12.cm}}
\caption{\small 
  The static $SU(N_c = 3)$ quark-antiquark potential
  $V_{\fundamental}(R) = V_3(R)$ as a function of the quark-antiquark
  separation $R$. The solid, dotted, and dashed lines indicate the
  full static potential and its perturbative and non-perturbative
  contributions, respectively. For small quark-antiquark separations,
  $R \ltsim 0.5\,\fm$, the perturbative contribution dominates and
  gives rise to the well-known color Coulomb behavior at small
  distances. For medium and large quark-antiquark separations, $R
  \gtsim 0.5\,\fm$, the non-perturbative contribution dominates and
  leads to the confining linear rise of the static potential. As our
  model is working in the quenched approximation, string breaking
  cannot be described, which is expected to stop the linear increase
  for $R\,\gtsim\,1\,\fm$~\cite{Laermann:1998gm,Bali:2001gf}.}
\label{Fig_Static_Quark-Antiquark_Potential_Components}
\end{figure}

The static $SU(N_c = 3)$ quark-antiquark potential
$V_{\fundamental}(R) = V_3(R)$ is shown as a function of the
quark-antiquark separation $R$ in
Fig.~\ref{Fig_Static_Quark-Antiquark_Potential_Components},
where the solid, dotted, and dashed lines indicate the full static
potential and its perturbative and non-perturbative contributions,
respectively.  For small quark-antiquark separations $R \ltsim
0.5\,\fm$, the perturbative contribution dominates giving rise to the
well-known color Coulomb behavior. For medium and large
quark-antiquark separations $R \gtsim 0.5\,\fm$, the non-perturbative
contribution dominates and leads to the confining linear rise of the
static potential. The transition from perturbative to string behavior
takes place at source separations of about $0.5\,\fm$ in agreement
with the recent results of L\"uscher and Weisz~\cite{Luscher:2002qv}.
This supports our value for the gluon mass $m_G=m_{\rho}=0.77\,\GeV$
which is important only around $R\approx 0.4\,\fm$, i.e.\ for the
interplay between perturbative and non-perturbative physics. For
$R\ltsim 0.3\,\fm$ and $R\gtsim 0.5\,\fm$, the effect of the gluon
mass, introduced as an IR regulator in our perturbative component, is
negligible. String breaking is expected to stop the linear increase
for $R\,\gtsim\,1\,\fm$ where lattice investigations show deviations
from the linear rise in full QCD~\cite{Laermann:1998gm,Bali:2001gf}.
As our model is working in the quenched approximation, string breaking
through dynamical quark-antiquark production is excluded.

\begin{figure}[t!]
\centerline{\epsfig{figure=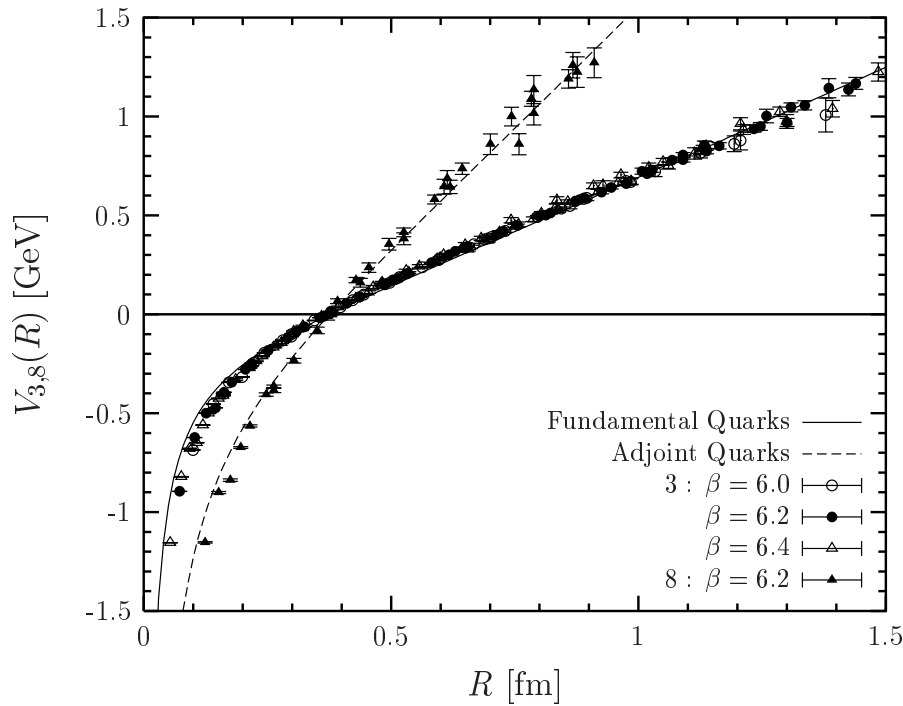,width=12.cm}}
\caption{\small 
  The static $SU(N_c = 3)$ potential of color dipoles in the
  fundamental representation $V_3(R)$ (solid line) and adjoint
  representation $V_8(R)$ (dashed line) as a function of the dipole
  size $R$ in comparison to $SU(3)$ lattice data for $\beta = 6.0$,
  6.2, and 6.4~\cite{Bali:2000un,Bali:2001gf}. The model results are
  in good agreement with the lattice data. This particularly holds for
  the obtained Casimir scaling behavior.}
\label{Fig_Static_Quark-Antiquark_Potential_F_vs_A}
\end{figure}

As can be seen from~(\ref{Eq_Vr(R)_Gaussian_approximation}), the
static potential shows {\em Casimir scaling} which emerges in our
approach as a trivial consequence of the Gaussian approximation used
to truncate the cumulant
expansion~(\ref{Eq_matrix_cumulant_expansion}). Indeed, the Casimir
scaling hypothesis~\cite{Ambjorn:1984dp} has been verified to high
accuracy for $SU(3)$ on the lattice~\cite{Deldar:1999vi,Bali:2000un}
(see also Fig.~\ref{Fig_Static_Quark-Antiquark_Potential_F_vs_A}).
These lattice results have been interpreted as a strong hint toward
Gaussian dominance in the QCD vacuum and thus as evidence for a strong
suppression of higher cumulant
contributions~\cite{Shevchenko:2000du,Shevchenko:2001ij}. In contrast
to our model, the instanton model can describe neither Casimir
scaling~\cite{Shevchenko:2001ij} nor the linear rise of the confining
potential~\cite{Chen:1999ct}.

Figure~\ref{Fig_Static_Quark-Antiquark_Potential_F_vs_A}
shows the static $SU(N_c = 3)$ potential for fundamental sources
$V_{\fundamental}(R) = V_3(R)$ (solid line) and adjoint sources
$V_{\adjoint}(R) = V_8(R)$ (dashed line) as a function of the dipole
size $R$ in comparison to $SU(3)$ lattice
data~\cite{Bali:2000un,Bali:2001gf}.  The model results are in good
agreement with the lattice data. In particular, the obtained Casimir
scaling behavior is strongly supported by $SU(3)$ lattice
data~\cite{Deldar:1999vi,Bali:2000un}.  This, however, points also to
a shortcoming of our model: From
Eq.~(\ref{Eq_Vr(R)_Gaussian_approximation}) and
Fig.~\ref{Fig_Static_Quark-Antiquark_Potential_F_vs_A} it is clear
that {\em string breaking} is described neither for fundamental nor
for adjoint dipoles in our model which indicates that not only
dynamical fermions (quenched approximation) but also some gluon
dynamics are missing.

An extension of the model that allows one to describe color screening
remains a major challenge.  Without such a modification, our model
unfortunately cannot contribute to the recent discussion on the
scaling behavior of $k$-string tensions $\sigma_k$, i.e.\ the tensions
of strings connecting sources with $\cal{N}$-ality $k \ge 1$. A source
of $\cal{N}$-ality $k\le N_c/2$ is defined as a source in the
representation constructed from the tensor product of quarks---objects
transforming under the fundamental representation---and
antiquarks---objects transforming under the conjugated
representation---where $k$ is the number of quarks minus the number of
antiquarks modulo $N_c$; see e.g.~\cite{Lucini:2001nv}.  For $SU(N_c)$
with $N_c \ge 4$, $k$ strings are particularly interesting since in
addition to the fundamental ($k=1$) string other strings also exist
that are stable against color screening. Based on lattice results for
$N_c = 4,5,6$~\cite{Lucini:2001nv,DelDebbio:2001kz+X}, the present
debate is whether the corresponding string tensions show Casimir
scaling behavior $\sigma_k \propto k(N_c-k)$ , the sine law behavior
$\sigma_k \propto \sin(\pi k /N_c)$ predicted from M-theory approaches
to QCD~\cite{Hanany:1997hr}, or simply the behavior of $k$
non-interacting fundamental strings $\sigma_k =
k\sigma_{\fundamental}$. Physical explanations of the lattice
results obtained are discussed, for example, in the center vortex confinement
mechanism~\cite{Greensite:2002yn,Greensite:2003bk}.

%
%
%
%
\section{Chromo-Field Distributions of Color Dipoles}
\label{Sec_Flux_Tube}

In this section we compute the chromo-electric fields generated by a
static color dipole in the fundamental and adjoint representation of
$SU(N_c)$. We find formation of a color flux tube that confines the
two color sources in the dipole. This confining string is analyzed
quantitatively. Its mean squared radius is calculated and transverse
and longitudinal energy density profiles are provided. The interplay
between perturbative and non-perturbative contributions to the
chromo-field distributions is investigated and exact Casimir scaling
is found for both contributions.

\begin{figure}[t]
\centerline{\epsfig{figure=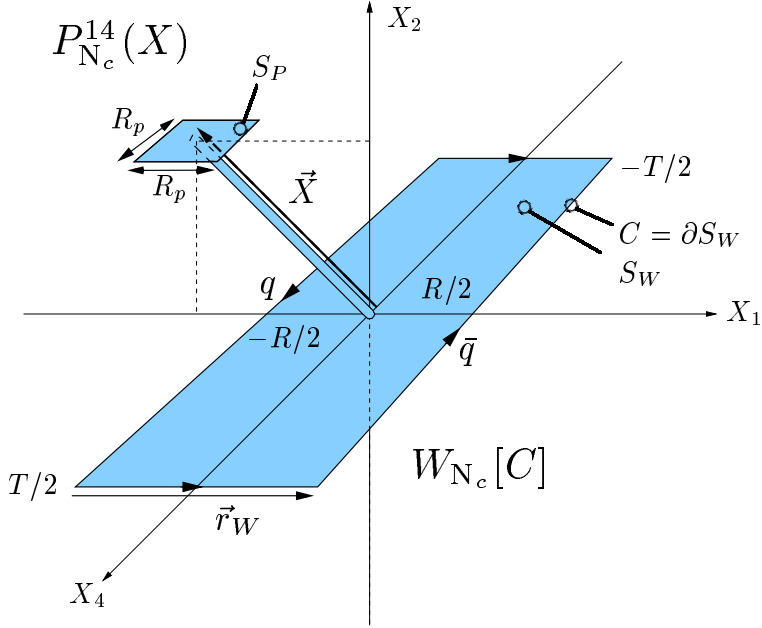,width=10.cm}}
\caption{\small
  The plaquette-loop geometry needed to compute the squared
  chromo-electric field $E_{\parallel}^2(X)$ generated by a static
  color dipole in the fundamental $SU(N_c)$ representation
  ($r=\Fundamental$).  The rectangular path $C$ indicates the
  world-line of the static dipole described the {\WW} loop
  $W_{\fundamental}[C]$. The square with side length $R_P$ illustrates
  the plaquette $P_{\fundamental}^{14}(X)$.  The shaded areas
  represent the minimal surfaces used in our computation of the
  chromo-field distributions.  The thin tube allows us to compare the
  gluon field strengths in surface $S_P$ with the gluon field
  strengths in surface $S_W$.}
\label{Fig_PW_arrangement}
\end{figure}

As already explained in Sec.~\ref{Sec_Static_Potential}, the static
color dipole---two static color sources separated by a distance $R$ in
a net color singlet state---is described by a {\WW} loop $W_r[C]$ with
a rectangular path $C$ of spatial extension $R$ and temporal extension
$T\to\infty$ (cf.\ Fig.~\ref{Fig_ONE_WWL}) where $r$ indicates the
$SU(N_c)$ representation of the sources considered. A second small
quadratic loop or plaquette in the fundamental representation placed
at the space-time point $X$ with side length $R_P\to 0$ and oriented
along the $\alpha\beta$ axes,
\be
        P_{\fundamental}^{\alpha \beta}(X) 
        = \tilde{\Tr}_{\fundamental}
        \exp\!\!\left[
        -i g \oint_{C_P}\!\!\!dZ_{\mu} \G_{\mu}^a(Z) t_{\fundamental}^a 
        \right] 
        = 1 
        - R_P^4\frac{g^2}{4N_c}\G_{\alpha\beta}^a(X)\G_{\alpha\beta}^a(X) 
        + \Order(R_P^6)
        \ ,
\label{Eq_plaquette}
\ee
is needed---as a ``Hall probe''---to calculate the chromo-field
distributions at the space-time point $X$ caused by the static
sources~\cite{Fukugita:1983du,Flower:gs}
\bea
        \Delta G_{r\,\alpha \beta}^2(X) 
        & := &
        \Big\langle 
        \frac{g^2}{4\pi^2}\G_{\alpha\beta}^a(X)\G_{\alpha\beta}^a(X)
        \Big\rangle_{W_r[C]}
        - 
        \Big\langle 
        \frac{g^2}{4\pi^2}\G_{\alpha\beta}^a(X)\G_{\alpha\beta}^a(X)
        \Big\rangle_{\mbox{\scriptsize vac}}
\label{Eq_DeltaG2_definition}\\
        & = &
        -\,\lim_{R_P \to 0}\inv{R_P^4} \frac{N_c}{\pi^2} 
        \left[
        \frac
        {\langle W_r[C] P_{\fundamental}^{\alpha \beta}(X) \rangle}
        {\langle W_r[C] \rangle}
        - \langle P_{\fundamental}^{\alpha \beta}(X) \rangle
        \right]
\label{Eq_DeltaG2_formula}
\eea
with {\em no} summation over $\alpha$ and $\beta$
in~(\ref{Eq_plaquette}), (\ref{Eq_DeltaG2_definition}),
and~(\ref{Eq_DeltaG2_formula}). In
definition~(\ref{Eq_DeltaG2_definition})
$\langle\ldots\rangle_{W_r[C]}$ indicates the VEV in the presence of
the static color dipole while $\langle\ldots\rangle_{\mbox{\scriptsize
    vac}}$ indicates the VEV in the absence of any color sources.
Depending on the plaquette orientation indicated by $\alpha$ and
$\beta$, one obtains from~(\ref{Eq_DeltaG2_formula}) the squared
components of the chromo-electric and chromo-magnetic fields at the
space-time point $X$:
\be
        \Delta G_{r\,\alpha \beta}^2(X) 
        = \frac{g^2}{4\pi^2}
        \left( \barray{cccc}
        0       & B_z^2 & B_y^2 & E_x^2 \\
        B_z^2   & 0     & B_x^2 & E_y^2 \\
        B_y^2   & B_x^2 & 0     & E_z^2 \\
        E_x^2   & E_y^2 & E_z^2 & 0
        \earray \right)(X)
        \ , 
\label{Eq_chromo-electromagnetic_fields}        
\ee
i.e.\ space-time plaquettes ($\alpha\beta=i4$) measure chromo-electric
fields and space-space plaquettes ($\alpha\beta=ij$) chromo-magnetic
fields. As shown in Fig.~\ref{Fig_PW_arrangement}, we place the static
color sources on the $X_1$ axis at $(X_1 = \pm R/2,0,0,X_4)$ and use
the following notation plausible from symmetry arguments
\be
        E_{\parallel}^2 = E_x^2
        \ ,\quad
        E_{\perp}^2 = E_y^2 = E_z^2
        \ ,\quad
        B_{\parallel}^2 = B_x^2
        \ ,\quad
        B_{\perp}^2 = B_y^2 = B_z^2
        \ .
\label{Eq_E_B_para_perp}
\ee
Figure~\ref{Fig_PW_arrangement} illustrates also the plaquette
$P_{\fundamental}^{14}(X)$ at $X = (X_1, X_2,0,0)$ needed to compute
$E_{\parallel}^2(X)$. Due to symmetry arguments, the complete
information on the chromo-field distributions is obtained from
plaquettes in ``transverse'' space $\mbox{$X = (X_1, X_2,0,0)$}$ with
four different orientations, $\alpha\beta = 14,\,24,\,13,\,23$ [cf.\ 
Eq.~(\ref{Eq_E_B_para_perp})].

The {\em energy} and {\em action density distributions} around a
static color dipole in the $SU(N_c)$ representation $r$ are given by
the squared chromo-field distributions
\bea
        \varepsilon_r(X) 
        & = & 
        \inv{2}\left[-\vec{E}^2(X)+\vec{B}^2(X)\right]
\label{Eq_energy_density}\\
        \actiondensity_r(X)
        & = &
        -\inv{2}\left[\vec{E}^2(X)+\vec{B}^2(X)\right]
\label{Eq_action_density}
\eea
with signs according to Euclidean space-time conventions. Low-energy
theorems that relate the energy and action stored in the chromo-fields
of the static color dipole to the corresponding ground state energy
are discussed in the next section.

For the chromo-field distributions of a static color dipole in the
{\em fundamental} representation of $SU(N_c)$, i.e.\ a static
quark-antiquark pair, we obtain with our results for the VEV of one
loop~(\ref{Eq_final_result_<W[C]>}) and the correlation of two loops
in the fundamental
representation~(\ref{Eq_final_Euclidean_result_<W[C1]W[C2]>_fundamental})
\bea
        &&\Delta G_{\fundamental\,\alpha\beta}^2(X) =
        -\lim_{R_P \to 0}\inv{R_P^4}\frac{N_c}{\pi^2} 
        \exp\left[-\frac{C_2(\!\Fundamental\!)}{2}\,\chi_{S_P S_P}\right]
\label{Eq_chromo_fields_F_1}\\
        &&\hspace*{2cm}\times 
        \Bigg(\frac{N_c+1}{2N_c}\exp\!\left[-\frac{N_c-1}{2N_c}\chi_{S_P S_W}\right]
        + \frac{N_c-1}{2N_c}\exp\!\left[ \frac{N_c+1}{2N_c}\chi_{S_P S_W}\right] - 1\Bigg)
\nonumber
\eea
where $\chi_{S_i S_j}$ is defined in~(\ref{Eq_chi_Si_Sj}). The
subscripts $P$ and $W$ indicate surface integrations to be performed
over the surfaces spanned by the plaquette and the Wegner-Wilson loop,
respectively. Choosing the surfaces---as illustrated by the shaded
areas in Fig.~\ref{Fig_PW_arrangement}---to be the minimal surfaces
connected by an infinitesimal thin tube (which gives no contribution
to the integrals) it is clear that $\chi_{S_P S_P} \propto R_P^4$ and
$\chi_{S_P S_W} \propto R_P^2$. Being interested in the chromo-fields
at the space-time point $X$, the extension of the quadratic plaquette
is taken to be infinitesimally small, $R_P \rightarrow 0$, so that one
can expand the exponential functions and keep only the term of lowest
order in $R_P$:
\be
        \Delta G_{\fundamental\,\alpha\beta}^2(X) = 
        -\,C_2(\!\Fundamental\!)\,\lim_{R_P \to 0}\inv{R_P^4}\,\inv{4\pi^2}\,\chi_{S_P S_W}^2
        \ .
\label{Eq_chromo_fields_F_final_result}
\ee
This result---obtained with the matrix cumulant expansion in a very
straightforward way---agrees exactly with the result derived
in~\cite{Rueter:1994cn} with the expansion method. Indeed, the
expansion method agrees for small $\chi$ functions with the matrix
cumulant expansion (Berger-Nachtmann approach) used in this work but
breaks down for large $\chi$ functions, where the matrix cumulant
expansion is still applicable.

The chromo-field distributions of a static color dipole in the {\em
  adjoint} representation of $SU(N_c)$, i.e.\ a static gluino pair,
are computed analogously. Using our
result~(\ref{Eq_final_Euclidean_result_<Wf[C1]Wa[C2]>}) for the
correlation of one loop in the fundamental representation (plaquette)
with one loop in the adjoint representation (static sources), one
obtains
\bea
        \!\!\!\!&&\!\!\!\!\!\!\!\!
        \Delta G_{\adjoint\,\alpha\beta}^2(X) =
        -\lim_{R_P \to 0}\inv{R_P^4}\frac{N_c}{\pi^2} 
        \exp\left[-\frac{C_2(\!\Fundamental\!)}{2}\chi_{S_P S_P}\right]\,
        \Bigg(\!\inv{N_c^2\!-\!1}\,\exp\!\Big[\frac{N_c}{2}\,\chi_{S_P S_W}\Big]
\nonumber\\
        && \hskip 1.5cm
        +\,\frac{N_c\!+\!2}{2(N_c\!+\!1)}\exp\!\Big[\!-\inv{2}\,\chi_{S_P S_W}\Big]
        +\frac{N_c\!-\!2}{2(N_c\!-\!1)}\exp\!\Big[\inv{2}\,\chi_{S_P S_W}\Big]
        -1\!\Bigg)
\label{Eq_chromo_fields_A_1}
\eea
which reduces---as explained for sources in the fundamental
representation---to
\be
        \Delta G_{\adjoint\,\alpha\beta}^2(X) = 
        -\,C_2(\!\Adjoint\!)\,\lim_{R_P \to 0}\inv{R_P^4}\,\inv{4\pi^2}\,\chi_{S_P S_W}^2
        \ .
\label{Eq_chromo_fields_A_final_result}
\ee
Thus, the squared chromo-electric fields of an adjoint dipole differ
from those of a fundamental dipole only in the eigenvalue of the
corresponding quadratic Casimir operator $C_2(r)$. In fact, {\em
  Casimir scaling} of the chromo-field distributions holds for dipoles
in any representation $r$ of $SU(N_c)$ in our model. As can be seen
with the low-energy theorems discussed below, this is in line with the
Casimir scaling of the static dipole potential found in the previous
section. In addition to lattice investigations that show Casimir
scaling of the static dipole potential to high accuracy in
$SU(3)$~\cite{Deldar:1999vi,Bali:2000un}, Casimir scaling of the
chromo-field distributions has been considered on the lattice as well
but only for $SU(2)$~\cite{Trottier:1995fx}. Here only slight
deviations from the Casimir scaling hypothesis have been found which
were interpreted as hints toward adjoint quark screening.

In our model the shape of the field distributions around the color
dipole is identical for all $SU(N_c)$ representations $r$ and given by
$\chi_{S_P S_W}^2$. This again illustrates the shortcoming of our
model discussed in the previous section. Working in the quenched
approximation, one expects a difference between fundamental and
adjoint dipoles: {\em string breaking} cannot occur in fundamental
dipoles as dynamical quark-antiquark production is excluded but should
be present for adjoint dipoles because of gluonic vacuum polarization.
Comparing~(\ref{Eq_chromo_fields_F_final_result})
with~(\ref{Eq_chromo_fields_A_final_result}) it is clear that this
difference is not described in our model. In fact, as shown in
Sec.~\ref{Sec_Static_Potential}, string breaking is described neither
for fundamental nor for adjoint dipoles. Interestingly, even on the
lattice there has been no striking evidence for adjoint quark
screening in quenched QCD~\cite{Kallio:2000jc}. It is even conjectured
that the {\WW} loop operator is not suited to studies of string
breaking~\cite{Gusken:1997sa+X}.

In the LLCM there are perturbative ($\pert$) and non-perturbative
($\nprt$) contributions to the chromo-electric fields according to the
structure of the gluon field strength correlator, (\ref{Eq_Ansatz})
and~(\ref{Eq_F_decomposition}),
\bea
        \Delta G_{r\,\alpha\beta}^2(X) 
        & = & 
        -C_2(r)\,\lim_{R_P \to 0}\inv{R_P^4}\,\inv{\pi^2}\,
\label{Eq_chromo_fields_F_no_interference}\\
        && \times\left(
          \left[\chi_{S_P S_W}^{\pert}(X)\right]_{\alpha\beta}^2 
        + \left\{
          \left[\chi_{S_P S_W}^{\nprt\,nc}(X)\right]_{\alpha\beta}
          +  \left[\chi_{S_P S_W}^{\nprt\,c}(X)\right]_{\alpha\beta}
        \right\}^2
      \right) 
      \ .
\nonumber
\eea
Interference of perturbative and non-perturbative correlations is not
considered to be in line with the applications of our model to
high-energy
scattering~\cite{Shoshi:2002in,Shoshi:2002ri,Shoshi:2002fq,Shoshi:2002mt}
with separate hard (perturbative) and soft (non-perturbative) Pomeron
exchanges.  The interferences do not change the qualitative picture.
Slight modifications occur in regions where fields originating from
perturbative and non-perturbative correlations are of similar size.
For the $\chi$ functions
in~(\ref{Eq_chromo_fields_F_no_interference}), we give directly in the
following the final results obtained with the minimal surfaces shown
in Fig.~\ref{Fig_PW_arrangement}.  Details of their derivation can be
found in Appendix~\ref{Sec_Chi_Computation}.

The {\em perturbative contribution} ($P$) described by massive gluon
exchange leads, of course, to the well-known {\em color Yukawa field}
which reduces to the {\em color Coulomb field} for $m_G=0$. It
contributes only to the chromo-electric fields, $E_{\parallel}^2 =
E_x^2$ ($\alpha\beta=14$) and $E_{\perp}^2 = E_y^2 = E_z^2$
($\alpha\beta=24$), and reads explicitly for $X = (X_1, X_2, 0, 0)$
\bea
\!\!\!\!\!\!\!\!\!\!\!\!
        \left[\chi_{S_P S_W}^{\pert}(X)\right]_{14}
        &\!\!=\!\!& -\,\frac{R_P^2}{2}\!\int_{-\infty}^{\infty}\!\!\!d\tau
        \left\{
        (X_1 - R/2)\,
        g^2(Z_{1A}^2)\,D_\pert(Z_{1A}^2)
        \right.
\nonumber\\
        && \hphantom{-\,\frac{R_P^2}{2}\!\int_{-\infty}^{\infty}\!\!\!d\tau\Big(}
        \left.       
        - \,(X_1 + R/2)\,
        g^2(Z_{1C}^2)\,D_\pert(Z_{1C}^2)
        \right\}
\label{Eq_Chi_PW_p_14}\\
\!\!\!\!\!\!\!\!\!\!\!\!
        \left[\chi_{S_P S_W}^{\pert}(X)\right]_{24}
        &\!\!=\!\!& -\,\frac{R_P^2}{2}\!\int_{-\infty}^{\infty}\!\!\!d\tau\,X_2
        \left\{
        g^2(Z_{1A}^2)\,D_\pert(Z_{1A}^2)
        - g^2(Z_{1C}^2)\,D_\pert(Z_{1C}^2)
        \right\}
\label{Eq_Chi_PW_p_24}
\eea
with the perturbative correlation function~(\ref{Eq_Dp(z,mg)}), the
running coupling~(\ref{Eq_g2(z_perp)}), and
\be
        Z_{1A}^2 = \left(X_1\!-\!\frac{R}{2}\right)^2+X_2^2+\tau^2
        \quad \mbox{and} \quad
        Z_{1C}^2 = \left(X_1\!+\!\frac{R}{2}\right)^2+X_2^2+\tau^2
        \ .
\label{Eq_|Z1A|_|Z1C|}
\ee

The {\em non-confining non-perturbative contribution} ($\nprt\,nc$)
has the same structure as the perturbative contribution---as expected
from the identical tensor structure---but differs, of course, in the
prefactors and the correlation function, $D_1 \neq D_p$. Its
contributions to the chromo-electric fields $E_{\parallel}^2 = E_x^2$
($\alpha\beta=14$) and $E_{\perp}^2 = E_y^2 = E_z^2$
($\alpha\beta=24$) read for $X = (X_1, X_2, 0, 0)$
\bea
        \left[\chi_{S_P S_W}^{\nprt\,\,nc}(X)\right]_{14}
        &=&
        -\,\frac{R_P^2 \pi^2 G_2 (1\!-\!\kappa)}{6\,(N_c^2\!-\!1)}
        \!\int_{-\infty}^{\infty}\!\!\!d\tau
        \Big\{
        (X_1 - R/2)\,D_1(Z_{1A}^2)
\nonumber\\
        && 
        \hphantom{-\,\frac{R_P^2 \pi^2 G_2}{6\,(N_c^2\!-\!1)}}
        - \,(X_1 + R/2)\,D_1(Z_{1C}^2)
        \Big\}
\label{Eq_Chi_PW_np_nc_14}\\
        \left[\chi_{S_P S_W}^{\nprt\,\,nc}(X)\right]_{24}
        &=&
        -\,\frac{R_P^2 \pi^2 G_2 (1\!-\!\kappa)}{6\,(N_c^2\!-\!1)}
        \!\int_{-\infty}^{\infty}\!\!\!d\tau\,X_2
        \Big\{D_1(Z_{1A}^2)-D_1(Z_{1C}^2)\Big\}
\label{Eq_Chi_PW_np_nc_24}
\eea
with the exponential correlation
function~(\ref{Eq_SVM_correlation_functions}) and $Z_{1A}^2$ and
$Z_{1C}^2$ as given in~(\ref{Eq_|Z1A|_|Z1C|}).

The {\em confining non-perturbative contribution} ($\nprt\,c$) has a
different structure that leads to confinement and flux-tube formation.
It gives contributions only to the chromo-electric field
$E_{\parallel}^2 = E_x^2$ ($\alpha\beta=14$) which read for $X = (X_1,
X_2, 0, 0)$
\bea
        \left[\chi_{S_P S_W}^{\nprt\,\,c}(X)\right]_{14}
        & = & 
        R_P^2 R 
        \frac{\pi^2 G_2 \kappa}{3\,(N_c^2\!-\!1)}
        \int_{0}^{1} d\rho\,
        D^{(3)}(\vec{Z_{\perp}}^2)
        \ ,
\label{Eq_Chi_PW_np_c_14}
\eea
with the correlation function given in~(\ref{Eq_D(3)np_c(z,a)}) as
derived from the exponential correlation
function~(\ref{Eq_SVM_correlation_functions}), and
\be
        \vec{Z}_{\perp}^2 = [X_1+(1/2-\rho)R]^2+X_2^2
        \ .
\label{Eq_|Z|}
\ee

In our model there are no contributions to the {\em chromo-magnetic
  fields}, i.e.\ the static color charges do not affect the magnetic
background field
\be
        B_{\parallel}^2 = B_x^2 = 0
        \quad \mbox{and} \quad
        B_{\perp}^2 = B_y^2 = B_z^2 = 0
        \ ,
\label{Eq_B^2=0}
\ee
which can be seen from the corresponding plaquette-loop geometries as
pointed out in Appendix~\ref{Sec_Chi_Computation}. Thus, the energy
and action densities are identical in our approach and completely
determined by the squared chromo-electric fields
\be
        \varepsilon_r(X) 
        \,\,=\,\,\actiondensity_r(X)
        \,\,=\,\, -\inv{2}\,\vec{E}^2(X)
        \ .
\label{Eq_energy=action_density}
\ee
This picture is in agreement with other effective theories of
confinement such as the 't~Hooft--Mandelstam
picture~\cite{Mandelstam:1976pi+X} or dual QCD~\cite{Baker:bc} and,
indeed, a relation between the dual Abelian Higgs model and the SVM
has been established~\cite{Baker:1998jw}.  In contrast, lattice
investigations work at scales at which the chromo-electric and
chromo-magnetic fields are of similar
magnitude~\cite{Bali:1994de,Haymaker:fm,Green:1996be,Pennanen:1997qm}.
Indeed, these simulations have been performed in $SU(N_c=2)$ at bare
couplings $g^2_0$ down to $\beta_0=4/g^2_0=2.74$ corresponding to a
minimum lattice cutoff of $0.04\,\fm$, which determines also the
minimum size of the plaquette used in the measurements of the color
fields.  Interestingly, as shown in Figs.~13 and 14
of~\cite{Bali:1994de}, the action density slightly decreases by
decreasing the lattice spacing from $0.08\,\fm$ ($\beta_0=2.5$) to
$0.05\,\fm$ ($\beta_0=2.635$). In our model the vanishing of the
chromo-magnetic fields determines the value of the Callan-Symanzik
$\beta$ function at the renormalization scale at which our
non-perturbative component is working. This is shown as a result of
low-energy theorems in the next section.

\begin{figure}[h]
\centerline{
\epsfig{figure=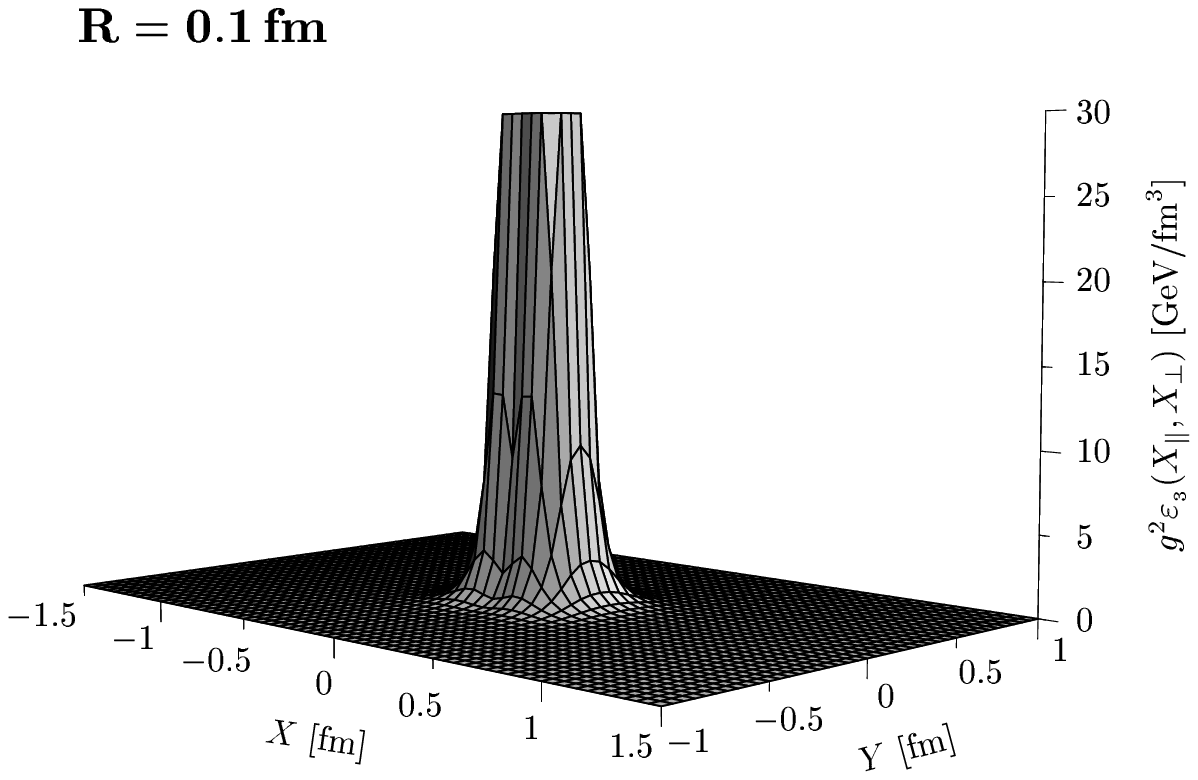,width=7.7cm}
\hspace*{-0.8cm}
\epsfig{figure=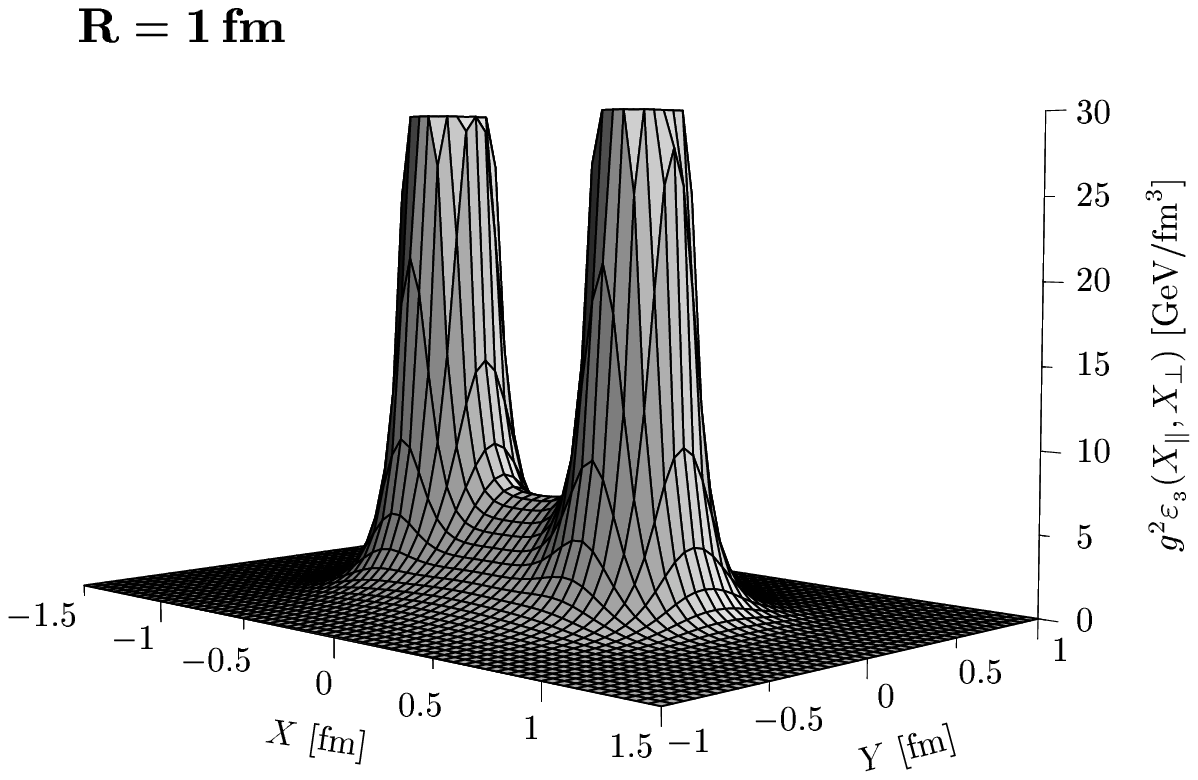,width=7.7cm}}
\centerline{
\epsfig{figure=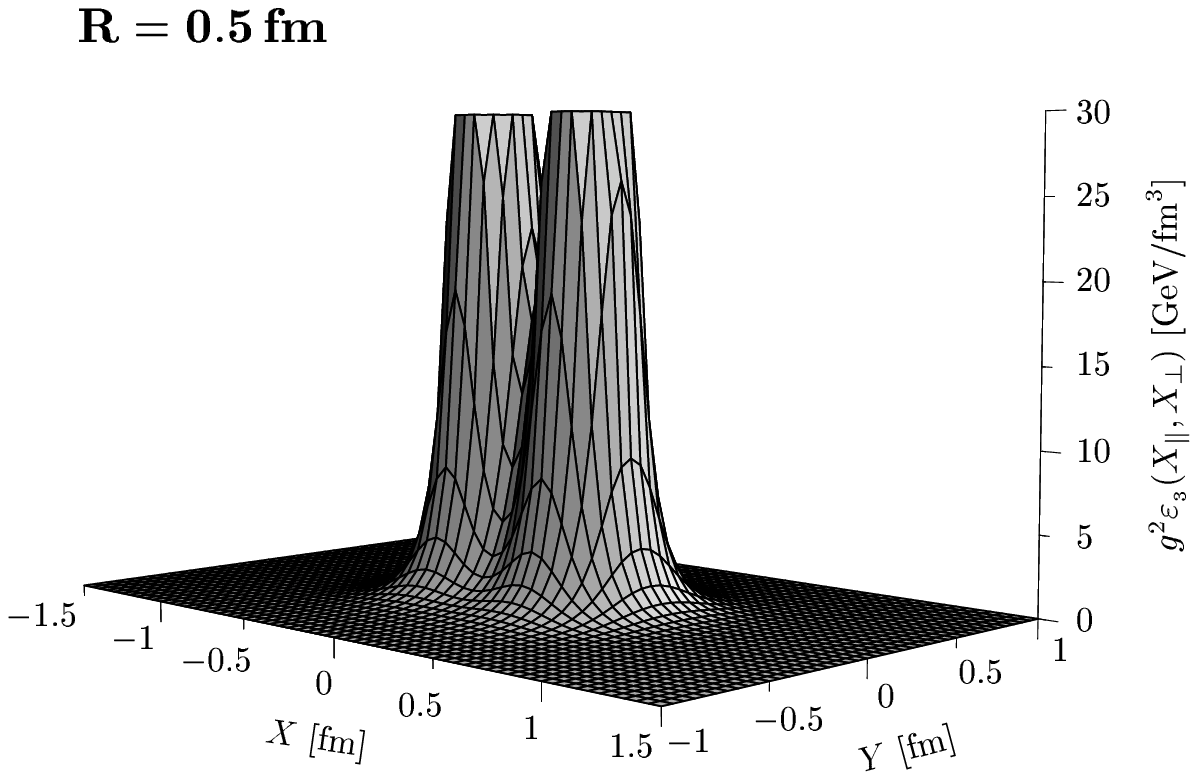,width=7.7cm}
\hspace*{-0.8cm}
\epsfig{figure=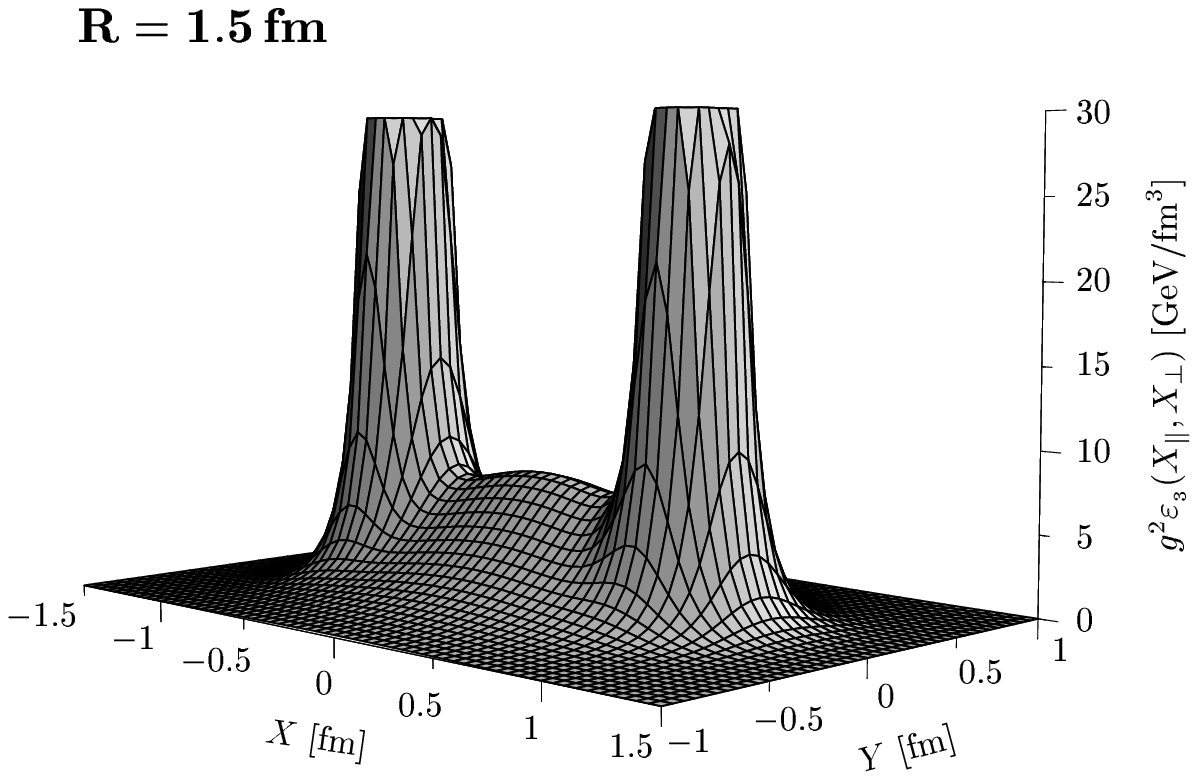,width=7.7cm}}
\caption{\small
  Energy density distributions $g^2\varepsilon_3(X_1,X_2\!=\!X_3)$
  generated by a color dipole in the fundamental $SU(3)$
  representation ($r\!=\!3$) for quark-antiquark separations of $R =
  0.1,\,0.5,\,1$ and $1.5\,\fm$. Flux-tube formation leads to the
  confining QCD string with increasing dipole size $R$.}
\label{Fig_3D_profiles}
\end{figure}

In Fig.~\ref{Fig_3D_profiles} the energy density distributions
$g^2\varepsilon_3(X_1,X_2\!=\!X_3)$ generated by a color dipole in the
fundamental $SU(N_c=3)$ representation ($r\!=\!3$) are shown for
quark-antiquark separations of $R = 0.1,\,0.5,\,1$ and $1.5\,\fm$.
With increasing dipole size $R$, one sees explicitly the formation of
the flux tube which represents the confining QCD string.

\begin{figure}[p]
\centerline{\epsfig{figure=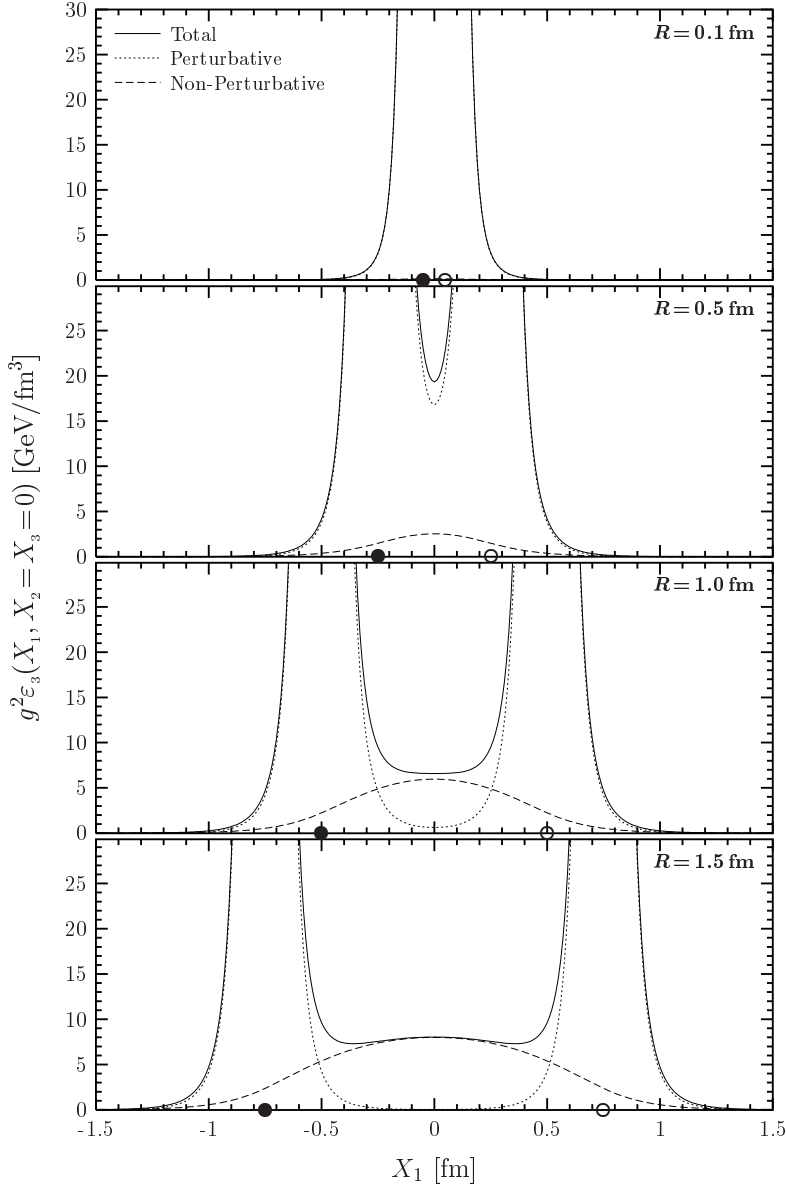,width=10cm}}
\caption{\small
  Longitudinal energy density profiles
  $g^2\varepsilon_3(X_1,X_2\!=\!X_3\!=\!0)$ generated by a color
  dipole in the fundamental $SU(3)$ representation ($r\!=\!3$) for
  quark-antiquark separations of $R = 0.1,\,0.5,\,1$ and $1.5\,\fm$.
  The dotted and dashed lines give the perturbative and
  non-perturbative contributions, respectively, and the solid lines
  the sum of both.  The open and filled circles indicate the quark and
  antiquark positions. For small dipoles, $R=0.1\,\fm$, perturbative
  physics dominates and non-perturbative correlations are negligible.
  For large dipoles, $R\gtsim 1\,\fm$, the formation the confining
  string (flux tube) can be seen which dominates the chromo-electric
  fields between the color sources.}
\label{Fig_L_profiles}
\end{figure}

\begin{figure}[p]
\centerline{\epsfig{figure=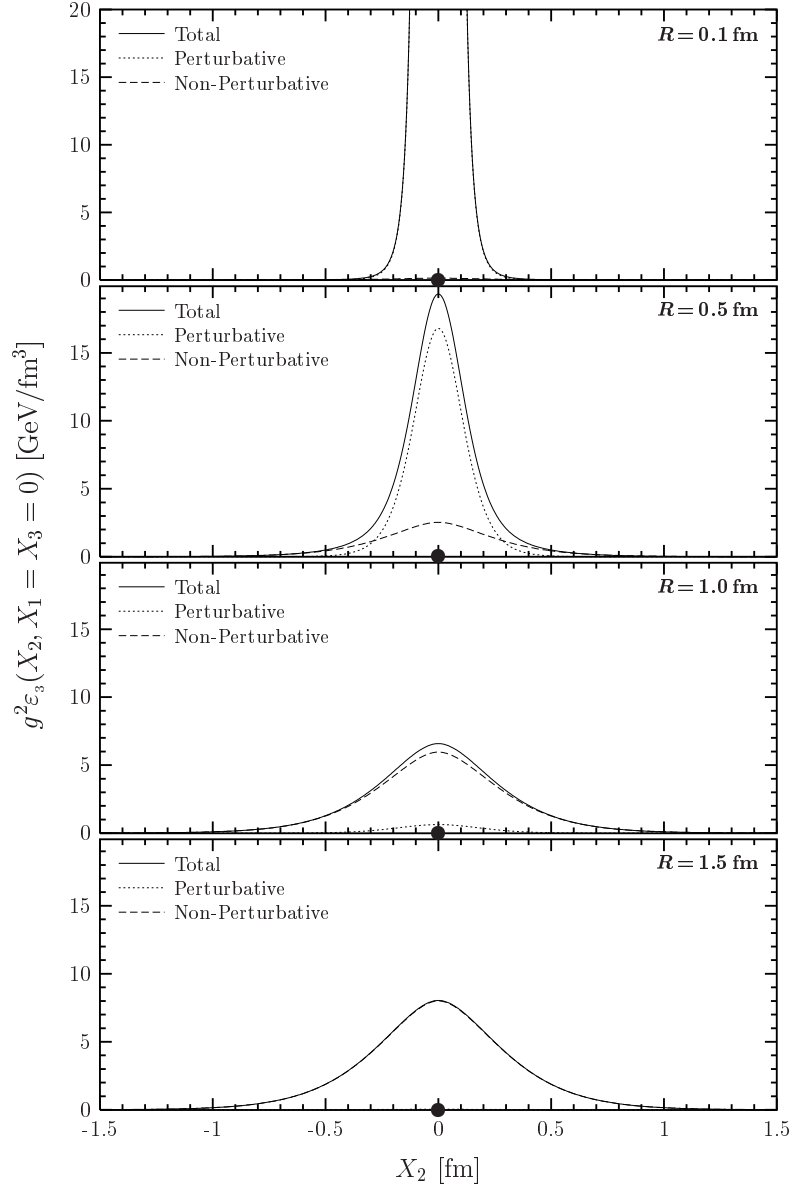,width=10cm}}
\caption{\small
  Transverse energy density profiles
  $g^2\varepsilon_3(X_2,X_1\!=\!X_3\!=\!0)$ generated by a color
  dipole in the fundamental $SU(3)$ representation ($r\!=\!3$) for
  quark-antiquark separations of $R = 0.1,\,0.5,\,1$ and $1.5\,\fm$.
  The dotted and dashed lines give the perturbative and
  non-perturbative contributions, respectively, and the solid lines
  the sum of both.  The filled circles indicate the positions of the
  color sources. For small dipoles, $R=0.1\,\fm$, perturbative physics
  dominates and non-perturbative correlations are negligible. For
  large dipoles, $R\gtsim 1\,\fm$, the formation the confining string
  (flux tube) can be seen which dominates the chromo-electric fields
  between the color sources.}
\label{Fig_T_profiles}
\end{figure}

The {\em longitudinal} and {\em transverse energy density profiles}
generated by a color dipole in the fundamental representation ($r=3$)
of $SU(N_c=3)$ are shown for quark-antiquark separations (dipole
sizes) of $R = 0.1,\,0.5,\,1$ and $1.5\,\fm$ in
Figs.~\ref{Fig_L_profiles} and~\ref{Fig_T_profiles}. The perturbative
and non-perturbative contributions are given by the dotted and dashed
lines, respectively, and the sum of both in the solid lines. The open
and filled circles indicate the quark and antiquark positions. As can
be seen from~(\ref{Eq_DeltaG2_formula})
and~(\ref{Eq_chromo-electromagnetic_fields}), we cannot compute the
energy density separately but only the product $g^2\varepsilon_r(X)$.
Nevertheless, a comparison of the total energy stored in
chromo-electric fields to the ground state energy of the color dipole
via low-energy theorems yields $g^2 = 10.2$ $(\equiv \alphaS=0.81)$
for the non-perturbative SVM component as shown in the next section.

In Figs.~\ref{Fig_L_profiles} and~\ref{Fig_T_profiles} the formation
of the confining string (flux tube) with increasing source separations
$R$ can again be seen explicitly: For small dipoles, $R=0.1\,\fm$,
perturbative physics dominates and non-perturbative correlations are
negligible. For large dipoles, $R\gtsim 1\,\fm$, the non-perturbative
correlations lead to formation of a narrow flux tube which dominates
the chromo-electric fields between the color sources.

\begin{figure}[t]
\centerline{\epsfig{figure=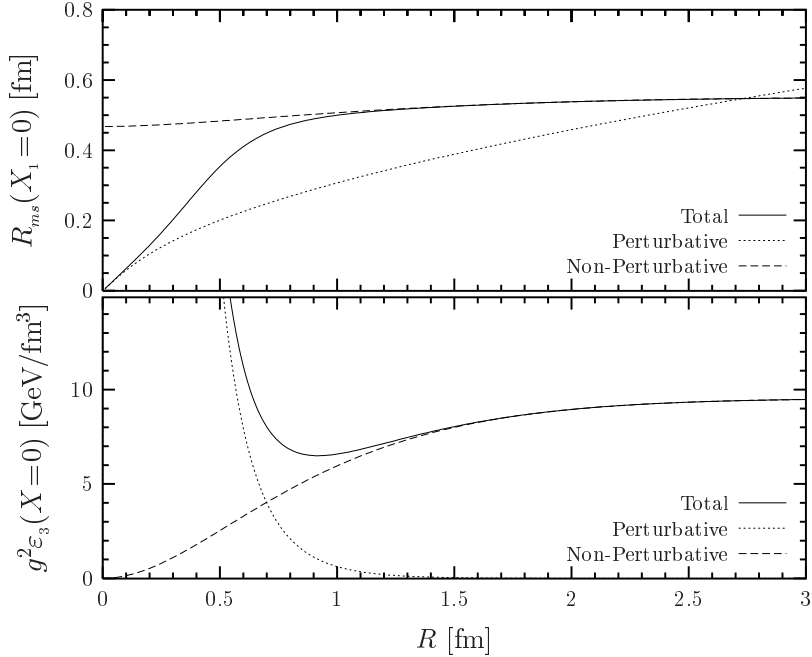,width=12cm}}
\caption{\small
  Root mean squared radius $R_{ms}$ of the flux tube and energy
  density in the center of a fundamental $SU(3)$ dipole
  $g^2\varepsilon_3(X=0)$ as a function of the dipole size $R$.
  Perturbative and non-perturbative contributions are given
  respectively by the dotted and dashed lines and the sum of both in
  the solid lines. For large $R$, both the width and height of the
  flux tube in the central region are governed completely by
  non-perturbative physics and saturate respectively at
  $R_{ms}^{R\to\infty}\approx 0.55\,\fm$ and
  $\varepsilon_3^{R\to\infty}(X=0)\approx 1\,\GeV/\fm^3$. The latter
  value is extracted with the result $g^2 = 10.2$ deduced from
  low-energy theorems in the next section.}
\label{Fig_R_ms_and_g2epsilon(0)}
\end{figure}

Figure~\ref{Fig_R_ms_and_g2epsilon(0)} 
shows the evolution of the transverse width (upper plot) and height
(lower plot) of the flux tube in the central region of the {\WW} loop
as a function of the dipole size $R$ where perturbative and
non-perturbative contributions are given by the dotted and dashed
lines, respectively, and the sum of both in the solid lines. The width
of the flux tube is best described by the root mean squared ($ms$)
radius
\be
        R_{ms}
        = \sqrt{\frac{\int dX_{\perp}\,X_{\perp}^3\,g^2\varepsilon_r(X_1=0,X_{\perp})}
        {\int dX_{\perp}\,X_{\perp}\,g^2\varepsilon_r(X_1=0,X_{\perp})}}
        \ ,
\label{Eq_R_ms}
\ee
which is universal for dipoles in all $SU(N_c)$ representations $r$ as
the Casimir factors divide out. The height of the flux tube is given
by the energy density in the center of the considered dipole,
$g^2\varepsilon_r(X=0)$. For large source separations, $R \gtsim
1\,\fm$, both the width and height of the flux tube in the central
region of the {\WW} loop are governed completely by non-perturbative
physics and saturate for a fundamental $SU(3)$ dipole
($r=\Fundamental=3$) at reasonable values of
\be
        R_{ms}^{R\to\infty}\approx 0.55\,\fm
        \quad \mbox{and} \quad 
        \varepsilon_3^{R\to\infty}(X=0)\approx 1\,\GeV/\fm^3
        \quad \mbox{with} \quad g^2 = 10.2
        \ .
\label{Eq_R_ms_and_g2epsilon(0)_saturation_values}
\ee

Note that the qualitative features of the non-perturbative SVM
component do not depend on the specific choice for the parameters,
surfaces, and correlation functions and have already been discussed
with the pyramid mantle choice of the surface and different
correlation functions in the first investigation of flux-tube
formation in the SVM~\cite{Rueter:1994cn}. The quantitative results,
however, are sensitive to the parameter values, the surface choice,
and the correlation functions and are presented above with the LLCM
parameters, the minimal surfaces, and the exponential correlation
function.

%
%
%
%
%
\section{Low-Energy Theorems}
\label{Sec_Low_Energy_Theorems}

In this section we use low-energy theorems to test the consistency of
the non-perturbative SVM component and to determine the value of the
Callan-Symanzik $\beta$ function and $\alphaS = g^2/(4\pi)$ at the
renormalization scale at which this component is working. The energy
and action sum rules considered allow us to confirm the consistency of
our loop-loop correlation result with the result obtained for the VEV
of one loop. Finally, we compare our results for $\beta$ and $\alphaS$
to model independent QCD results for the Callan-Symanzik $\beta$
function.

Many low-energy theorems have been derived in continuum theory by
Novikov, Shifman, Vainshtein, and Zakharov~\cite{Novikov:xi+X} and in
lattice gauge theory by Michael~\cite{Michael:1986yi}. Here we
consider the energy and action sum rules---known in lattice QCD as
{\em Michael sum rules}---that relate the energy and action stored in
the chromo-fields of a static color dipole to the corresponding ground
state energy~\cite{Wilson:1974sk,Brown:1979ya}
\be
        E_r(R) = 
        - \lim_{T \to \infty} \inv{T} 
        \ln \langle W_r[C] \rangle
        \ .
\label{Eq_Er(R)_def}        
\ee
In their original form~\cite{Michael:1986yi}, however, the Michael sum
rules are incomplete~\cite{Dosch:1995fz,Rothe:1995hu+X}. In
particular, significant contributions to the energy sum rule from the
trace anomaly of the energy-momentum tensor have been
found~\cite{Rothe:1995hu+X} that modify the naively expected relation
in line with the importance of the trace anomaly found for hadron
masses~\cite{Ji:1995sv}. Taking all these contributions into account,
the {\em energy} and {\em action sum rule} read
respectively~\cite{Rothe:1995hu+X,Michael:1995pv,Green:1996be}
\bea
        && 
        E_r(R) 
        = \int d^3X\,\varepsilon_r(X)
        - \inv{2}\frac{\beta(g)}{g}
        \int d^3X\,\actiondensity_r(X)
        \ ,
\label{Eq_energy_sum_rule}\\
        &&
        E_r(R) + R\,\frac{\partial E_r(R)}{\partial R}
        = - \frac{2\beta(g)}{g}
        \int d^3X\,\actiondensity_r(X)
        \ ,
\label{Eq_action_sum_rule}
\eea
where the Callan-Symanzik function is denoted by $\beta(g)=\mu
\partial g/\partial\mu$ with the renormalization scale $\mu$.

Inserting~(\ref{Eq_action_sum_rule}) into~(\ref{Eq_energy_sum_rule}),
we find the following relation between the total energy stored in the
chromo-fields $E_r^{\mbox{\scriptsize tot}}(R)$ and the ground state
energy $E_r(R)$:
\be
        E_r^{\mbox{\scriptsize tot}}(R) 
        := \int d^3X\,\varepsilon_r(X)
        = \inv{4}\left(3\,E_r(R) - R\frac{\partial E_r(R)}{\partial R} \right)
        \ .
\label{Eq_Etot-Er(R)_relation}
\ee
The difference from the naive classical expectation that the full
ground state energy of the static color sources is stored in the
chromo-fields is due to the trace anomaly
contribution~\cite{Rothe:1995hu+X} described by the second term on the
RHS of~(\ref{Eq_energy_sum_rule}). Indeed, for the Coulomb potential,
obtained in tree-level perturbation theory, the action sum
rule~(\ref{Eq_action_sum_rule}) shows explicitly that the trace
anomaly contribution vanishes on the classical level as expected.

With the low energy theorems~(\ref{Eq_action_sum_rule})
and~(\ref{Eq_Etot-Er(R)_relation}) the ratio of the integrated squared
chromo-magnetic to the integrated squared chromo-electric field
distributions can be derived
\be
        Q(R) := \frac{\int d^3X \vec{B}^2(X)}{\int d^3X \vec{E}^2(X)}
        = \frac{\left[2+3\,\beta(g)/g\right]\,E_r(R) 
        + \left[2-\beta(g)/g\right]\,R\,\frac{\partial E_r(R)}{\partial R}}
        {\left[2-3\,\beta(g)/g\right]\,E_r(R) 
        + \left[2+\beta(g)/g\right]\,R\,\frac{\partial E_r(R)}{\partial R}}
      \ .
\label{Eq_Q_ratio_general}
\ee
This ratio can be used, for example, to determine non-perturbatively
the Callan-Symanzik $\beta(g)$ function. For $SU(N_c=2)$ lattice
investigations along these lines have already been
performed~\cite{Bali:1995gm,Michael:1996aj,Green:1996be,Pennanen:1997qm}.\footnote{In~\cite{Bali:1994de}
  the $\beta$ function was determined similarly based on a
  high-statistics study of chromo-field distributions in $SU(N_c=2)$
  but unfortunately without taking the trace anomaly contribution into
  account.}

In the large $R$ region, the static color dipole potential can be
approximated by the linear potential $V_r(R)=\sigma_r R = E_r(R) -
E_{\self}$ with string tension $\sigma_r$ in the considered
representation $r$. In this approximation, the
ratio~(\ref{Eq_Q_ratio_general}) becomes the simple form
\be
        Q(R) \Big|_{V_r(R)=\sigma_r R}
        = \frac{2+\beta(g)/g}{2-\beta(g)/g}
      \ .
\label{Eq_Q_ratio_linear_potential}
\ee
Since the non-perturbative SVM component of our model describes the
confining linear potential for large source separations $R$, we can
use~(\ref{Eq_Q_ratio_linear_potential}) together with the vanishing of
the chromo-magnetic fields~(\ref{Eq_B^2=0}) to determine the value of
the Callan-Symanzik $\beta$ function at the scale $\mu_{\nprt}$ at
which the non-perturbative component is working
\be
        \frac{\beta(g)}{g}\Big|_{\mu =\mu_{\nprt}} = -2
        \ .
\label{Eq_beta/g=-2}
\ee
Here one should emphasize that this value is strictly valid only at
asymptotically large values of $R$, while perturbative correlations
must be taken into account to extend this investigation to smaller
values of $R$.

Concentrating on the confining non-perturbative component ($\nprt c$)
we now use (\ref{Eq_Etot-Er(R)_relation}) to determine the value of
$\alphaS = g^2/(4\pi)$ at which the non-perturbative SVM component is
working. The RHS of~(\ref{Eq_Etot-Er(R)_relation}) is obtained
directly from the confining contribution to the static potential
$E_r^{\nprt c}(R)=V_r^{\nprt c}(R)$ given in~(\ref{Eq_Vr(R)_NP_c}) in
Sec.~\ref{Sec_Static_Potential}. The left-hand side (LHS)
of~(\ref{Eq_Etot-Er(R)_relation}), however, involves a division by the
{\em a priori} unknown value of $g^2$ after integrating
$g^2\varepsilon_r(X)$ for the chromo-electric field of the confining
non-perturbative component~(\ref{Eq_Chi_PW_np_c_14}). As discussed in
the previous section, we cannot compute the energy density separately
but only the product $g^2\varepsilon_r(X)$.  Adjusting the value of
$g^2$ such that~(\ref{Eq_Etot-Er(R)_relation}) is exactly satisfied
for source separations of $R=1.5\,\fm$, we find that the
non-perturbative component is working at the scale $\mu_{\nprt}$ at
which
\be
        g^2(\mu_{\nprt}) = 10.2
        \quad \equiv \quad
        \alphaS(\mu_{\nprt}) = 0.81
        \ .
\label{Eq_alphaS=0.81}
\ee
As already mentioned in Sec.~\ref{Sec_QCD_Components}, we use this
value as a practical asymptotic limit for the simple one-loop
coupling~(\ref{Eq_g2(z_perp)}) used in our perturbative component.
Note that earlier SVM investigations along these lines found a smaller
value of $\mbox{$\alphaS(\mu_{\nprt}) = 0.57$}$ with the pyramid
mantle choice for the surface~\cite{Rueter:1994cn,Dosch:1995fz} but
were incomplete since only the contribution from the traceless part of
the energy-momentum tensor was considered in the energy sum rule.

\begin{figure}[t]
\centerline{\epsfig{figure=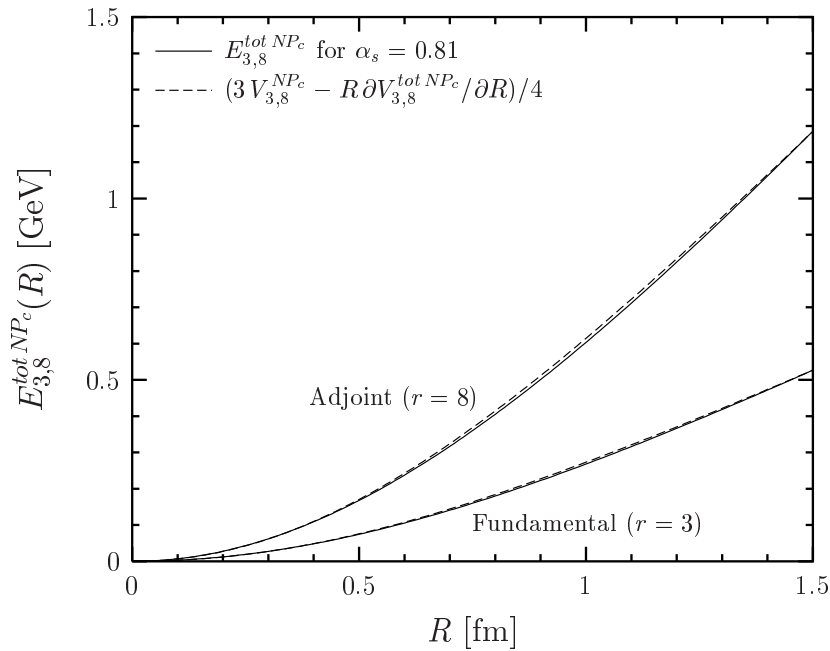,width=12cm}}
\caption{\small
  The total energy stored in the chromo-field distributions around a
  static color dipole of size $R$ in the fundamental ($r=3$) and
  adjoint ($r=8$) representation of $SU(3)$ from the confining
  non-perturbative SVM component, $E_{3,8}^{\mbox{\scriptsize
      tot}\,\nprt_c}(R)$, for $\alphaS = 0.81$ (solid lines) compared
  with the relation to the corresponding ground state energy (dashed
  lines) given by the low-energy
  theorem~(\ref{Eq_Etot-Er(R)_relation}). Good consistency is found
  even down to very small values of $R$.}
\label{Fig_NP_c_consistency}
\end{figure}

In Fig.~\ref{Fig_NP_c_consistency}
we show the total energy stored in the chromo-field distributions
around a static color dipole in the fundamental ($r=3$) and adjoint
($r=8$) representation of $SU(3)$ from the confining non-perturbative
SVM component, $E_{3,8}^{\mbox{\scriptsize tot}\,\nprt_c}(R)$, for
$\alphaS = 0.81$ (solid lines) as a function of the dipole size $R$.
Comparing this total energy, which appears on the LHS
of~(\ref{Eq_Etot-Er(R)_relation}), with the corresponding RHS
of~(\ref{Eq_Etot-Er(R)_relation}) (dashed lines), we find good
consistency even down to very small values of $R$. This is a
nontrivial and important result as it confirms the consistency of our
loop-loop correlation result---needed to compute the chromo-electric
field---with the result obtained for the VEV of one loop---needed to
compute the static potential $V_r^{\nprt_c}(R)$. Moreover, it shows
that the minimal surfaces ensure the consistency of our
non-perturbative component. The good consistency found for the pyramid
mantle choice of the surface relies on the naively expected energy sum
rule~\cite{Rueter:1994cn,Dosch:1995fz} in which the contribution from
the traceless part of the energy-momentum tensor is not taken into
account.

Let us discuss the values of $\beta/g$ and $g^2$ at the
renormalization scale $\mu_{\nprt}$---given respectively in
(\ref{Eq_beta/g=-2}) and (\ref{Eq_alphaS=0.81})---in comparison with
the perturbative expansion~\cite{Hagiwara:fs} and lattice
computations~\cite{Luscher:1993gh} of the Callan-Symanzik function in
pure $SU(N_c=3)$ gauge theory. We obtained (\ref{Eq_beta/g=-2}) and
(\ref{Eq_alphaS=0.81}) such that the renormalization scales
$\mu_{\nprt}$ appearing in the two equations should be in good
agreement.  Considering $\beta/g$ as a function $g^2$, one thus can
compare our combination with the perturbative
expansion~\cite{Hagiwara:fs}. This comparison shows that our result is
close to the curve obtained on the two-loop level in perturbation
theory. In contrast, the non-perturbative lattice results for the
$\beta$ function of L\"uscher {\it et al.}~\cite{Luscher:1993gh} are
in good agreement with the perturbative three-loop result computed in
the minimal modified subtraction ($\overline{\mathrm{MS}}$)
scheme~\cite{Tarasov:au}. However, it must be stressed that in the
lattice investigation the considered values of the running coupling
$g^2$ stay below $3.5$ while our comparison requires values up to
$g^2(\mu_{\nprt}) = 10.2$. Thus, relying on a large extrapolation of
the model independent QCD results, our comparison provides at best an
orientation. For a meaningful consistency check, we have to map out
the Callan-Symanzik function at smaller values of $R$, where also
perturbative correlations must be taken into account and thus
refinements of our treatment of renormalization are needed. The
low-energy theorems will provide crucial criteria for the success of
such improvements.

%
%
%
%
%
\section{Euclidean Approach to High-Energy Scattering}
\label{Sec_DD_Scattering}

In this section we present a Euclidean approach to high-energy
reactions of color dipoles in the eikonal approximation. After a short
review of the functional integral approach to high-energy
dipole-dipole scattering in Minkowski space-time, we generalize the
analytic continuation introduced by
Meggiolaro~\cite{Meggiolaro:1996hf+X} from parton-parton scattering to
dipole-dipole scattering. This shows how one can access high-energy
reactions directly in lattice QCD. We apply this approach to compute
the scattering of dipoles in the fundamental and adjoint
representation of $SU(N_c)$ at high-energy in the Euclidean LLCM. The
result shows the consistency with the analytic continuation of the
gluon field strength correlator used in all earlier applications of
the SVM and LLCM to high-energy scattering. Finally, we comment on the
QCD van der Waals potential which appears in the limiting case of two
static color dipoles.

In {\em Minkowski space-time}, high-energy reactions of color dipoles
in the eikonal approximation have been considered---as basis for
hadron-hadron, photon-hadron, and photon-photon reactions---in the
functional integral approach to high-energy collisions developed
originally for parton-parton
scattering~\cite{Nachtmann:1991ua+X,Nachtmann:ed.kt} and then extended
to gauge-invariant dipole-dipole
scattering~\cite{Kramer:1990tr,Dosch:1994ym,Dosch:RioLecture}. The
corresponding $T$-matrix element for the elastic scattering of two
color dipoles at transverse momentum transfer ${\vec q}_{\!\perp}$ ($t
= -{\vec q}_{\!\perp}^{\,\,2}$) and c.m.\ energy squared~$s$ reads
\be
        T^M_{r_1 r_2}(s,t,z_1,\vec{r}_{1\perp},z_2,\vec{r}_{2\perp}) =
        2is \int \!\!d^2b_{\!\perp} 
        e^{i {\vec q}_{\!\perp} {\vec b}_{\!\perp}}
        \left[1-S^M_{r_1 r_2}(s,{\vec b}_{\!\perp},z_1,\vec{r}_{1\perp},z_2,\vec{r}_{2\perp})\right]
\label{Eq_model_T_amplitude}
\ee
with the $S$-matrix element ($M$ refers to Minkowski space-time)
\be
        S^M_{r_1 r_2}(s,{\vec b}_{\!\perp},z_1,\vec{r}_{1\perp},z_2,\vec{r}_{2\perp})
        = \lim_{T \rightarrow \infty}
        \frac{\langle W_{r_1}[C_1] W_{r_2}[C_2]\rangle_M}
        {\langle W_{r_1}[C_1]\rangle_M \langle W_{r_2}[C_2]\rangle_M}
        \ .
\label{Eq_S_DD_Minkowski}
\ee
The color dipoles are considered in the $SU(N_c)$ representation $r_i$
and have transverse size and orientation ${\vec r}_{i\perp}$. The
longitudinal momentum fraction carried by the quark of dipole $i$ is
$z_i$. [Here and in the following we use several times the term quark
generically for color sources in an arbitary $SU(N_c)$
representation.]  The impact parameter between the dipoles
is~\cite{Dosch:1997ss}
\be
        {\vec b}_{\!\perp} 
        \,=\, {\vec r}_{1q} + (1-z_1) {\vec r}_{1\perp} 
            - {\vec r}_{2q} - (1-z_2) {\vec r}_{2\perp} 
        \,=\, {\vec r}_{1\,cm} - {\vec r}_{2\,cm} 
        \ ,
\label{Eq_impact_vector}
\ee
where ${\vec r}_{iq}$ (${\vec r}_{i\qbar}$) is the transverse position
of the quark (antiquark), ${\vec r}_{i\perp} = {\vec r}_{i\qbar} -
{\vec r}_{iq}$, and ${\vec r}_{i\,cm} = z_i {\vec r}_{iq} +
(1-z_i){\vec r}_{i\qbar}$ is the center of light-cone momenta.
Figure~\ref{Fig_loop_loop_scattering_surfaces} illustrates the (a)
space-time and (b) transverse arrangement of the dipoles.
\befig[p!]
  \begin{center}
        \epsfig{file=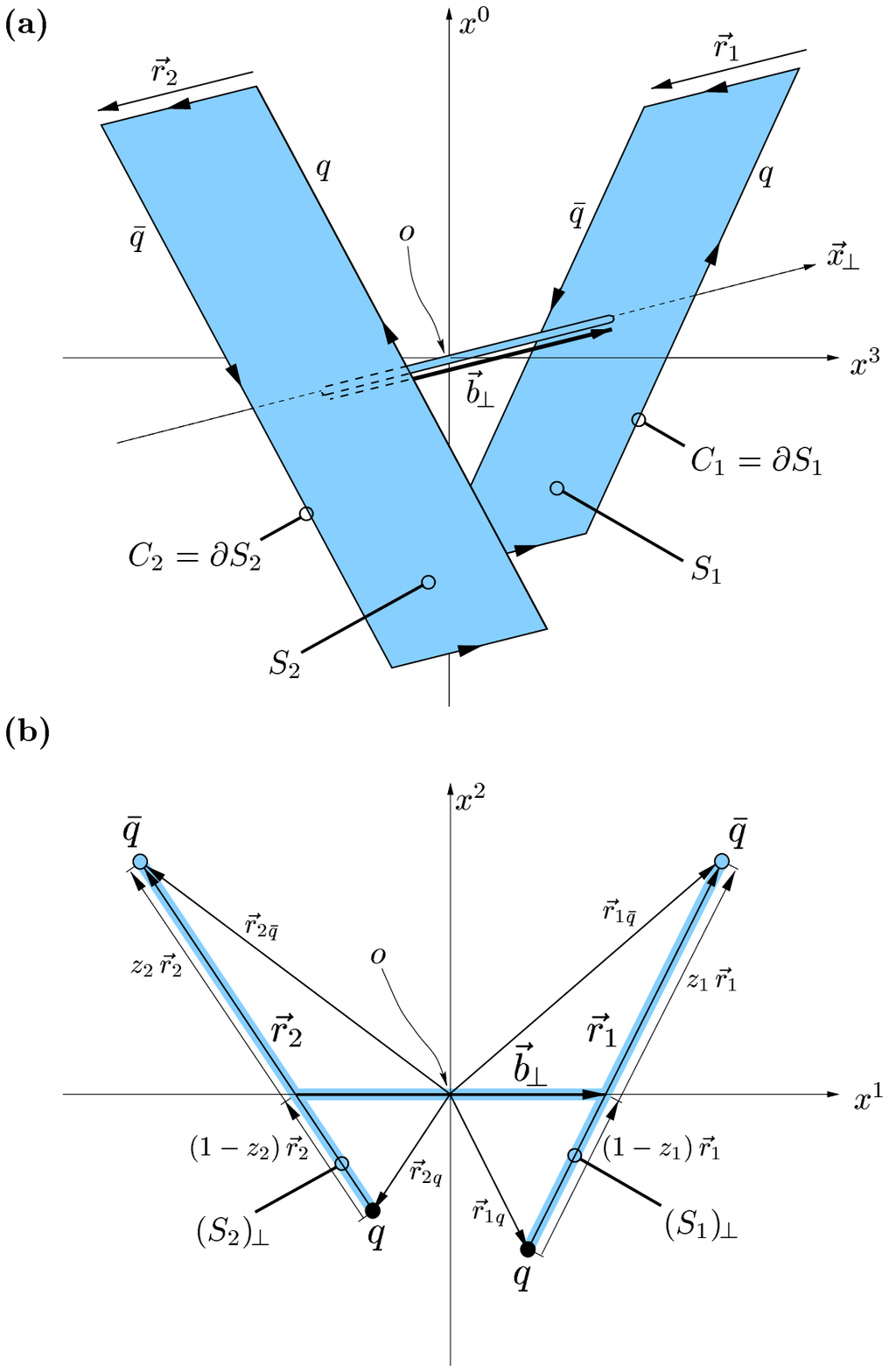,width=10.cm}
  \end{center}
\caption{\small High-energy dipole-dipole scattering in the eikonal
  approximation represented by Wegner-Wilson loops in the fundamental
  representation of $SU(N_c)$: (a) space-time and (b) transverse
  arrangement of the Wegner-Wilson loops. The shaded areas represent
  the strings extending from the quark to the antiquark path in each
  color dipole.  The thin tube allows us to compare the field
  strengths in surface $S_1$ with the field strengths in surface
  $S_2$. The impact parameter $\vec{b}_{\perp}$ connects the centers
  of light-cone momenta of the dipoles.}
\label{Fig_loop_loop_scattering_surfaces}
\efig
The dipole trajectories $C_i$ are described as straight lines. This is
a good approximation as long as the kinematical assumption behind the
eikonal approximation, $s \gg -t$, holds which allows us to neglect
the change of the dipole velocities $v_i = p_i/m$ in the scattering
process, where $p_i$ is the momentum and $m$ the mass of the
considered dipole.  Moreover, the paths $C_i$ are considered
light-like\footnote{In fact, exactly light-like trajectories ($\gamma
  \to \infty$) are considered in most applications of the functional
  integral approach to high-energy
  collisions~\cite{Kramer:1990tr,Dosch:1994ym,Dosch:RioLecture,Rueter:1996yb,Dosch:1998nw,Rueter:1998qy+X,Rueter:1998up,Dosch:1997ss,Berger:1999gu,Dosch:2001jg,Kulzinger:2002iu,Shoshi:2002in,Shoshi:2002ri,Shoshi:2002fq,Shoshi:2002mt}.
  A detailed investigation of the more general case of finite rapidity
  $\gamma$ can be found in~\cite{Kulzinger:2002iu}.} in line with the
high-energy limit, $m^2 \ll s \to \infty$. For the {\em hyperbolic
  angle} or {\em rapidity gap} between the dipole trajectories $\gamma
= (v_1 \cdot v_2)$---which is the central quantity in the analytic
continuation discussed below and also defined through $s =
4m^2\cosh^2(\gamma/2)$---the high-energy limit implies
\be
        \lim_{m^2\ll s\to\infty} \gamma \approx \ln(s/m^2) \to \infty 
        \ .
\label{Eq_rapidity_light-like_loops}
\ee
The QCD VEV's $\langle\ldots\rangle_M$ in the $S$-matrix
element~(\ref{Eq_S_DD_Minkowski}) represent {\em Minkowskian}
functional integrals~\cite{Nachtmann:ed.kt} in which---as in the
Euclidean case discussed above---the functional integration over the
fermion fields has already been carried out.

The Euclidean approach to the described elastic scattering of dipoles
in the eikonal approximation is based on {\em Meggiolaro's analytic
  continuation} of the high-energy parton-parton scattering
amplitude~\cite{Meggiolaro:1996hf+X}. Meggiolaro's analytic
continuation has been derived in the functional integral approach to
high-energy collisions~\cite{Nachtmann:1991ua+X,Nachtmann:ed.kt} in
which parton-parton scattering is described in terms of {\WW} lines:
The Minkowskian amplitude, $g^M(\gamma,T,t)$, given by the expectation
value of two {\WW} lines, forming a hyperbolic angle $\gamma$ in
Minkowski space-time, and the Euclidean ``amplitude,''
$g^E(\Theta,T,t)$, given by the expectation value of two {\WW} lines,
forming an angle $\Theta \in [0,\pi]$ in Euclidean space-time, are
connected by the following analytic continuation in the angular
variables and the temporal extension $T$, which is needed as an IR
regulator in the case of {\WW} lines,
\bea
        g^E(\Theta,T,t) & = & g^M(\gamma\to i\Theta,T\to -iT,t) 
        \ ,
\label{Eq_gE=gM}\\
        g^M(\gamma,T,t) & = & g^E(\Theta\to -i\gamma, T\to iT,t)
        \ .
\label{Eq_gM=gE}
\eea
Generalizing this relation to {\em gauge-invariant} dipole-dipole
scattering described in terms of {\WW} loops, the IR divergence known
from the case of {\WW} lines vanishes and no finite IR regulator $T$
is necessary. Thus, the Minkowskian $S$-matrix
element~(\ref{Eq_S_DD_Minkowski}), given by the expectation values of
two {\WW} loops, forming an hyperbolic angle $\gamma$ in Minkowski
space-time, can be computed from the Euclidean ``$S$-matrix element''
\be
        S^E_{r_1 r_2}(\Theta,{\vec b}_{\!\perp},z_1,\vec{r}_{1\perp},z_2,\vec{r}_{2\perp})
        = \lim_{T \rightarrow \infty}
        \frac{\langle W_{r_1}[C_1] W_{r_2}[C_2]\rangle_E}
        {\langle W_{r_1}[C_1]\rangle_E \langle W_{r_2}[C_2]\rangle_E}
\label{Eq_S_DD_Euclidean}
\ee
given by the expectation values of two {\WW} loops, forming an angle
$\Theta \in [0,\pi]$ in Euclidean space-time, via an analytic
continuation in the angular variable
\be
        S^M_{r_1 r_2}(\gamma\approx\ln[s/m^2],{\vec b}_{\!\perp},z_1,\vec{r}_{1\perp},z_2,\vec{r}_{2\perp})
        = S^E_{r_1 r_2}(\Theta\to -i\gamma,{\vec b}_{\!\perp},z_1,\vec{r}_{1\perp},z_2,\vec{r}_{2\perp})
        \ ,
\label{Eq_SM=SE(theta->-igamma)}
\ee
where $E$ indicates Euclidean space-time and the QCD VEV's
$\langle\ldots\rangle_E$ represent Euclidean functional integrals that
are equivalent to the ones denoted by $\langle\ldots\rangle_G$ in the
preceding sections, i.e.\ in which the functional integration over the
fermion fields has already been carried out.

\begin{figure}[t!]
\centerline{\epsfig{figure=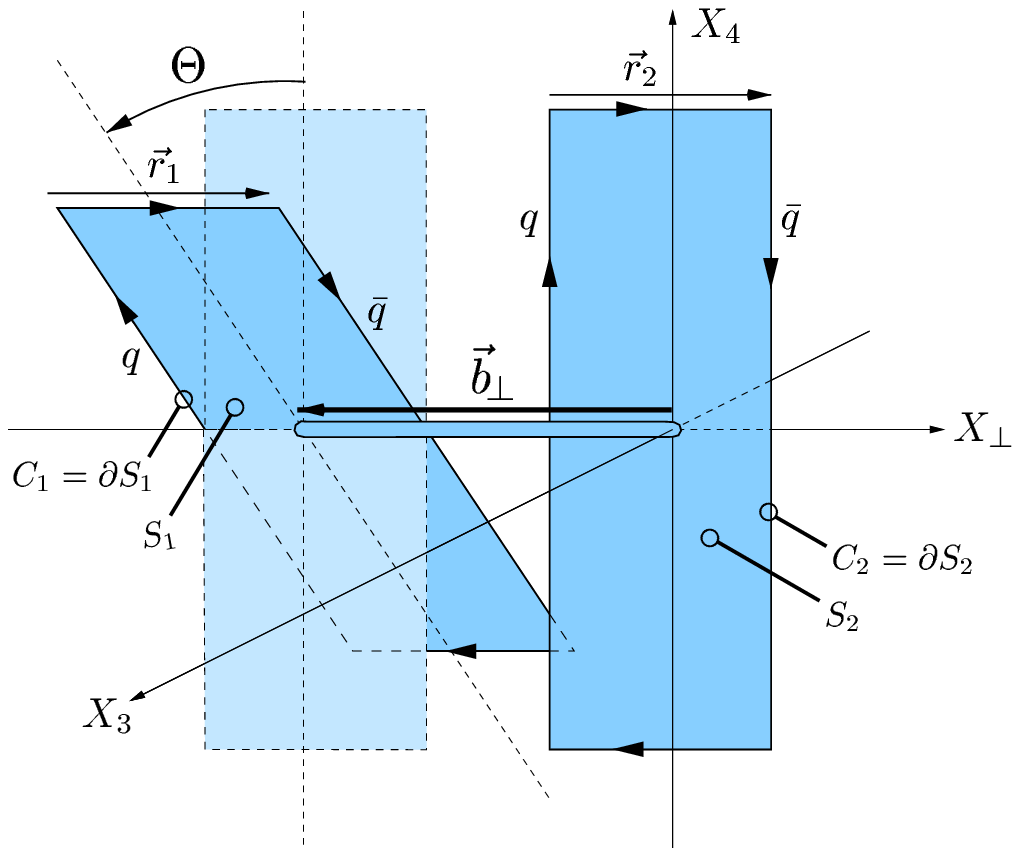,width=10.cm}}
\caption{\small 
  The loop-loop geometry necessary to compute $S^E_{r_1
    r_2}(\Theta\neq 0, \cdots)$ illustrated as a generalization of the
  geometry relevant for the computation of the van der Waals potential
  between two static dipoles ($\Theta=0$). While the potential between
  two static dipoles is computed from two loops along parallel
  ``temporal'' unit vectors, $t_1 = t_2 = (0,0,0,1)$, the Euclidean
  $S$-matrix element~(\ref{Eq_S_DD_Euclidean}) involves the tilting of
  one of the two loops, e.g.\ the tilting of $t_1$ by the angle
  $\Theta$ toward the $X_3$ axis, $t_1
  =(0,0,-\sin\Theta,\cos\Theta)$.}
\label{Fig_tilted_loops}
\end{figure}

The angle $\Theta$ is best illustrated in the relation of the
Euclidean $S$-matrix element~(\ref{Eq_S_DD_Euclidean}) to the van der
Waals potential between two static dipoles, $V_{r_1 r_2}(\Theta=0,
\vec{b}, z_1, \vec{r}_1, z_2, \vec{r}_2)$, discussed at the end of
this section,
\be
        S^E_{r_1 r_2}(\Theta,\vec{b}_{\!\perp},z_1,\vec{r}_{1\perp},z_2,\vec{r}_{2\perp})
        = \lim_{T \rightarrow \infty}
        \exp\!\left[-\,T\,V_{r_1 r_2}(\Theta,\vec{b}_{\!\perp},z_1,\vec{r}_{1\perp},z_2,\vec{r}_{2\perp})\right]
        \ .
\label{Eq_S_DD_<->_V_DD}
\ee
Figure~\ref{Fig_tilted_loops} shows the loop-loop geometry necessary
to compute $S^E_{r_1 r_2}(\Theta\neq 0, \cdots)$ and how it is
obtained by generalizing the geometry relevant for the computation of
the potential between two static dipoles ($\Theta=0$): While the
potential between two static dipoles is computed from two loops along
parallel ``temporal'' unit vectors, $t_1 = t_2 = (0,0,0,1)$, the
Euclidean $S$-matrix element~(\ref{Eq_S_DD_Euclidean}) involves the
tilting of one of the two loops, e.g.\ the tilting of $t_1$ by the
angle $\Theta$ toward the $X_3$ axis, $t_1 =
(0,0,-\sin\Theta,\cos\Theta)$. The ``temporal'' unit vectors $t_i$ are
also discussed in Appendix~\ref{Sec_Parameterizations} together with
another illustration of the tilting angle $\Theta$.

Since the Euclidean $S$-matrix element~(\ref{Eq_S_DD_Euclidean})
involves only configurations of {\WW} loops in Euclidean space-time
and {\em Euclidean} functional integrals, it can be computed directly
on a Euclidean lattice. With~(\ref{Eq_S_DD_Euclidean}) evaluated
numerically for many different values of $\Theta \in [0,\pi]$, one
needs to find the function that describes the angular dependence
obtained. If this function is analytic in $\Theta$, the analytic
continuation $\Theta\to -i\gamma$ leads immediately to the desired
Minkowskian $S$-matrix element~(\ref{Eq_S_DD_Minkowski}). An obvious
difficulty in this proposal is the breaking of rotational invariance
by the lattice.  Moreover, first attempts in the direction described
have shown that the signal size for~(\ref{Eq_S_DD_Euclidean})
decreases significantly with increasing $\Theta$ so that it is already
covered for small values of $\Theta$ by the statistical
fluctuations~\cite{DiGiacomo:2002PC}. At present, it is not clear how
to overcome these technical difficulties but the stakes are high: Once
precise results are available, the analytic
continuation~(\ref{Eq_SM=SE(theta->-igamma)}) could allow us to access
hadronic high-energy reactions directly in lattice QCD, i.e.\ within a
non-perturbative description of QCD from first principles.

More generally, the presented gauge-invariant analytic
continuation~(\ref{Eq_SM=SE(theta->-igamma)}) makes any approach
limited to a Euclidean formulation of the theory applicable for
investigations of high-energy reactions.  Indeed, Meggiolaro's
approach has already been used to access high-energy scattering from
the supergravity side of the AdS/CFT
correspondence~\cite{Janik:2000zk+X}, which requires a positive
definite metric in the definition of the minimal
surface~\cite{Rho:1999jm}, and to examine the effect of instantons on
high-energy scattering~\cite{Shuryak:2000df+X}.

Let us now perform the analytic continuation explicitly in our
Euclidean model. For the scattering of two color dipoles in the {\em
  fundamental representation} of $SU(N_c)$, the Euclidean $S$-matrix
element becomes with the VEV's~(\ref{Eq_final_result_<W[C]>})
and~(\ref{Eq_final_Euclidean_result_<W[C1]W[C2]>_fundamental})
\bea
        S^E_{DD}(\Theta,\vec{b}_{\!\perp},z_1,\vec{r}_{1\perp},z_2,\vec{r}_{2\perp})
         && \!\!\!\!\!\!
        :=\,\,S^E_{\fundamental\fundamental}
        (\Theta,\vec{b}_{\!\perp},z_1,\vec{r}_{1\perp},z_2,\vec{r}_{2\perp})
\nonumber\\
        && \hspace{-4cm} = \lim_{T \rightarrow \infty}
        \left(
        \frac{N_c\!+\!1}{2N_c}\exp\!\left[-\frac{N_c\!-\!1}{2 N_c}\chi_{S_1 S_2}\right]
        + \frac{N_c\!-\!1}{2N_c}\exp\!\left[ \frac{N_c\!+\!1}{2 N_c}\chi_{S_1 S_2}\right]
        \right)
        \ ,
\label{Eq_S_DD_1}
\eea
where $\chi_{S_i S_j}$, defined in~(\ref{Eq_chi_Si_Sj}), decomposes
into a perturbative ($\pert$) and non-perturbative ($\nprt$) component
according to our decomposition of the gluon field strength
correlator~(\ref{Eq_F_decomposition}):
\be
        \chi_{S_1 S_2} 
        \,\,=\,\, 
        \chi_{S_1 S_2}^{\pert} 
        \,+\, \chi_{S_1 S_2}^{\nprt}
        \,\,=\,\, 
        \chi_{S_1 S_2}^{\pert} 
        \,+\, \left(\chi_{S_1 S_2}^{\nprt\,\,nc} 
          \,+\, \chi_{S_1 S_2}^{\nprt\,\,c}\right)
        \ .
\label{Eq_chi_decomposition}        
\ee
In the limit $T_1=T_2=T\to\infty$ and for $\Theta \in [0,\pi]$, the
components read
\be
        \chi_{S_1 S_2}^{\pert} 
        =\cot\Theta\,\,\chi^{\pert}
        \,\, , \quad
        \chi_{S_1 S_2}^{\nprt\,\,nc} 
        =\cot\Theta\,\,\chi^{\nprt\,\,nc}
        \,\, , \quad
        \chi_{S_i S_j}^{\nprt\,\,c} 
        =\cot\Theta\,\,\chi^{\nprt\,\,c}
\label{Eq_S_DD_p_npc_npnc_E}
\ee
with
\bea
        \!\!\!\!\!\!\!\!\!\!\!\!\!
        \chi^{\pert} &\!\!=\!\!& 
        \left[ 
        g^2 D^{\prime\,(2)}_{\pert}
        \left(|\vec{r}_{1q}-\vec{r}_{2\qbar}|\right)
        +g^2 D^{\prime\,(2)}_{\pert}
        \left(|\vec{r}_{1\qbar}-\vec{r}_{2q}|\right)
        \right.
\nonumber \\
        \!\!\!\!\!\!\!\!\!\!\!\!\!
        &&
        \left.
        -\,g^2 D^{\prime\,(2)}_{\pert}
        \left(|\vec{r}_{1q}-\vec{r}_{2q}|\right)
        -g^2 D^{\prime\,(2)}_{\pert}
        \left(|\vec{r}_{1\qbar}-\vec{r}_{2\qbar}|\right)
        \right]
        \ ,
\label{Eq_S_DD_chi_p_M}\\ 
        \!\!\!\!\!\!\!\!\!
        \chi^{\nprt\,\,nc} &\!\!=\!\!& 
        \frac{\pi^2 G_2 (1-\kappa)}{3(N_c^2-1)} 
        \left[ 
        D^{\prime\,(2)}_1
        \left(|\vec{r}_{1q}-\vec{r}_{2\qbar}|\right)
        +D^{\prime\,(2)}_1
        \left(|\vec{r}_{1\qbar}-\vec{r}_{2q}|\right)
         \right.
\nonumber \\
        \!\!\!\!\!\!\!\!\!\!\!\!\!
        &&
        \hphantom{-\frac{\pi^2 G_2 (1-\kappa)}{3(N_c^2-1)}}
        \left.
        -\,D^{\prime\,(2)}_1
        \left(|\vec{r}_{1q}-\vec{r}_{2q}|\right) 
        -D^{\prime\,(2)}_1
        \left(|\vec{r}_{1\qbar}-\vec{r}_{2\qbar}|\right)
       \right]
        \ ,
\label{Eq_S_DD_chi_np_nc_M}\\
        \!\!\!\!\!\!\!\!\!\!\!\!\!
        \chi^{\nprt\,\,c} &\!\!=\!\!& 
        \frac{\pi^2 G_2 \kappa}{3(N_c^2-1)}\,
        \left(\vec{r}_1\cdot\vec{r}_2\right)
        \int_0^1 \! dv_1 \int_0^1 \! dv_2 \,\, 
        D^{(2)}\left(|\vec{r}_{1q}\! +\! v_1\vec{r}_{1\perp} 
        \!-\! \vec{r}_{2q}\! -\! v_2\vec{r}_{2\perp}|\right)
\label{Eq_S_DD_chi_np_c_M}
\eea
as derived explicitly in Appendix~\ref{Sec_Chi_Computation} with the
minimal surfaces illustrated in Fig.~\ref{Fig_tilted_loops}. In
Eq.~(\ref{Eq_S_DD_chi_p_M}) the shorthand notation $g^2
D^{\prime\,(2)}_{\pert}(|\vec{Z_\perp}|) =
g^2(|\vec{Z_\perp}|)\,D^{\prime\,(2)}_{\pert}(|\vec{Z_\perp}|)$ is
used with $g^2(|\vec{Z_\perp}|)$ again understood as the running
coupling~(\ref{Eq_g2(z_perp)}). The transverse Euclidean correlation
functions
\be
        D_x^{(2)}(\vec{Z}^2)      
        := \int \frac{d^4K}{(2\pi)^2}\,e^{iKZ}\,
        \tilde{D}_x(K^2)\,\delta(K_3)\,\delta(K_4)
\label{Eq_D(2)x}
\ee
are obtained from the (massive) gluon
propagator~(\ref{Eq_massive_gluon_propagator}) and the exponential
correlation function~(\ref{Eq_SVM_correlation_functions})
\bea
        \!\!\!\!\!\!\!\!\!\!\!\!\!\!\!\!
        D^{\prime\,(2)}_{\pert}(\vec{Z}_{\!\perp}^2)
        & \!\!=\!\! &
        \inv{2\pi} K_0\left(m_G |\vec{Z}_{\!\perp}|\right)
\label{Eq_D'(2)p(z,mg)}\\ 
        \!\!\!\!\!\!\!\!\!\!\!\!\!\!\!\!
        D^{\prime\,(2)}_1(\vec{Z}_{\!\perp}^2)
        & \!\!=\!\! &
        \pi a^4  \Big(
        3 \!+\! 3\frac{|\vec{Z}_{\!\perp}|}{a} \!+\! \frac{|\vec{Z}_{\!\perp}|^2}{a^2} 
        \Big)
        \exp\!\Big(\!-\frac{|\vec{Z}_{\!\perp}|}{a}\Big)
\label{Eq_D'(2)np_nc(z,a)}\\
        \!\!\!\!\!\!\!\!\!\!\!\!\!\!\!\!
        D^{(2)}(\vec{Z}_{\!\perp}^2)      
        & \!\!=\!\! &
        2 \pi a^2 
        \Big(1\!+\!\frac{|\vec{Z}_{\!\perp}|}{a}\Big) 
        \exp\!\Big(\!-\frac{|\vec{Z}_{\!\perp}|}{a}\Big)
\label{Eq_D(2)np_c(z,a)}
\eea
With the full $\Theta$ dependence exposed
in~(\ref{Eq_S_DD_p_npc_npnc_E}), the analytic
continuation~(\ref{Eq_SM=SE(theta->-igamma)}) reads
\be
        \chi_{S_1 S_2} = \cot\Theta\,\,\chi
        \quad\underrightarrow{\,\,\Theta\to -i\gamma\,\,\,\,}\quad
        \cot(-i\gamma)\,\chi
        \quad\underrightarrow{\,\,s\to\infty\,\,\,\,}\quad
        i\chi
\label{Eq_analytic_continuation_of_chi}
\ee
and leads to the desired Minkowskian $S$-matrix element for elastic
dipole-dipole ($DD$) scattering in the high-energy limit in which the
dipoles move on the light cone:
\bea
        \lim_{s \rightarrow \infty} 
        S_{DD}^{M}(s,{\vec b}_{\!\perp},z_1,\vec{r}_{1\perp},z_2,\vec{r}_{2\perp}) 
        &&\!\!\!\!\!\!:=\,\, 
        \lim_{s \rightarrow \infty} 
        S^M_{\fundamental\fundamental}(s,{\vec b}_{\!\perp},z_1,\vec{r}_{1\perp},z_2,\vec{r}_{2\perp})
\nonumber\\
        && \hspace{-4cm} = 
        S^E_{DD}(\cot\Theta \to i,{\vec b}_{\!\perp},z_1,\vec{r}_{1\perp},z_2,\vec{r}_{2\perp})
\nonumber\\
        && \hspace{-4cm} = 
        \lim_{T \rightarrow \infty}
        \left(
        \frac{N_c\!+\!1}{2N_c}\exp\!\left[-i\frac{N_c\!-\!1}{2N_c}\chi\right]
        + \frac{N_c\!-\!1}{2N_c}\exp\!\left[i\frac{N_c\!+\!1}{2N_c}\chi\right]
        \right)
        ,
\label{Eq_S_DD_1_M}
\eea
where $\chi =\chi^{\pert}+\chi^{\nprt\,\,nc}+\chi^{\nprt\,\,c}$
with~(\ref{Eq_S_DD_chi_p_M}), (\ref{Eq_S_DD_chi_np_nc_M}),
and~(\ref{Eq_S_DD_chi_np_c_M}).

It is striking that exactly the same result was obtained\footnote{To
  see this identity, recall that $\langle W[C]\rangle = 1$ for
  light-like loops and consider in~\cite{Shoshi:2002in} the
  result~(2.30) for the loop-loop correlation function (2.3) together
  with the $\chi$ function (2.40) and its components given in (2.49),
  (2.54), and (2.57) with the transverse Minkowskian correlation
  functions (2.50), (2.55), and (2.58). Note that all these equation
  numbers refer to Ref.~\cite{Shoshi:2002in}.}
in~\cite{Shoshi:2002in} with the alternative analytic continuation
introduced for applications of the SVM to high-energy
reactions~\cite{Kramer:1990tr,Dosch:1994ym,Dosch:RioLecture}. In this
complementary approach the gauge-invariant bilocal gluon field
strength correlator is analytically continued from Euclidean to
Minkowskian space-time by the substitution $\delta_{\mu\rho}
\rightarrow - g_{\mu\rho}$ and the analytic continuation of the
Euclidean correlation functions to real time $D^E_x(Z^2) \rightarrow
D^M_x(z^2)$. In the subsequent steps, one finds $\langle W[C]\rangle_M
= 1$ due to the light-likeness of the loops and that the longitudinal
correlations can be integrated out $\langle
W_{r_1}[C_1]W_{r_2}[C_2]\rangle_M = f(s,{\vec b}_{\!\perp},\cdots)$.
One is left with exactly the Euclidean correlations in transverse
space that have been obtained above. This confirms the analytic
continuation used in the earlier LLCM investigations in Minkowski
space-time~\cite{Shoshi:2002in,Shoshi:2002ri,Shoshi:2002fq,Shoshi:2002mt}
and in all earlier SVM applications to high-energy
scattering~\cite{Kramer:1990tr,Dosch:1994ym,Dosch:RioLecture,Rueter:1996yb,Dosch:1998nw,Rueter:1998qy+X,Rueter:1998up,Dosch:1997ss,Berger:1999gu,Donnachie:2000kp+X,Dosch:2001jg,Kulzinger:2002iu}.

In the limit of small $\chi$ functions, $|\chi^{\pert}| \ll 1$ and
$|\chi^{\nprt}| \ll 1$, Eq.~(\ref{Eq_S_DD_1_M}) reduces to
\be
        \lim_{s \rightarrow \infty} 
        S_{DD}^{M}(s,{\vec b}_{\!\perp},z_1,\vec{r}_{1\perp},z_2,\vec{r}_{2\perp})
        \approx 1 + \frac{N_c^2-1}{8 N_c^2}\,\chi^2 
        = 1 + \frac{C_2(\Fundamental)}{4 N_c}\,\chi^2 
        \ .
\label{Eq_S_DD_M_small_chi}
\ee
The perturbative correlations, $(\chi^{\pert})^2$, describe the
well-known {\em two-gluon exchange}
contribution~\cite{Low:1975sv+X,Gunion:iy} to dipole-dipole
scattering, which is, of course, an important successful cross-check
of the presented Euclidean approach to high-energy scattering. The
non-perturbative correlations, $(\chi^{\nprt})^2$, describe the
corresponding non-perturbative two-point interactions that contain
contributions of the confining QCD string to dipole-dipole scattering.
We analyzed these string contributions systematically as
manifestations of confinement in high-energy scattering reactions in
our previous work~\cite{Shoshi:2002fq}.

From the small-$\chi$ limit, one sees that the full $S$-matrix
element~(\ref{Eq_S_DD_1_M}) describes multiple gluonic interactions.
Indeed, the higher order terms in the expansion of the exponential
functions ensure the fundamental $S$-matrix unitarity condition in
impact parameter space as discussed
in~\cite{Berger:1999gu,Shoshi:2002in}.

Concerning the energy dependence, the $S$-matrix
element~(\ref{Eq_S_DD_1_M}) leads to energy-independent cross sections
in contradiction to the experimental observation. Although
disappointing from the phenomenological point of view, this is not
surprising since our approach does not describe the explicit gluon
radiation needed for a non-trivial energy dependence. However, based
on the $S$-matrix element~(\ref{Eq_S_DD_1_M}), a phenomenological
energy dependence can be constructed that allows a unified description
of high-energy hadron-hadron, photon-hadron, and photon-photon
reactions and an investigation of saturation effects in hadronic cross
sections manifesting the $S$-matrix
unitarity~\cite{Shoshi:2002in,Shoshi:2002ri,Shoshi:2002mt}. This, of
course, can only be an intermediate step. For a more fundamental
understanding of hadronic high-energy reactions in our model, gluon
radiation and quantum evolution have to be implemented explicitly.

Although the scattering of two color dipoles in the fundamental
representation of $SU(N_c)$ is, of course, the most relevant case, we
can derive immediately also the Minkowskian $S$-matrix element for the
scattering of a fundamental ($D$) and an adjoint dipole (``glueball''
$\glueball$) in the Euclidean LLCM.
Using~(\ref{Eq_final_Euclidean_result_<Wf[C1]Wa[C2]>}) and proceeding
otherwise as above, we find in the high-energy limit
\bea
        && \!\!\!\!\!\!\!\!
        \lim_{s \rightarrow \infty} 
        S^M_{D\,\glueball}(s, \vec{b}, z_1, \vec{r}_1, z_2, \vec{r}_2) 
        \,\,:=\,\, \lim_{s \rightarrow \infty} 
        S^M_{\fundamental\,\adjoint}(\Theta, \vec{b}, z_1, \vec{r}_1, z_2, \vec{r}_2) 
\label{Eq_S_fa_final_result}\\
        && \!\!\!\!\!\!\!\!
        = \lim_{T \rightarrow \infty}
        \Bigg(\,\inv{N_c^2\!-\!1}\,\exp\!\Big[i\,\frac{N_c}{2}\,\chi\Big]
        +\frac{N_c\!+\!2}{2(N_c\!+\!1)}\exp\!\Big[\!-\,i\,\inv{2}\,\chi\Big]
        +\frac{N_c\!-\!2}{2(N_c\!-\!1)}\exp\!\Big[i\,\inv{2}\,\chi\Big]
        \Bigg)
        \ ,
\nonumber
\eea
where $\chi =\chi^{\pert}+\chi^{\nprt\,nc}+\chi^{\nprt\,c}$
with~(\ref{Eq_S_DD_chi_p_M}), (\ref{Eq_S_DD_chi_np_nc_M}),
and~(\ref{Eq_S_DD_chi_np_c_M}).

Finally, we would like to comment on the {\em van der Waals
  interaction} of two color dipoles, which is, as already mentioned,
related to the Euclidean $S$-matrix element in the limiting case of
$\Theta=0$ as can be seen from~(\ref{Eq_S_DD_<->_V_DD}): The QCD van
der Waals potential between two static dipoles can be expressed in
terms of {\WW} loops~\cite{Appelquist:1978rt,Bhanot:1979af}
\be
        V_{r_1 r_2}(\Theta=0, \vec{b}, z_1=1/2, \vec{r}_1, z_2=1/2, \vec{r}_2) =
        - \lim_{T \rightarrow \infty} \frac{1}{T} 
        \ln \frac{\langle W_{r_1}[C_1] W_{r_2}[C_2] \rangle}
        {\langle W_{r_1}[C_1] \rangle \langle W_{r_2}[C_2] \rangle}
        \ .
\label{Eq_V_DD}        
\ee
In this limit ($\Theta=0$) intermediate octet states and their limited
lifetime become important as is well known from perturbative
computations of the QCD van der Waals potential between two static
color dipoles~\cite{Appelquist:1978rt,Bhanot:1979af,Peskin:1979va+X}:
Working with static dipoles, i.e.\ infinitely heavy color sources,
there is an energy degeneracy between the intermediate octet states
and the initial (final) singlet states that leads for perturbative
two-gluon exchange to a linear divergence in $T$ as $T\to\infty$. This
IR divergence can be lifted by manually introducing an energy gap
between the singlet ground state and the excited octet state and thus
a limit on the lifetime of the intermediate octet
state~\cite{Appelquist:1978rt,Bhanot:1979af,Peskin:1979va+X}.

In the perturbative limit of $g^2\to 0$ and $T$ large but finite,
i.e.\ $\chi^{\pert} \ll 1$, the perturbative component of our model
describes the two-gluon exchange contribution to the van der Waals
potential which is plagued by this IR divergence due to the static
limit. In the more general case of $g^2$ finite and $T\to\infty$,
which is applicable also for the non-perturbative component of our
model, one cannot use the small-$\chi$ limit and multiple gluonic
interactions become important. Here, our perturbative component
describes multiple gluon exchanges that reduce to an effective
one-gluon exchange contribution to the van der Waals potential whose
interaction range ($\propto 1/m_G$) contradicts the common
expectations. Indeed, it is also in contradiction to our results for
the glueball mass $M_{\glueball}$ which determines the interaction
range ($\propto 1/M_{\glueball}$) between two color dipoles for large
dipole separations. As already mentioned in
Sec.~\ref{Sec_QCD_Components}, we find for the perturbative component,
$M_{\glueball}^{\pert} = 2 m_G$, i.e.\ half of the interaction range
of one-gluon exchange, by computing the exponential decay of the
correlation of two small quadratic loops $P^{\alpha \beta}_{r_i}$ for
large Euclidean times $\tau\to\infty$,
\be
         M_{\glueball} :=
        - \lim_{\tau \rightarrow \infty} \frac{1}{\tau} 
        \ln \frac{\langle P_{r_1}^{\alpha \beta}(0) P_{r_2}^{\alpha \beta}(\tau)\rangle}
        {\langle P_{r_1}^{\alpha\beta}(0) \rangle 
         \langle P_{r_2}^{\alpha\beta}(\tau) \rangle}
        \ .
\label{Eq_Glueball_mass}
\ee
Note that we find for the non-perturbative component,
$M_{\glueball}^{\nprt} = 2/a$, which is smaller than
$M_{\glueball}^{\pert} = 2 m_G$ with the LLCM parameters and thus
governs the long range correlations in the LLCM.

Thus, for a meaningful investigation of the QCD van der Waals forces
within our model, one has to go beyond the static limit in order to
describe the limited lifetime of the intermediate octet states
appropriately. This we postpone for future work since the focus in
this work is on high-energy scattering where the gluons are always
exchanged within a short time interval due to the light-likeness of
the scattered particles and the finite correlation lengths.
Nevertheless, going beyond the static limit in the dipole-dipole
potential means going beyond the eikonal approximation in high-energy
scattering and it is, of course, of utmost importance to see how such
generalizations alter our results.

%
%
%
%
%
\section{Conclusion}
\label{Sec_Conclusion}

We have introduced the Euclidean version of the loop-loop correlation
model \cite{Shoshi:2002in} in which the QCD vacuum is described by
perturbative gluon exchange and the non-perturbative stochastic vacuum
model \cite{Dosch:1987sk+X}.  This combination leads to a static
quark-antiquark potential with color Coulomb behavior for small and
confining linear rise for large source separations in good agreement
with lattice QCD results. We have computed in the LLCM the vacuum
expectation value of one Wegner-Wilson loop, $\langle W_{r}[C]
\rangle$, and the correlation of two Wegner-Wilson loops, $\langle
W_{r_1}[C_1] W_{r_2}[C_2] \rangle$, for arbitrary loop geometries and
general representations $r_{(i)}$ of $SU(N_c)$.  Specifying the loop
geometries, these results allow us to compute the static
quark-antiquark potential, the glueball mass, the chromo-field
distributions of static color dipoles, the QCD van der Waals potential
between two static color dipoles, and the $S$-matrix element for
high-energy dipole-dipole scattering.

We have applied the LLCM to compute the potential and the
chromo-electric fields of a static color dipole in the fundamental and
adjoint representation of $SU(N_c)$. The formation of a confining
color flux tube is described by the non-perturbative SVM
correlations~\cite{Rueter:1994cn} and the color Coulomb field is
obtained from perturbative gluon exchange. We have found Casimir
scaling for both the perturbative and non-perturbative contributions
to the chromo-electric fields in agreement with recent lattice
data~\cite{Bali:2000un}. String breaking is described neither for
sources in the fundamental representation nor for sources in the
adjoint representation, which indicates that in our approach not only
dynamical fermions (quenched approximation) are missing but also some
gluon dynamics. Transverse and longitudinal energy density profiles
have been provided. For small dipoles, $R=0.1\,\fm$, perturbative
physics dominates and non-perturbative correlations are negligible.
For large dipoles, $R\gtsim 1\,\fm$, the non-perturbative confining
string dominates the chromo-electric fields between the color sources.
The transition from perturbative to string behavior takes place at
source separations of about $0.5\,\fm$ in agreement with the recent
results of L\"uscher and Weisz~\cite{Luscher:2002qv}.  The root mean
squared radius $R_{ms}$ of the confining string and the energy density
in the center of a fundamental $SU(3)$ dipole $\varepsilon_3(X=0)$ are
governed completely by non-perturbative physics for large $R$ and
saturate as $R$ increases at $R_{ms}^{R\to\infty}\approx 0.55\,\fm$
and $\varepsilon_3^{R\to\infty}(X=0)\approx 1\,\GeV/\fm^3$.

We have presented the low-energy
theorems~\cite{Rothe:1995hu+X,Michael:1995pv,Green:1996be}, known in
lattice QCD as Michael sum rules~\cite{Michael:1986yi}, in their
complete form in continuum theory taking into account the important
contributions found in~\cite{Dosch:1995fz,Rothe:1995hu+X} that are
missing in the original formulation~\cite{Michael:1986yi}. We have
used the complete theorems to compare the energy and action stored in
the confining string with the confining part of the static
quark-antiquark potential. The comparison shows consistency of the
model results and indicates that the non-perturbative SVM component is
working at the renormalization scale at which $\beta(g)/g=-2$ and
$\alphaS = 0.81$.  Earlier SVM investigations along these lines have
found a different value of $\alphaS = 0.57$ with the pyramid mantle
choice for the surface~\cite{Rueter:1994cn,Dosch:1995fz} but were
incomplete since only the contribution from the traceless part of the
energy-momentum tensor has been considered in the energy sum.

A Euclidean approach to high-energy dipole-dipole scattering has been
established by generalizing Meggiolaro's analytic
continuation~\cite{Meggiolaro:1996hf+X} from parton-parton scattering
to gauge-invariant dipole-dipole scattering. The generalized analytic
continuation allows us to derive $S$-matrix elements for high-energy
reactions from configurations of {\WW} loops in Euclidean space-time
with Euclidean functional integrals. It thus shows how one can access
high-energy reactions directly in lattice QCD. First attempts in this
direction have already been carried out but only very few signals
could be extracted, while most of the data were dominated by
noise~\cite{DiGiacomo:2002PC}. We have applied this approach to
compute in the Euclidean LLCM the scattering of dipoles at
high-energy. The result derived in the Minkowskian version of the
LLCM~\cite{Shoshi:2002in} has been exactly recovered including the
well-known two-gluon exchange contribution to dipole-dipole
scattering~\cite{Low:1975sv+X,Gunion:iy}. This confirms the analytic
continuation of the gluon field strength correlator used in all
earlier applications of the SVM to high-energy
scattering~\cite{Kramer:1990tr,Dosch:1994ym,Dosch:RioLecture,Rueter:1996yb,Dosch:1998nw,Rueter:1998qy+X,Rueter:1998up,Dosch:1997ss,Berger:1999gu,Donnachie:2000kp+X,Dosch:2001jg,Kulzinger:2002iu}.

The $S$-matrix element obtained in our approach has already been used
to investigate manifestations of the confining QCD string in
high-energy reactions of photons and hadrons~\cite{Shoshi:2002fq} but
leads to energy-independent cross sections in contradiction to the
experimental observation~\cite{Shoshi:2002in}. The missing energy
dependence is disappointing but not surprising since our approach does
not describe explicit gluon radiation needed for a non-trivial energy
dependence. In our previous work we have introduced a phenomenological
energy dependence into the $S$-matrix element that allows a unified
description of hadron-hadron, photon-hadron, and photon-photon
reactions and respects the $S$-matrix unitarity condition in impact
parameter space~\cite{Shoshi:2002in,Shoshi:2002ri,Shoshi:2002mt}.
However, for a more fundamental understanding of hadronic high-energy
reactions in our model, one faces the highly ambitious task to
implement gluon radiation and quantum evolution explicitly.

More generally, the presented Euclidean approach to high-energy
scattering makes any method limited to a Euclidean formulation of the
theory applicable for investigations of high-energy reactions. Here
encouraging new results have been obtained with
instantons~\cite{Shuryak:2000df+X} and within the AdS/CFT
correspondence~\cite{Janik:2000zk+X} and it will be interesting to see
precise results from the lattice. A promising complementary Euclidean
approach has been proposed in~\cite{Hebecker:1999pb+X} where the
structure functions of deep inelastic scattering at small Bjorken $x$
are related to an effective Euclidean field theory. Here one hopes
that the limit $x\to 0$ corresponds to critical behavior in the
effective theory. The aim is again to provide a framework in which
structure functions can be calculated from first principles using
genuine non-perturbative methods such as lattice computations. In
another recent attempt, the energy dependence of the proton structure
function has been related successfully to critical properties of an
effective near light-cone Hamiltonian in a non-perturbative lattice
approach~\cite{Pirner:2001pv+X}. It will be interesting to see further
developments along these lines aiming at an understanding of hadronic
high-energy scattering from the QCD Lagrangian.

\section*{Acknowledgment}

We would like to thank N.~Brambilla, A.~Di Giacomo, C.~Ewerz,
H.~Forkel, M.~Jamin, E.~Meggiolaro, P.~Minkowski, O.~Nachtmann,
Yu.~Simonov, I.~Stamatescu, and A.~Vairo for stimulating discussions
and G.~Bali for helpful comments and providing us with lattice data.
We thank F.~Schwab for the careful reading of the manuscript and
J.~Behrend for the prototype of Fig.~\ref{Fig_Surface_Ordering}. This
research is partially funded by the INTAS project ``Non-Perturbative
QCD'' and the European TMR Contract HPRN-CT-2000-00130.

%
%
%
%
\begin{appendix}
%
%
%
\section{The non-Abelian Stokes Theorem}
\label{Sec_Surface_Ordering}

In this appendix we review briefly the derivation of the non-Abelian
Stokes theorem~\cite{Arefeva:dp+X} and explain the emerging surface
ordering. We follow the lucid presentation given
in~\cite{Nachtmann:ed.kt}.

\begin{figure}[htb]
\centerline{\epsfig{figure=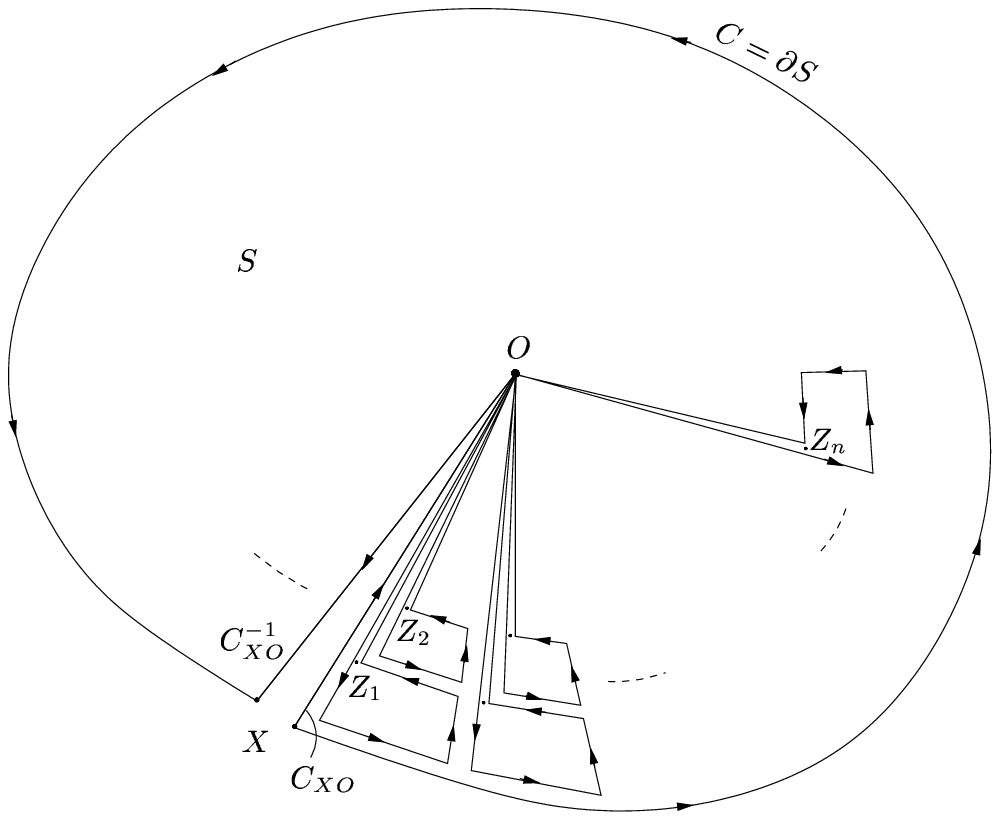,width=10.cm}}
\caption{\small
  A surface $S$ with boundary $C=\partial S$ in Euclidean space-time,
  the reference point $O$ on $S$, and the fan-type net with center $O$
  spanned over $S$.}
\label{Fig_Surface_Ordering}
\end{figure}

Let us consider a surface $S$ in Euclidean space-time with boundary $C
= \partial S$ and the QCD Schwinger string $\Phi_r(X,X;C)$ defined
according to~(\ref{Eq_parallel_transport}) that starts at some point
$X$ on the boundary and evolves along the path $C$ back to the point
$X$ as illustrated in Fig.~\ref{Fig_Surface_Ordering}. We now explain
how the non-Abelian line integral over $C$ associated with the QCD
Schwinger string is transformed into the non-Abelian surface integral
over $S$ which involves the surface ordering $\Ps$.

First, we choose an arbitrary reference point $O$ on the surface $S$
and draw a fan-type net on $S$ as a spider could do\footnote{Note that
  a real spider draws its net in a sequence different from the one
  described. The final result however is very similar.}
(cf.~Fig.~\ref{Fig_Surface_Ordering}). This net is spanned over $S$
and is given by the following curve: $C_{XO}$ running from $X$ to $O$,
followed by $C_{Z_1O}^{-1}$ running from $O$ to $Z_1$, where the path
around the infinitesimal small square $\Delta\sigma_{\mu\nu}(Z_1)$,
i.e.\ the plaquette at $Z_1$, is attached before it goes back to $O$
along $C_{Z_1O}$ and so on. The net is completed with $C_{XO}^{-1}$
that runs from $O$ to $X$.  Apart from the initial and final elements
of the net, $C_{XO}$ and $C_{XO}^{-1}$, we have many plaquettes with
``handles'' connecting them to $O$. With the following basic
properties of the QCD Schwinger string,
\bea
        \Phi_r(O,X;C_{XO})\,\Phi_r(O,X;C_{XO})^{-1} 
        & = & \Identity\ ,
\label{Eq_QCD_SS_1}\\
        \Phi_r(O,X;C_{XO})^{-1} 
        \,\,=\,\,\, \Phi_r^{\dagger}(O,X;C_{XO}) 
        & = & \Phi_r(X,O;C_{OX}=C^{-1}_{XO})\ ,
\label{Eq_QCD_SS_2}\\
        \Phi_r(Z,X;C_{XZ})\,\Phi_r(X,O;C_{OX}) 
        & = & \Phi_r(Z,O;C_{OX}+C_{XZ})\ ,
\label{Eq_QCD_SS_3}
\eea
one sees immediately that the QCD Schwinger string along the net
spanned on $S$ is equivalent to the QCD Schwinger string
$\Phi_r(X,X;C)$ along the path $C$:
\bea
        \Phi_r(X,X;C) 
        &=& \Phi_r(O,X;C_{XO}) \cdot \big(\mbox{product of QCD Schwinger strings} 
\nonumber \\
        &&\mbox{for the plaquettes with handles}\big) \cdot \Phi_r(O,X;C_{XO})^{-1} \ .  
\label{Eq_product_of_plaquettes}
\eea

Next, we consider the contribution of a single plaquette with handle.
The QCD Schwinger string for one plaquette, say the one at $Z_n$
singled out in Fig.~\ref{Fig_Surface_Ordering},
reads~\cite{Nachtmann:ed.kt}
\be
        \Phi_r(\mbox{plaquette at $Z_n$}) 
        = \Identity
        -ig\inv{2}\Delta\sigma_{\mu\nu}(Z_n)\,\G^a_{\mu\nu}(Z_n)\,t_r^a 
        + \,\cdots\, \ .
\ee
where $\Delta\sigma_{\mu\nu}(Z_n)$ denotes the surface element at the
point $Z_n$.  Taking into account the handles, the contribution of
this plaquette to~(\ref{Eq_product_of_plaquettes}) becomes
\bea
        &&
        \Phi_r(O,Z_n;C_{Z_nO})
        \Phi_r(\mbox{plaquette at $Z_n$})
        \Phi_r(O,Z_n;C_{Z_nO})^{-1}
\nonumber \\
        && 
        = \Identity 
        -ig\inv{2}\Delta\sigma_{\mu\nu}(Z_n)\,\G^a_{\mu \nu}(O,Z_n;C_{Z_nO})\,t_r^a 
        +\,\cdots\, \ ,
\label{Eq_one_plaquette}
\eea
where~(\ref{Eq_QCD_SS_2}) and the parallel transported gluon field
strength as defined in~(\ref{Eq_gluon_field_strength_tensor}) have
been used.

Finally, inserting~(\ref{Eq_one_plaquette})
into~(\ref{Eq_product_of_plaquettes}) for all $n$, i.e.\ summing up
the contributions of all plaquettes with handles, while respecting
the ordering, one obtains in the limit of an infinitesimally fine net
\bea
        && \Phi_r(X,X;C) = \Phi_r(O,X;C_{XO}) 
\label{Eq_infinitesimal_plaquettes}\\
        && \quad \quad \quad \quad 
        \cdot \,\Ps\exp\left[-i\,\frac{g}{2} 
               \int_{S}\!d\sigma_{\mu\nu}(Z) 
               \G^a_{\mu\nu}(O,Z;C_{ZO})\,t_r^a 
               \right] 
        \cdot \Phi_r(O,X;C_{XO})^{-1} \ .
\nonumber
\eea
Here $\Ps$ denotes the ordering on the whole surface $S$ as implied by
the net shown in Fig.~\ref{Fig_Surface_Ordering}. Taking the trace
in~(\ref{Eq_infinitesimal_plaquettes}) and exploiting its cyclic
property leads ultimately to the non-Abelian version of Stokes
theorem
\be
        \rTr\,\Phi_r(X,X;C) 
        =  \rTr\,\Ps\exp\left[-i\,\frac{g}{2} 
          \int_{S} \! d\sigma_{\mu\nu}(Z) 
          \G^a_{\mu\nu}(O,Z;C_{ZO})\,t_r^a 
        \right] \ .
\label{Eq_non-Abelian}
\ee

%
%
\section{Loop and Minimal Surface Parametrizations}
\label{Sec_Parameterizations}

\begin{figure}[p]
\centerline{\epsfig{figure=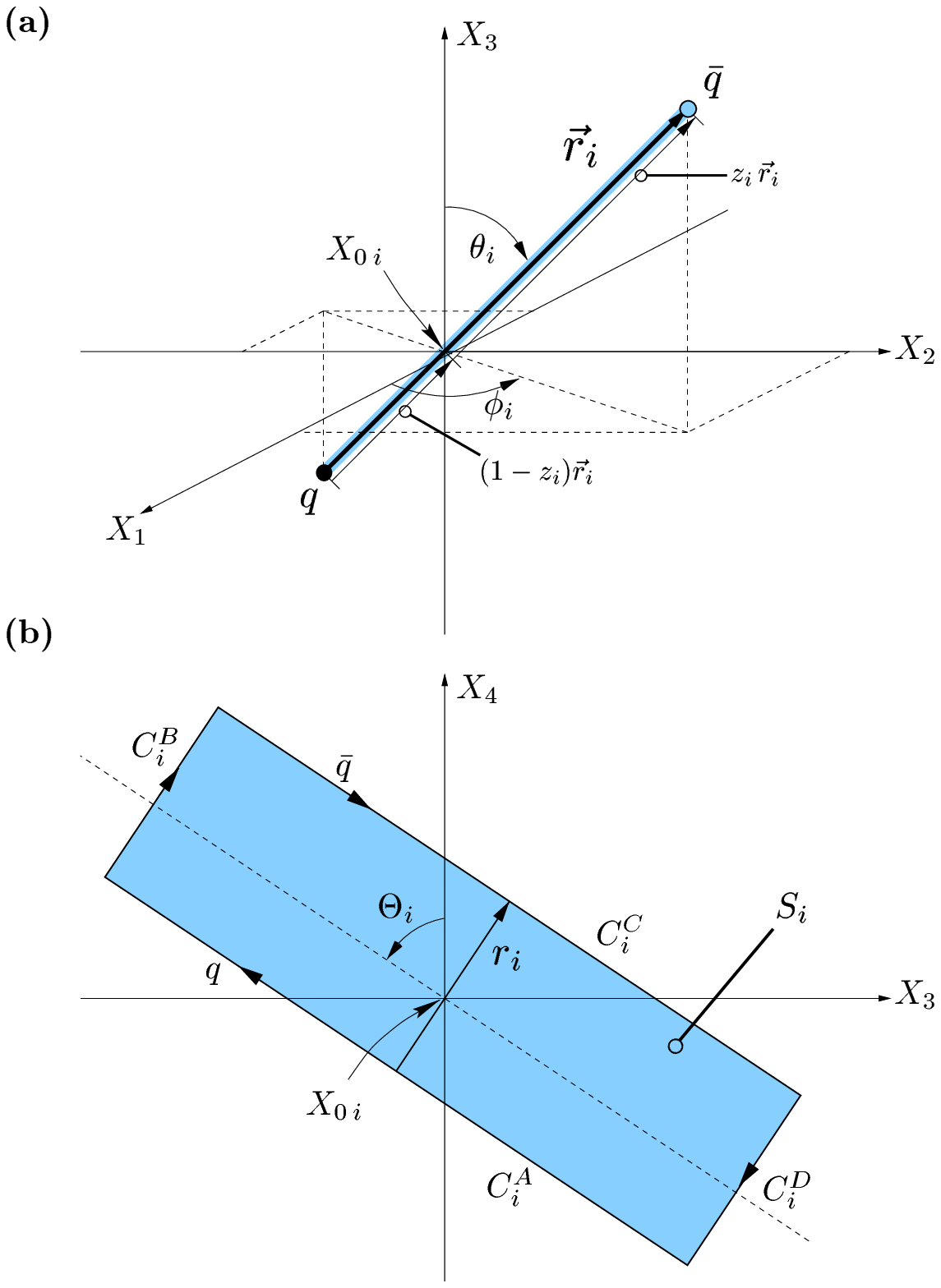,width=10.cm}}\hskip 0.5cm
\caption{\small
  (a) Spatial arrangement of a color dipole and (b) its world-line in
  Euclidean ``longitudinal'' space given by the rectangular loop $C_i$
  that defines the {\em minimal surface} $S_i$ with $\partial S_i =
  C_i$. The minimal surface is represented by the shaded area.  In our
  model, it is interpreted as the world-sheet of the QCD string that
  confines the quark and antiquark in the dipole.}
\label{Fig_OneLoop_MinimalSurface}
\end{figure}

A rectangular loop $C_i$ with ``spatial'' extension $R_i$ and
``temporal'' extension $2T_i$ placed in four-dimensional Euclidean
space, as shown in Fig.~\ref{Fig_OneLoop_MinimalSurface}, has the
following parameter representation:
\be
        C_i 
        \,\,=\,\, 
        C_i^A \,\cup\, C_i^B \,\cup\, C_i^C \,\cup\, C_i^D
\label{Eq_Ci_parameterization}
\ee
with
\bea
        C_i^A \,\,=\,\,  
        \Big\{ 
        X_i^A(u_i) 
        & = & 
        X_{\!0\,i} - (1-z_i)\,r_i + u_i\,t_i,\quad\hphantom{v_i\,r_i}
        u_i \in [-T_i,T_i]
        \Big\} 
\label{Eq_Ci^A_parameterization}\\
        C_i^B \,\,=\,\, 
        \Big\{ 
        X_i^B(v_i) 
        & = & 
        X_{\!0\,i} - (1-z_i)\,r_i + v_i\,r_i + T_i\,t_i,\;\quad
        v_i \in [0,1]
        \Big\} 
\label{Eq_Ci^B_parameterization}\\
        C_i^C \,\,=\,\, 
        \Big\{ 
        X_i^C(u_i) 
        & = & 
        X_{\!0\,i} + z_i\,r_i + u_i\,t_i,\quad\hphantom{(1-z_i)\,r_i}
        u_i \in [T_i,-T_i]
        \Big\} 
\label{Eq_Ci^C_parameterization}\\
        C_i^D \,\,=\,\,
        \Big\{ 
        X_i^D(v_i) 
        & = & 
        X_{\!0\,i} - (1-z_i)\,r_i + v_i\,r_i - T_i\,t_i,\;\quad
        v_i \in [1,0]
        \Big\} 
\label{Eq_Ci^D_parameterization}
\eea
where
\be
        r_i
        := \left( \barray{c} 
        R_i\,\sin\theta_i\,\cos\phi_i \\ 
        R_i\,\sin\theta_i\,\sin\phi_i \\ 
        R_i\,\cos\theta_i\,\cos\Theta_i \\ 
        R_i\,\cos\theta_i\,\sin\Theta_i
        \earray \right)
        \quad \mbox{and} \quad
        t_i
        := \left( \!\! \barray{c} 
        0 \\ 
        0 \\
        -\sin\Theta_i \\
        \hphantom{-}\cos\Theta_i
        \earray \right)
        \ .
\label{Eq_ri_ti_four_vectors}
\ee
The ``center'' of the loop $C_i$ is given by $X_{\!0\,i}$. The
parameters $z_i$, $R_i$, $\theta_i$, $\phi_i$, and $\Theta_i$ are
defined in Fig.~\ref{Fig_OneLoop_MinimalSurface} which illustrates (a)
the spatial arrangement of a color dipole and (b) its world-line $C_i$
in Euclidean ``longitudinal'' space. The tilting angle $\Theta_i\neq
0$ is the central quantity in the analytic continuation presented in
Sec.~\ref{Sec_DD_Scattering}.  Moreover, $\Theta_1 =\pi/2$ together
with $\Theta_2 = 0$ allows us to compute conveniently the
chromo-magnetic field distributions in
Appendix~\ref{Sec_Chi_Computation}.

The {\em minimal surface} $S_i$ is the planar surface bounded by the
loop $C_i=\partial S_i$ given in~(\ref{Eq_Ci_parameterization}). It
can be parametrized as follows
\be
        S_i =  
        \Big\{ 
        X_i(u_i,v_i) 
        = X_{\!0\,i} - (1-z_i)\,r_i + v_i\,r_i + u_i\,t_i,
        \;u_i \in [-T_i,T_i], \;v_i \in [0,1] 
        \Big\}
\label{Eq_Si_parameterization}
\ee
with $r_i$ and $t_i$ given in~(\ref{Eq_ri_ti_four_vectors}). The
corresponding infinitesimal surface element reads
\be
        d\sigma_{\mu\nu}(X_i)
        = \Bigg( 
        \frac{\partial X_{\!i\mu}}{\partial u_i} 
        \frac{\partial X_{\!i\nu}}{\partial v_i}
        - \frac{\partial X_{\!i\mu}}{\partial v_i} 
        \frac{\partial X_{\!i\nu}}{\partial u_i} 
        \Bigg)\,du_i\,dv_i
        = \Bigg( 
        t_{i\mu} r_{i\nu} - r_{i\mu} t_{i\nu} 
        \Bigg)\,du_i\,dv_i
        \ .
\label{Eq_Si_surface_element}
\ee
%
%
%
%
%
\section[$\chi$ Computations with Minimal Surfaces]{\boldmath$\chi$ Computations with Minimal Surfaces}
\label{Sec_Chi_Computation}

The quantities considered in the main text are computed from the VEV
of one loop $\langle W[C] \rangle$ and the loop-loop correlation
function $\langle W[C_1] W[C_2] \rangle$. Using the Gaussian
approximation in the gluon field strengths, both are expressed in
terms of $\chi_{S_i S_j}$ functions~(\ref{Eq_chi_SS})
and~(\ref{Eq_chi_Si_Sj}) as shown in Secs.~\ref{Sec_<W[C]>}
and~\ref{Sec_<W[C_1]W[C_2]>}. These $\chi$ functions are central
quantities since here the ansatz of the gauge-invariant bilocal gluon
field strength correlator and the surface choice enter the model. In
this appendix, these functions are computed explicitly for minimal
surfaces~(\ref{Eq_Si_parameterization}) and the $F_{\mu\nu\rho\sigma}$
ansatz given in~(\ref{Eq_F_decomposition}), (\ref{Eq_PGE_Ansatz_F}),
and~(\ref{Eq_MSV_Ansatz_F}). Note that the contributions from the
infinitesimally thin tube---which allows us to compare the field
strengths in surface $S_1$ with the field strength in surface
$S_2$---cancel mutually.

Depending on the geometries and the relative arrangement of the loops,
the $\chi$ functions determine the physical quantities investigated
within the LLCM such as the static $q\qbar$
potential~(\ref{Eq_Vr(R)_Gaussian_approximation}), the chromo-field
distributions of a
color dipole~(\ref{Eq_chromo_fields_F_final_result}), and the
$S$-matrix element for elastic dipole-dipole
scattering~(\ref{Eq_S_DD_1}).

We compute the three components $\chi_{S_1 S_2}^{\pert}$, $\chi_{S_1
  S_2}^{\nprt\,nc}$, and $\chi_{S_1 S_2}^{\nprt\,c}$ separately for
general loop arrangements from which the considered quantities are
obtained as special cases. Without loss of generality, the center of
the loop $C_2$ is placed at the origin of the coordinate system,
$X_{\!0\,2}=(0,0,0,0)$. Moreover, $C_2$ is kept untilted, $\Theta_2 =
0$, and $\Theta := \Theta_1$ is used to simplify notation. We limit
our general computation to loops with $r_{1,2} =
(\vec{r}_{1,2\perp},0,0) \,\,\equiv\,\, \theta_{1,2} = \pi/2$ and
transverse ``impact parameters'' $b = X_{\!0\,1} - X_{\!0\,2} =
X_{\!0\,1} = (b_1, b_2, 0, 0) = (\vec{b}_{\!\perp},0,0)$ which allows
us to compute all of the considered quantities.

\subsection*{\boldmath $\chi_{S_1 S_2}^{\nprt\,c}$ Computation}

Starting with the definition
\bea
        \chi_{S_1 S_2}^{\nprt\,c} 
        & := & \frac{\pi^2}{4} 
        \int_{S_1} \!\! d\sigma_{\mu\nu}(X_1) 
        \int_{S_2} \!\! d\sigma_{\rho\sigma}(X_2)\,
        F_{\mu\nu\rho\sigma}^{\nprt\,c}(Z = X_1 - X_2)
\label{Eq_chi_S1S2_NP_c_def_E}\\ 
        & = & 
        \frac{\pi^2 G_2 \kappa}{12(N_c^2-1)}
        \int_{S_1} \!\! d\sigma_{\mu\nu}(X_1) 
        \int_{S_2} \!\! d\sigma_{\rho\sigma}(X_2)
        \left(\delta_{\mu\rho}\delta_{\nu\sigma}
         -\delta_{\mu\sigma}\delta_{\nu\rho}\right) D(Z^2)
        \ ,
\nonumber
\eea
one exploits the anti-symmetry of the surface elements,
$d\sigma_{\mu\nu} = - d\sigma_{\nu\mu}$, and applies the surface
parametrization~(\ref{Eq_Si_parameterization}) with the corresponding
surface elements~(\ref{Eq_Si_surface_element}) to obtain
\be
        \chi_{S_1 S_2}^{\nprt\,c} = 
        \cos\Theta\,
        \frac{\pi^2 G_2 \kappa}{3(N_c^2-1)}\,
        (r_1\cdot r_2)
        \int_0^1 \!\! dv_1 \int_0^1 \!\! dv_2
        \int_{-T_1}^{T_1} \!\! du_1 
        \int_{-T_2}^{T_2} \!\! du_2 \,D(Z^2)
\label{Eq_chi_S1S2_NP_c_T<infty_result_E}
\ee
with
\be
        Z = X_{1} - X_{2}
        = \left( \barray{l}
   \!\!\hphantom{-\,}
        \vec{b}_{\!\perp} - (1-z_1)\,\vec{r}_{1\perp} + v_1\,\vec{r}_{1\perp} 
        + (1-z_2)\,\vec{r}_{2\perp} - v_2\,\vec{r}_{2\perp} \\
   \!\!
        -\,u_1 \sin\Theta \\
   \!\!\hphantom{-\,}
        u_1 \cos\Theta - u_2
        \earray \right)
        \ ,
\label{Eq_Z_S1_S2_E}
\ee
where the identities $t_1\cdot r_2 = r_1 \cdot t_2= 0$ and $t_1 \cdot
t_2 = \cos\Theta$, evident from~(\ref{Eq_ri_ti_four_vectors}) with the
mentioned specification of the loop geometries, have been used. In the
limit $T_2 \rightarrow \infty$, the $u_2$ integration can be performed
\bea
        && \lim_{T_2\to\infty}\int_{-T_2}^{T_2}\!\!du_2\,D(Z^2)
        \,\, = \,\, 
        \int \frac{d^4K}{(2\pi)^4}\,\tilde{D}(K^2) 
                \lim_{T_2\to\infty}\int_{-T_2}^{T_2}\!\!du_2\,e^{iKZ}
        \nonumber \\
        && = \int \frac{d^4K}{(2\pi)^4}\,\tilde{D}(K^2)\, 
                2\pi\,\delta(K_4)\,
                \exp\!\left[
                i\vec{K}_{\!\perp}\vec{Z}_{\!\perp} + iK_3 u_1\sin\Theta + iK_4 u_1\cos\Theta
                \right]
        \nonumber \\
        && = \int \frac{d^3K}{(2\pi)^3}\,\tilde{D}^{(3)}(\vec{K}^2)\,
                \exp\!\left[
                i\vec{K}_{\!\perp}\vec{Z}_{\!\perp} + iK_3 u_1\sin\Theta
                \right]
        \,\, = \,\, 
        D^{(3)}(\vec{Z}^2)
        \ ,
\label{Eq_chi_S1S2_NP_c_u2_integration}
\eea
which leads to
\be
        \lim_{T_2 \to \infty}
        \chi_{S_1 S_2}^{\nprt\,c} = 
        \cos\Theta\,
        \frac{\pi^2 G_2 \kappa}{3(N_c^2-1)}\,
        (\vec{r}_{1\perp}\cdot\vec{r}_{2\perp})
        \int_0^1 \!\! dv_1 \int_0^1 \!\! dv_2
        \int_{-T_1}^{T_1} \!\! du_1 \,
        D^{(3)}(\vec{Z}^2)
        \ .
\label{Eq_chi_S1S2_NP_c_T2->infty_result_E}
\ee
Taking in addition the limit $T_{1} \rightarrow \infty$, the $u_1$
integration can be performed as well
\be
        \lim_{T_1 \to \infty} 
        \int_{-T_1}^{T_1} \!\! du_1 
        e^{iK_3 u_1\sin\Theta}
        = \left\{\begin{array}{lc}
        2\pi\delta(K_3\sin\Theta)
        = \frac{2\pi\delta(K_3)}{|\sin\Theta|}
        & \quad \mbox{for} \quad \sin\Theta \neq 0 \, , \\
        \lim_{T_1\to\infty}\,2\,T_1
        & \quad \mbox{for} \quad \sin\Theta = 0 \, . \\
        \end{array}\right.
\label{Eq_chi_S1S2_NP_c_u1_integration}
\ee
With $T_1 = T_2 = T/2 \to \infty$, one obtains for $\sin\Theta\neq 0$
\be
        \lim_{T\to\infty}
        \chi_{S_1 S_2}^{\nprt\,c} = 
        \frac{\cos\Theta}{|\sin\Theta|}\,
        \frac{\pi^2 G_2 \kappa}{3(N_c^2-1)}\,
        (\vec{r}_{1\perp}\cdot \vec{r}_{2\perp})
        \int_0^1 \!\! dv_1 \int_0^1 \!\! dv_2\,
        D^{(2)}(\vec{Z}_{\!\perp}^2)
\label{Eq_chi_S1S2_NP_c_T->infty_S_E}
\ee
and for $\sin\Theta = 0$ 
\be
        \lim_{T\to\infty}
        \chi_{S_1 S_2}^{\nprt\,c} = 
        \lim_{T\to\infty}\,T\,
        \cos\Theta\,
        \frac{\pi^2 G_2 \kappa}{3(N_c^2-1)}\,
        (\vec{r}_{1\perp}\cdot \vec{r}_{2\perp})
        \int_0^1 \!\! dv_1 \int_0^1 \!\! dv_2\,
        D^{(3)}(\vec{Z}^2)
        \ .
\label{Eq_chi_S1S2_NP_c_T->infty_V_E}
\ee
Evidently,~(\ref{Eq_chi_S1S2_NP_c_T->infty_S_E}) is the result given
in~(\ref{Eq_S_DD_p_npc_npnc_E}) and~(\ref{Eq_S_DD_chi_np_c_M}) which
describes the confining contribution to the dipole-dipole scattering
matrix element $S_{DD}$.

From~(\ref{Eq_chi_S1S2_NP_c_T->infty_V_E}), one obtains the confining
contribution to the static color dipole potential for $S_1 = S_2 = S$
which implies $T_1 = T_2 = T/2$, $\Theta = 0$, $z_1 = z_2$, $r_1 = r_2
= r$, and $\vec{r}_{1\perp}\cdot \vec{r}_{2\perp} = r^2 = R^2$ so that
\bea
        \lim_{T\to\infty}
        \chi_{SS}^{\nprt\,c} 
        & = &
        \lim_{T\to\infty}\,T\,
        \frac{\pi^2 G_2 \kappa}{3(N_c^2-1)}\,
        R^2
        \int_0^1 \!\! dv_1 \int_0^1 \!\! dv_2\,
        D^{(3)}(\vec{Z}^2=(v_1-v_2)^2R^2)
\nonumber\\
        & = &
        \lim_{T\to\infty}\,T\,
        \frac{2 \pi^2 G_2 \kappa}{3(N_c^2-1)}\,
        R^2
        \int_0^1 \!\! d\rho\,
        (1-\rho)\,
        D^{(3)}(\rho^2 R^2)
        \ ,
\label{Eq_chi_SS_NP_c_T->infty_V_E}
\eea
which leads directly to~(\ref{Eq_Vr(R)_NP_c}).
 
From~(\ref{Eq_chi_S1S2_NP_c_T2->infty_result_E}) the confining
contribution to the chromo-field distributions $\Delta G_{\alpha
  \beta}^2(X)$ can be computed conveniently.
Equation~(\ref{Eq_chi_S1S2_NP_c_T2->infty_result_E}) reads for $S_1 =
S_P$, $T_1 = R_P/2$ and $R_1 = R_P$, and $S_2 = S_W$, $T_2 = T/2$ and
$R_2 = R$
\be
        \lim_{T \to \infty}
        \chi_{S_P S_W}^{\nprt\,c} =
        \cos \Theta\, 
        \frac{\pi^2 G_2 \kappa}{3(N_c^2-1)}\,
        (\vec{r}_{1\perp}\cdot\vec{r}_{2\perp})
        \int_0^1 \!\! dv_1 \int_0^1 \!\! dv_2
        \int_{-R_P/2}^{R_P/2} \!\! du_1 \,
        D^{(3)}(\vec{Z}^2)
\label{Eq_chi_SPSW_NP_c_T2->infty_result_E}
\ee
with
\be
        \vec{Z} = \vec{X}_{1} - \vec{X}_{2}
        = \left( \barray{l}
   \!\!\hphantom{-\,}
        \vec{b}_{\!\perp} - (1-z_1)\,\vec{r}_{1\perp} + v_1\,\vec{r}_{1\perp} 
        + (1-z_2)\,\vec{r}_{2\perp} - v_2\,\vec{r}_{2\perp} \\
   \!\!
        -\,u_1 \sin\Theta 
        \earray \right)
        \ .
\label{Eq_vec3_Z_S1_S2_E}
\ee
The confining non-perturbative contribution to the chromo-magnetic
fields vanishes as it is obtained for plaquettes with $\Theta =
\pi/2$. The corresponding contribution to the chromo-electric fields
can be computed with $\Theta=0$ as follows: Due to $R_1 = R_p\to 0$, the $u_1$ and
$v_1$ integrations in~(\ref{Eq_chi_SPSW_NP_c_T2->infty_result_E}) can
be performed with the mean value theorem. Keeping only terms up to
$\Order(R_p^2)$, the confining non-perturbative contribution to the
chromo-field distributions $\Delta G_{\alpha \beta}^2(X)$ is obtained
as given in~(\ref{Eq_Chi_PW_np_c_14}).

\subsection*{\boldmath$\chi_{S_1 S_2}^{\nprt\,nc}$ Computation}

We start again with the definition
\bea
        \chi_{S_1 S_2}^{\nprt\,nc}
        & := & \frac{\pi^2}{4} 
        \int_{S_1} \!\! d\sigma_{\mu\nu}(X_1) 
        \int_{S_2} \!\! d\sigma_{\rho\sigma}(X_2)\,
        F_{\mu\nu\rho\sigma}^{\nprt\,nc}(Z = X_1 - X_2)
\nonumber \\ 
        & = & 
        \frac{\pi^2 G_2 (1\!-\!\kappa)}{12(N_c^2-1)} \,
        \int_{S_1} \!\! d\sigma_{\mu\nu}(X_1) 
        \int_{S_2} \!\! d\sigma_{\rho\sigma}(X_2)
\label{Eq_chi_S1S2_NP_nc_def_E}\\ 
        && \times
        \inv{2}\Bigl[
        \frac{\partial}{\partial Z_\nu}
        \left(Z_\sigma \delta_{\mu\rho}
          -Z_\rho \delta_{\mu\sigma}\right)
        +\frac{\partial}{\partial Z_\mu}
        \left(Z_\rho \delta_{\nu\sigma}
          -Z_\sigma \delta_{\nu\rho}\right)\Bigr]\,
        D_1(Z^2)
\nonumber
\eea
and use the anti-symmetry of both surface elements to obtain
\bea
        \!\!\!\!\!\!\!\!\!\!\!\!\!\!\!\!\!\!
        \chi_{S_1 S_2}^{\nprt\,nc}\!\!
        & = & 
        \frac{\pi^2 G_2 (1\!-\!\kappa)}{6(N_c^2-1)} \,
        \int_{S_1} \!\! d\sigma_{\mu\nu}(X_1) 
        \int_{S_2} \!\! d\sigma_{\rho\sigma}(X_2)\,
        \frac{\partial}{\partial Z_\nu}\,Z_\sigma\,
        \delta_{\mu\rho}\,D_1(Z^2)
\label{Eq_chi_S1S2_NP_nc_1_E}\\ 
        & = & 
        \frac{\pi^2 G_2 (1\!-\!\kappa)}{3(N_c^2-1)}
        \!\int_{S_1} \!\! d\sigma_{\mu\nu}(X_1) 
        \!\int_{S_2} \!\! d\sigma_{\rho\sigma}(X_2)\,
        \frac{\partial}{\partial Z_\nu}\,
        \frac{\partial}{\partial Z_\sigma}\,
        \delta_{\mu\rho}\,D_1^{\prime}(Z^2)
\label{Eq_chi_S1S2_NP_nc_2_E}\\
        & = & 
        \!\!\!\!
        -\,\frac{\pi^2 G_2 (1\!-\!\kappa)}{3(N_c^2-1)}
        \!\!\int_{S_1} \!\!\! d\sigma_{\mu\nu}(X_1)
        \frac{\partial}{\partial X_{1\nu}}
        \!  \int_{S_2} \!\!\! d\sigma_{\rho\sigma}(X_2)
        \frac{\partial}{\partial X_{2\sigma}}\,
        \delta_{\mu\rho}\,D_1^{\prime}(Z^2)
\label{Eq_chi_S1S2_NP_nc_3_E}
\eea
with
\be
        D_1^{\prime}(Z^2)
        = \int\!\!\frac{d^4K}{(2\pi)^4}\,e^{iKZ}\,
        \tilde{D}_1^{\prime}(K^2)
        = \int\!\!\frac{d^4K}{(2\pi)^4}\,e^{iKZ}\,
        \frac{d}{dK^2}\,\tilde{D}_1(K^2)
        \ .
\label{Eq_D1'(Z^2)_def_E}
\ee
As evident from~(\ref{Eq_chi_S1S2_NP_nc_3_E}), Stokes theorem can be
used to transform each of the surface integrals in $\chi_{S_1
  S_2}^{\nprt\,nc}$ into a line integral:
\bea
        \chi_{S_1 S_2}^{\nprt\,nc}\!\!
        & = & 
        -\,\frac{\pi^2 G_2 (1\!-\!\kappa)}{3(N_c^2-1)}
        \int_{S_1} \!\!\! d\sigma_{\mu\nu}(X_1)
        \frac{\partial}{\partial Z_{\nu}}
        \!  \oint_{C_2} \!\!\! dZ_{\rho}(X_2)\,
        \delta_{\mu\rho}\,D_1^{\prime}(Z^2)
\label{Eq_chi_S1C2_NP_nc_D1'_E}\\
         & = & 
        -\,\frac{\pi^2 G_2 (1\!-\!\kappa)}{6(N_c^2-1)}
        \int_{S_1} \!\!\! d\sigma_{\mu\nu}(X_1)
        \!  \oint_{C_2} \!\!\! dZ_{\rho}(X_2)\,
        \delta_{\mu\rho}\,Z_{\nu}\,D_1(Z^2)
\label{Eq_chi_S1C2_NP_nc_D1_E}\\
        & = & 
        -\,\frac{\pi^2 G_2 (1\!-\!\kappa)}{3(N_c^2-1)}\,
        \oint_{C_1} \!\!\! dZ_{\mu}(X_1)
        \oint_{C_2} \!\!\! dZ_{\rho}(X_2)\,
        \delta_{\mu\rho}\,D_1^{\prime}(Z^2)
\label{Eq_chi_C1C2_NP_nc_D1'_E}
        \ .
\eea
With the line parametrizations of $C_1$ and $C_2$ given
in~(\ref{Eq_Ci_parameterization}) and the specification of the loop
geometries mentioned at the beginning of this appendix,
(\ref{Eq_chi_C1C2_NP_nc_D1'_E}) becomes
\bea
        && 
        \!\!\!\!\!\!\!\!
        \chi_{S_1 S_2}^{\nprt\,nc}
        = -\,\frac{\pi^2 G_2 (1-\kappa)}{3(N_c^2-1)}
\label{Eq_chi_C1C2_NP_nc_D1_1_E}\\
        &&
        \!\!\!\!\!\!\!\!
        \times
        \Bigg\{
        \cos\Theta\!\!
        \int_{-T_1}^{T_1} \!\!\! du_1 \!\!
        \int_{-T_2}^{T_2} \!\!\! du_2
        \Big[
        D_1^{\prime}(Z_{AA}^2) \!-\! D_1^{\prime}(Z_{AC}^2) 
        \!-\! D_1^{\prime}(Z_{CA}^2) \!+\! D_1^{\prime}(Z_{CC}^2) 
        \Big]
\nonumber\\
        \!\!\!\!\!\!\!\!
        &&\!\!\!\!\!\!
        +\,(\vec{r}_{1\perp}\cdot\vec{r}_{2\perp})\!\!
        \int_0^1 \!\!\! dv_1 \!\!
        \int_0^1 \!\!\! dv_2
        \Big[
        D_1^{\prime}(Z_{BB}^2) \!-\! D_1^{\prime}(Z_{BD}^2) 
        \!-\!D_1^{\prime}(Z_{DB}^2) \!+\! D_1^{\prime}(Z_{DD}^2) 
        \Big]
        \Bigg\}
\nonumber
\eea
where the following shorthand notation is used:
\be
        Z_{XY} := X_{1}^{X} - X_{2}^{Y}
        \quad \mbox{with} \quad
        X_{2}^X \in C_2^X
        \quad \mbox{and} \quad
        X_{2}^Y \in C_2^Y
        \ .
\label{Eq_Z_C1X_C2Y_E}
\ee
In the limit $R_{1,2} \ll T_{1,2} \to \infty$, the term proportional
to $\vec{r}_{1\perp}\cdot\vec{r}_{2\perp}$ on the RHS
of~(\ref{Eq_chi_C1C2_NP_nc_D1_1_E}) can be neglected and thus
(\ref{Eq_chi_C1C2_NP_nc_D1_1_E}) reduces to
\bea
        &&
        \lim_{T_1\to\infty\atop T_2\to\infty}
        \chi_{S_1 S_2}^{\nprt\,nc}
        =
        -\,\cos\Theta\,
        \frac{\pi^2 G_2 (1-\kappa)}{3(N_c^2-1)}\,
        \lim_{T_1\to\infty}\int_{-T_1}^{T_1} \!\! du_1
        \lim_{T_2\to\infty}\int_{-T_2}^{T_2} \!\! du_2
\label{Eq_chi_C1C2_NP_nc_D1_T->infty_E}\\
        &&\hspace{2.5cm}
        \times
        \Big[
        D_1^{\prime}(Z_{AA}^2) - D_1^{\prime}(Z_{AC}^2) 
        - D_1^{\prime}(Z_{CA}^2) + D_1^{\prime}(Z_{CC}^2) 
        \Big] 
        \ .
\nonumber
\eea
Here, the integrations over $u_1$ and $u_2$ can be performed
analytically proceeding analogously
to~(\ref{Eq_chi_S1S2_NP_c_u2_integration})
and~(\ref{Eq_chi_S1S2_NP_c_u1_integration}). With $T_1 = T_2 =
T/2\to\infty$, one obtains for $\sin\Theta \neq 0$
\bea
        &&\!\!\!\!\!\!
        \lim_{T\to\infty}
        \chi_{S_1 S_2}^{\nprt\,nc}
        = 
        -\,\frac{\cos\Theta}{|\sin\Theta|}\,
        \frac{\pi^2 G_2 (1-\kappa)}{3(N_c^2-1)}\,
\label{Eq_chi_C1C2_NP_nc_T->infty_S_E}\\
        && \!\!\!\!\!\!\hspace{0.5cm}
        \times
        \Big[
        D_1^{\prime\,(2)}(\vec{Z}_{AA\perp}^2) 
        - D_1^{\prime\,(2)}(\vec{Z}_{AC\perp}^2) 
        - D_1^{\prime\,(2)}(\vec{Z}_{CA\perp}^2) 
        +  D_1^{\prime\,(2)}(\vec{Z}_{CC\perp}^2) 
        \Big] 
\nonumber
\eea
and for $\sin\Theta = 0$ 
\bea
        &&\!\!\!\!\!\!
        \lim_{T\to\infty}
        \chi_{S_1 S_2}^{\nprt\,nc}
        = 
        -\,\lim_{T\to\infty}T\,
        \cos\Theta\,
        \frac{\pi^2 G_2 (1-\kappa)}{3(N_c^2-1)}\,
\label{Eq_chi_C1C2_NP_nc_T->infty_V_E}\\
        && \!\!\!\!\!\!\hspace{0.5cm}
        \times
        \Big[
        D_1^{\prime\,(3)}(\vec{Z}_{AA}^2) 
        - D_1^{\prime\,(3)}(\vec{Z}_{AC}^2) 
        - D_1^{\prime\,(3)}(\vec{Z}_{CA}^2) 
        +  D_1^{\prime\,(3)}(\vec{Z}_{CC}^2) 
        \Big]
        \ . 
\nonumber
\eea
With the identities
\be
        \vec{Z}_{AA\perp}  =  \vec{r}_{1q}-\vec{r}_{2q}\ ,\,\,\,\,
        \vec{Z}_{AC\perp}  =  \vec{r}_{1q}-\vec{r}_{2\qbar}\ ,\,\,\,\,
        \vec{Z}_{CA\perp}  =  \vec{r}_{1\qbar}-\vec{r}_{2q}\ ,\,\,\,\,
        \vec{Z}_{CC\perp}  =  \vec{r}_{1\qbar}-\vec{r}_{2\qbar}\ ,\,\,\,\,
\label{Eq_Z's_E}
\ee
one sees immediately that~(\ref{Eq_chi_C1C2_NP_nc_T->infty_S_E}) is
the result given in~(\ref{Eq_S_DD_p_npc_npnc_E})
and~(\ref{Eq_S_DD_chi_np_nc_M}) that describes the non-confining
non-perturbative contribution to the dipole-dipole scattering matrix
element $S_{DD}$.

From~(\ref{Eq_chi_C1C2_NP_nc_T->infty_V_E}), one obtains the
non-confining contribution to the static potential for $S_1 = S_2 =
S$, i.e.\ $T_1 = T_2 = T/2$, $\Theta = 0$, $r_1 = r_2 = r$,
\bea
        &&\!\!\!\!\!\!
        \lim_{T\to\infty}
        \chi_{S S}^{\nprt\,nc}
        = 
        -\,\lim_{T\to\infty}T\,\,
        \frac{\pi^2 G_2 (1-\kappa)}{3(N_c^2-1)}\,\,
\label{Eq_chi_SS_NP_nc_T->infty_V_E}\\
        && \!\!\!\!\!\!\hspace{0.5cm}
        \times
        \Big[
        D_1^{\prime\,(3)}(\vec{Z}_{AA}^2) - D_1^{\prime\,(3)}(Z_{AC}^2) 
        - D_1^{\prime\,(3)}(\vec{Z}_{CA}^2) +  D_1^{\prime\,(3)}(\vec{Z}_{CC}^2) 
        \Big]
        \ ,
\nonumber
\eea
which contributes to the self-energy of the color sources with
\bea
        \lim_{T\to\infty}
        \chi_{S S\,\mbox{\scriptsize self}}^{\nprt\,nc}
        &=& 
        -\,\lim_{T\to\infty}T\,\,
        \frac{\pi^2 G_2 (1-\kappa)}{3(N_c^2-1)}\,\,
        \Big[
        D_1^{\prime\,(3)}(\vec{Z}_{AA}^2) +  D_1^{\prime\,(3)}(\vec{Z}_{CC}^2) 
        \Big]
\nonumber\\
        &=& 
        -\,\lim_{T\to\infty}T\,\,
        \frac{2\pi^2 G_2 (1-\kappa)}{3(N_c^2-1)}\,\,
                D_1^{\prime\,(3)}(\vec{Z}_{AA}^2)
\label{Eq_chi_SS_NP_nc_T->infty_self_E}
\eea
and to the potential energy between the color sources with
\bea
        \lim_{T\to\infty}
        \chi_{S S\,\mbox{\scriptsize pot}}^{\nprt\,nc}
        &=& 
        \lim_{T\to\infty}T\,\,
        \frac{\pi^2 G_2 (1-\kappa)}{6(N_c^2-1)}\,\,
        \Big[
        D_1^{\prime\,(3)}(\vec{Z}_{AC}^2) +  D_1^{\prime\,(3)}(\vec{Z}_{CA}^2) 
        \Big]
\nonumber\\
        &=& 
        \lim_{T\to\infty}T\,\,
        \frac{\pi^2 G_2 (1-\kappa)}{3(N_c^2-1)}\,\,
        D_1^{\prime\,(3)}(\vec{Z}_{AC}^2)
        \ .
\label{Eq_chi_SS_NP_nc_T->infty_pot_E}
\eea
The last gives the non-confining contribution to the static
potential~(\ref{Eq_Vr(R)_NP_nc}).

The non-confining non-perturbative contribution to the chromo-electric
fields [$\Delta G_{\alpha \beta}^2(X)$ with $\alpha\beta=i4=4i$] can
be computed most conveniently from~(\ref{Eq_chi_S1C2_NP_nc_D1_E}) with
zero plaquette tilting angle $\Theta = 0$. The corresponding
contribution to the chromo-magnetic fields [$\Delta G_{\alpha
  \beta}^2(X)$ with $\alpha\beta=ij=ji$] is obtained for plaquette
tilting angle $\Theta = \pi/2$ and thus vanishes which can be seen
most directly from the surface
integrals~(\ref{Eq_chi_S1S2_NP_nc_1_E}). Now, we set $\Theta = 0 $ to
compute the contribution to the chromo-electric fields: Using the
surface $S_1 = S_P$ and loop $C_2 = \partial S_W$ parametrizations,
(\ref{Eq_Si_parameterization}) and~(\ref{Eq_Ci_parameterization}),
with our specification of the loop geometries, one obtains
from~(\ref{Eq_chi_S1C2_NP_nc_D1_E})
\bea
        \chi_{S_P S_W}^{\nprt\,nc}\!\!
        & = & 
        -\,\frac{\pi^2 G_2 (1\!-\!\kappa)}{3(N_c^2-1)}
        \int_{-R_P/2}^{R_P/2} \!\! du_1 
        \int_0^1 \!\! dv_1 
\label{Eq_chi_SPCW_NP_nc_D1_1_E}\\
        &&
        \times
        \left\{
        \int_{-T/2}^{T/2} \!\! du_2\,
        \left[
        (\vec{r}_{1\perp} \cdot \vec{Z}_{1A\perp})\,D_1(Z_{1A}^2)
        - (\vec{r}_{1\perp} \cdot \vec{Z}_{1C\perp})\,D_1(Z_{1C}^2)
        \right]
        \right.
\nonumber\\
        &&\hphantom{\times\,}
        \left.
        -\,
        (\vec{r}_{1\perp} \cdot \vec{r}_{2\perp})
        \int_0^1 \!\! dv_2\,
        \left[
        (\vec{r}_{1\perp} \cdot \vec{Z}_{1B\perp})\,D_1(Z_{1B}^2)
        - (\vec{r}_{1\perp} \cdot \vec{Z}_{1D\perp})\,D_1(Z_{1D}^2)
        \right]
        \right\}
\nonumber
\eea
with $T_1 = R_P/2$, $R_1 = R_P$, $T_2 = T/2$, $R_2 = R$, and the shorthand notation
\be
        Z_{1X} := X_{1} - X_{2}^{X}
        \quad \mbox{with} \quad
        X_1 \in S_1 = S_P
        \quad \mbox{and} \quad
        X_{2}^{X} \in C_2^X = \partial S_W^X
        \ .
\label{Eq_Z_SP_CWX_E}
\ee
In the limit $R \ll T \to \infty$, the term proportional to
$\vec{r}_{1\perp}\cdot\vec{r}_{2\perp}$ on the RHS
of~(\ref{Eq_chi_SPCW_NP_nc_D1_1_E}) can be neglected,
\bea
        \lim_{T \to \infty}
        \chi_{S_P S_W}^{\nprt\,nc}\!\!
        & = & 
        -\,\frac{\pi^2 G_2 (1\!-\!\kappa)}{3(N_c^2-1)}
        \int_{-R_P/2}^{R_P/2} \!\! du_1
        \int_0^1 \!\! dv_1 
        \lim_{T\to \infty}
        \int_{-T/2}^{T/2} \!\! du_2\,
\label{Eq_chi_SPCW_NP_nc_D1_T2->infty_E}\\
        &&
        \times
        \left[
        (\vec{r}_{1\perp} \cdot \vec{Z}_{1A\perp})\,D_1(Z_{1A}^2)
        - (\vec{r}_{1\perp} \cdot \vec{Z}_{1C\perp})\,D_1(Z_{1C}^2)
        \right]
        \ .
\nonumber
\eea
With an infinitesimal plaquette used to measure the chromo-electric
field, $R_1 = R_p\to 0$, the mean value theorem can be used to perform
the $u_1$ and $v_1$ integrations
in~(\ref{Eq_chi_SPCW_NP_nc_D1_T2->infty_E}). Keeping only terms up to
$\Order(R_p^2)$, this leads directly to the non-confining
non-perturbative contribution to the chromo-field distributions
$\Delta G_{\alpha \beta}^2(X)$ as given in~(\ref{Eq_Chi_PW_np_nc_14})
and~(\ref{Eq_Chi_PW_np_nc_24}).

\subsection*{\boldmath$\chi^{\pert}$ Computation}

Comparing the definition of the perturbative component
\bea
        \chi_{S_1 S_2}^{\pert} 
        & := & \frac{\pi^2}{4} 
        \int_{S_1} \!\! d\sigma_{\mu\nu}(X_1) 
        \int_{S_2} \!\! d\sigma_{\rho\sigma}(X_2)\,
        F_{\mu\nu\rho\sigma}^{\pert}(Z = X_1 - X_2)
\nonumber \\ 
        & = & 
        \frac{g^2}{4} \,
        \int_{S_1} \!\! d\sigma_{\mu\nu}(X_1) 
        \int_{S_2} \!\! d\sigma_{\rho\sigma}(X_2)
\label{Eq_chi_S1S2_P_def_E}\\ 
        && \times
        \inv{2}\Bigl[
        \frac{\partial}{\partial Z_\nu}
        \left(Z_\sigma \delta_{\mu\rho}
          -Z_\rho \delta_{\mu\sigma}\right)
        +\frac{\partial}{\partial Z_\mu}
        \left(Z_\rho \delta_{\nu\sigma}
          -Z_\sigma \delta_{\nu\rho}\right)\Bigr]\,
        D_{\pert}(Z^2)
\nonumber
\eea
with that of the non-confining non-perturbative component $\chi_{S_1
  S_2}^{\nprt\,nc}$ given in~(\ref{Eq_chi_S1S2_NP_nc_def_E}), one
finds an identical structure. Thus, accounting for the different
prefactors and the different correlation function, the results for
$\chi_{S_1 S_2}^{\pert}$ can be read off directly from the results for
$\chi_{S_1 S_2}^{\nprt\,nc}$ given above.

With $T_1 = T_2 = T/2\to\infty$ and our specification of the loop
geometries, one obtains the result for $\sin\Theta \neq 0$
from~(\ref{Eq_chi_C1C2_NP_nc_T->infty_S_E})
\bea
        &&\!\!\!\!\!\!
        \lim_{T\to\infty}
        \chi_{S_1 S_2}^{\pert}
        = 
        -\,
        \frac{\cos\Theta}{|\sin\Theta|}\,
        g^2\,
\label{Eq_chi_C1C2_P_T->infty_S_E}\\
        && \!\!\!\!\!\!\hspace{0.5cm}
        \times
        \Big[
        D_{\pert}^{\prime\,(2)}(\vec{Z}_{AA\perp}^2) - D_{\pert}^{\prime\,(2)}(Z_{AC\perp}^2) 
        - D_{\pert}^{\prime\,(2)}(\vec{Z}_{CA\perp}^2) +  D_{\pert}^{\prime\,(2)}(\vec{Z}_{CC\perp}^2) 
        \Big] 
\nonumber
\eea
and the result for $\sin\Theta = 0$
from~(\ref{Eq_chi_C1C2_NP_nc_T->infty_V_E})
\bea
        &&\!\!\!\!\!\!
        \lim_{T\to\infty}
        \chi_{S_1 S_2}^{\pert}
        = 
        -\,
        \lim_{T\to\infty}T\,
        \cos\Theta\,
        g^2\,        
\label{Eq_chi_C1C2_P_T->infty_V_E}\\
        && \!\!\!\!\!\!\hspace{0.5cm}
        \times
        \Big[
        D_{\pert}^{\prime\,(3)}(\vec{Z}_{AA}^2) - D_{\pert}^{\prime\,(3)}(Z_{AC}^2) 
        - D_{\pert}^{\prime\,(3)}(\vec{Z}_{CA}^2) +  D_{\pert}^{\prime\,(3)}(\vec{Z}_{CC}^2) 
        \Big]
        \ , 
\nonumber
\eea
where $Z_{XY}$ is defined in~(\ref{Eq_Z_C1X_C2Y_E}) and $Z_{XY\perp}$
is given explicitly in~(\ref{Eq_Z's_E}). Evidently,
(\ref{Eq_chi_C1C2_P_T->infty_S_E}) is the final result given
in~(\ref{Eq_S_DD_p_npc_npnc_E}) and~(\ref{Eq_S_DD_chi_p_M}) which
describes the perturbative contribution the dipole-dipole scattering
matrix element $S_{DD}$.

The perturbative contribution to the static potential is obtained from the expression corresponding to~(\ref{Eq_chi_SS_NP_nc_T->infty_V_E}),
\bea
        &&\!\!\!\!\!\!
        \lim_{T\to\infty}
        \chi_{S S}^{\pert}
        = 
        -\,\lim_{T\to\infty}T\,\,
        g^2\,\,
\label{Eq_chi_SS_P_T->infty_V_E}\\
        && \!\!\!\!\!\!\hspace{0.5cm}
        \times
        \Big[
        D_\pert^{\prime\,(3)}(\vec{Z}_{AA}^2) - D_\pert^{\prime\,(3)}(Z_{AC}^2) 
        - D_\pert^{\prime\,(3)}(\vec{Z}_{CA}^2) +  D_\pert^{\prime\,(3)}(\vec{Z}_{CC}^2) 
        \Big]
        \ ,
\nonumber
\eea
which contributes to the self-energy of the color sources with
\bea
        \lim_{T\to\infty}
        \chi_{S S\,\mbox{\scriptsize self}}^{\pert}
        &=& 
        -\,\lim_{T\to\infty}T\,\,
        g^2\,\,
        \Big[
        D_\pert^{\prime\,(3)}(\vec{Z}_{AA}^2) +  D_\pert^{\prime\,(3)}(\vec{Z}_{CC}^2) 
        \Big]
\nonumber\\
        &=& 
        -\,\lim_{T\to\infty}T\,\,
        2\,g^2\,\,
                D_\pert^{\prime\,(3)}(\vec{Z}_{AA}^2)
\label{Eq_chi_SS_P_T->infty_self_E}
\eea
and to the potential energy between the color sources with
\bea
        \lim_{T\to\infty}
        \chi_{S S\,\mbox{\scriptsize pot}}^{\pert}
        &=& 
        -\,\lim_{T\to\infty}T\,\,
        g^2\,\,
        \Big[
        D_\pert^{\prime\,(3)}(\vec{Z}_{AC}^2) +  D_\pert^{\prime\,(3)}(\vec{Z}_{CA}^2) 
        \Big]
\nonumber\\
        &=& 
        -\,\lim_{T\to\infty}T\,\,
        2\,g^2\,\,
        D_\pert^{\prime\,(3)}(\vec{Z}_{AC}^2)
        \ .
\label{Eq_chi_SS_P_T->infty_pot_E}
\eea
The last gives the perturbative contribution to the static
potential~(\ref{Eq_Vr(R)_color-Yukawa}).

The perturbative contribution to the chromo-magnetic fields [$\Delta
G_{\alpha \beta}^2(X)$ with $\alpha\beta=ij=ji$] vanishes while the
one to the chromo-electric fields [$\Delta G_{\alpha \beta}^2(X)$ with
$\alpha\beta=i4=4i$], for which a plaquette with $\Theta = 0$ is
needed, is obtained from the expression corresponding
to~(\ref{Eq_chi_SPCW_NP_nc_D1_T2->infty_E}),
\bea
        \lim_{T \to \infty}
        \chi_{S_P S_W}^{\pert}\!\!
        & = & 
        -\,g^2
        \int_{-R_P/2}^{R_P/2} \!\! du_1
        \int_0^1 \!\! dv_1 
        \lim_{T\to \infty}
        \int_{-T/2}^{T/2} \!\! du_2\,
\label{Eq_chi_SPCW_P_T2->infty_E}\\
        &&
        \times
        \left[
        (\vec{r}_{1\perp} \cdot \vec{Z}_{1A\perp})\,D_\pert(Z_{1A}^2)
        - (\vec{r}_{1\perp} \cdot \vec{Z}_{1C\perp})\,D_\pert(Z_{1C}^2)
        \right]
\nonumber
\eea
with $Z_{1X}$ as defined in~(\ref{Eq_Z_SP_CWX_E}). To perform the
$u_1$ and $v_1$ integrations in~(\ref{Eq_chi_SPCW_P_T2->infty_E}),
again the mean value theorem can be used since the plaquette has
infinitesimally small extensions, $R_1 = R_p\to 0$. Keeping only terms
up to $\Order(R_p^2)$, this leads directly to the perturbative
contribution to the chromo-field distribution $\Delta G_{\alpha
  \beta}^2(X)$ as given in~(\ref{Eq_Chi_PW_p_14})
and~(\ref{Eq_Chi_PW_p_24}).

\end{appendix}
%
%
%
  
%
%
%
\end{document}